\documentclass[11pt, twoside, sort, a4paper]{article}

\newcommand{\dom}{{\scalebox{0.7}{$d$}}}
\newcommand{\f}{{\scalebox{0.7}{$f$}}}

\usepackage{enumitem}
\usepackage{graphicx}
\usepackage{subcaption}
\usepackage{hyperref, bm}
\usepackage{mathtools}
\usepackage{float}

\newcommand{\myblue}[1]{\color{black}{#1}}

\usepackage{footnote}
\makesavenoteenv{tabular}
\makesavenoteenv{table}
\usepackage{geometry}
\usepackage{empheq}
 \usepackage{mathrsfs}
 \usepackage{tikz}

\AtBeginDocument{
\addtolength{\abovedisplayskip}{-0.5ex}
\addtolength{\abovedisplayshortskip}{-0.5ex}
\addtolength{\belowdisplayskip}{-0.5ex}
\addtolength{\belowdisplayshortskip}{-0.5ex}
}
\makeatletter
\newcommand{\subalign}[1]{%
  \vcenter{%
    \Let@ \restore@math@cr \default@tag
    \baselineskip\fontdimen10 \scriptfont\tw@
    \advance\baselineskip\fontdimen12 \scriptfont\tw@
    \lineskip\thr@@\fontdimen8 \scriptfont\thr@@
    \lineskiplimit\lineskip
    \ialign{\hfil$\m@th\scriptstyle##$&$\m@th\scriptstyle{}##$\crcr
      #1\crcr
    }%
  }
}

\usepackage{amsmath, amsthm, amssymb}
\usepackage{caption}
\usepackage{tikz}
\usetikzlibrary{matrix, positioning, calc}
\usepackage{multirow, nccmath}
\usepackage{algorithm}
\usepackage{algorithmic}

\usepackage{comment}
\usepackage{color}
\usepackage[titletoc,title]{appendix}

\newcommand{\red}[1]{{\color{black}{#1}}}

\newcommand{\apurple}[1]{\color{black}{#1}}
\newcommand{\dpurple}[1]{\color{black}{#1}}

\usepackage[running, mathlines]{lineno}

\newcommand{\EQ}{\begin{equation}}
\newcommand{\EN}{\end{equation}}
\newcommand{\EQS}{\begin{equation*}}
\newcommand{\ENS}{\end{equation*}}
\newcommand{\EQA}{\begin{eqnarray}}
\newcommand{\ENA}{\end{eqnarray}}
\newcommand{\EQAS}{\begin{eqnarray*}}
\newcommand{\ENAS}{\end{eqnarray*}}

\usepackage{accents}
\newlength{\dhatheight}

\newcommand{\Ebb}{\mathbb{E}}

\numberwithin{equation}{section}
\numberwithin{table}{section}
\numberwithin{figure}{section}

\newtheorem{remark}{Remark}

\numberwithin{definition}{section}
\numberwithin{theorem}{section}
\numberwithin{lemma}{section}
\numberwithin{remark}{section}
\numberwithin{assumption}{section}
\numberwithin{condition}{section}
\numberwithin{property}{section}
\numberwithin{proposition}{section}
\numberwithin{corollary}{section}
\numberwithin{algorithm}{section}

\makeatletter
\renewcommand*\env@matrix[1][c]{\hskip -\arraycolsep
  \let\@ifnextchar\new@ifnextchar
  \array{*\c@MaxMatrixCols #1}}
\makeatother

\makeatletter
\def\thm@space@setup{\thm@preskip=3pt
\thm@postskip=3pt}
\makeatother
\usepackage{titlesec}
\titlespacing*{\section}
  {0pt}{0.6\baselineskip}{0.2\baselineskip}
\titlespacing*{\subsection}
  {0pt}{0.6\baselineskip}{0.2\baselineskip}
\titlespacing*{\subsubsection}
  {0pt}{0.6\baselineskip}{0.2\baselineskip}

\usepackage[normalem]{ulem}
\usepackage{cancel}

\DeclareMathOperator{\T}{{\scalebox{0.6}{T}}}
\begin{document}

\title{Money-Back Tontines for Retirement Decumulation: Neural-Network Optimization under Systematic Longevity Risk}
\author{
German Nova Orozco \thanks{School of Mathematics and Physics, The University of Queensland, St Lucia, Brisbane 4072, Australia,
email: \texttt{g.novaorozco@student.uq.edu.au}
}
\and
Duy-Minh Dang\thanks{School of Mathematics and Physics, The University of Queensland, St Lucia, Brisbane 4072, Australia,
email: \texttt{duyminh.dang@uq.edu.au}
}
\and
Peter A. Forsyth\thanks{David R. Cheriton School of Computer Science, University of Waterloo, Waterloo ON, N2L 3G1, Canada,
email: \texttt{paforsyt@uwaterloo.ca}
}
}
\date{\today}
\maketitle

\begin{abstract}
Money-back guarantees (MBGs) address bequest concerns in pooled
retirement income products by returning the initial purchase price through
withdrawals or, after early death, through a benefit to the member's
beneficiaries or estate. We study the distinct actuarial problem created by
adding an MBG to an individual-account tontine with dynamic withdrawals,
investment in domestic and foreign assets, and systematic longevity risk.
The retiree trades expected withdrawals (EW) against the lower-tail Conditional
Value-at-Risk (CVaR) of terminal wealth under a fixed-horizon plan-to-live
convention. Neural networks parameterize admissible withdrawal and
rebalancing controls; the MBG is then valued ex post under the learned policy
through an equivalent up-front load combining the expected payout with an
upper-tail CVaR prudential buffer. We also approximate the effect of
contract pooling on per-contract tail risk and pricing.

Using long-horizon market and mortality data calibrated for an
Australian retiree, we find that expected MBG payouts are modest,
while the representative-contract payout has a severe upper tail. Contract pooling substantially reduces the upper-tail exposure on a
per-contract basis and lowers the approximate pooled-contract load.
International diversification materially improves the EW--CVaR
retirement-income trade-off, while affecting the MBG payout
distribution and equivalent load only modestly. Stochastic mortality likewise
has a modest effect on the efficient frontier and MBG pricing.

\vspace{.1in}


\vspace{.1in}

\end{abstract}
\section{Introduction}
\label{sec:intro}
The worldwide shift from Defined Benefit (DB) pensions to Defined Contribution (DC)
arrangements is well documented \cite{OECD2019PensionMarkets, ThinkAhead2023GlobalPensionAssets}.
While DC plans offer flexibility and portability, they also transfer much of the responsibility for managing key retirement risks to individuals.
In particular, upon retirement, a DC member faces the \emph{decumulation problem}:
how to invest and draw down accumulated savings to
support sustainable real (inflation-adjusted) spending under longevity risk and uncertain market returns \cite{MacDonaldEtAl2013, bernhardt2019modern, bar2023products}. Rapid
global population ageing further intensifies these challenges
\cite{un2024datasources}.

To manage longevity risk in practice, many retirees rely on simple spending and asset allocation heuristics, most notably the ``4\% rule'' \cite{bengen1994}
(often paired with a fixed stock--bond mix) and performance-based adjustments
to withdrawals and/or portfolio weights (e.g.\ \cite{GuytonKlinger2006Withdrawal}).
These rules are transparent and easy to implement, and this reliance on rule-based decumulation is consistent with the evidence in \cite{Anarkulova_2022_a},
which reports a revealed preference for spending rules among retirees and wealth advisors in DC drawdown settings.
However, because these rules are not derived from a risk--reward optimization, they can be far from efficient under realistic market and wealth conditions (see, e.g.\
\cite{ARVA2020,forsyth2022stochastic}).

As an alternative to rules-based drawdown strategies, retirees can transfer longevity risk to an insurer by purchasing a life annuity. In practice, voluntary annuitization remains limited (e.g.\ \cite{PeijnenburgEtAl2016}),
and low demand for life annuities can be rational once bequest motives, illiquidity costs, and product design are taken into account~\cite{MacDonaldEtAl2013}.

In parallel, there has been renewed interest in \emph{pooled} retirement income products that share longevity risk among members while offering greater transparency and flexibility than traditional annuities. Modern tontines and related survivor pools achieve longevity pooling by redistributing the accounts of deceased members to survivors, generating \emph{mortality credits} that can support higher sustainable real (inflation-adjusted) withdrawals for surviving members \cite{donnelly2014bringing, donnelly2015actuarial, milevsky2015optimal, Fullmer2019Tontines}. The appeal comes with a clear trade--off: mortality credits raise payouts conditional on survival, but wealth is \mbox{typically forfeited at death.}

More broadly, mean-field game and mean-field control formulations
provide a population-level approach in which individual decisions interact with
the evolving population distribution and may accommodate heterogeneity across
members
\cite{carmona2021deepmfg,phamwarin2023meanfieldnn}. By contrast, the
homogeneous large-pool framework considered in
\cite{forsyth2024optimal} reduces the pooling mechanism to a
representative-member problem. In that setting, a stochastic
optimal control framework is developed for a pure individual-account tontine
with deterministic mortality and a two-asset stock--bond investment universe,
but without a \mbox{death-benefit guarantee.}

Within the representative-member framework of
\cite{forsyth2024optimal}, the retiree has full control over both the
withdrawal amount (subject to minimum and maximum constraints) and the asset
allocation in the account. The optimization objective is defined using a
risk--reward criterion: reward is measured by expected cumulative
real (inflation-adjusted) withdrawals (EW) over a fixed 30-year
retirement horizon, while risk is measured by the lower-tail
Conditional Value-at-Risk (CVaR) of terminal real
(inflation-adjusted) wealth. The formulation assumes that the retiree
survives to the horizon, following the ``plan to live, not to die'' convention;
see, e.g.\ \cite{pfau2018overview}. The resulting control problem is solved numerically using a grid-based method for the associated partial integro-differential equation (PIDE).

The results in \cite{forsyth2024optimal} show that longevity
pooling can materially improve the withdrawal--risk trade--off relative to
non--pooled drawdown strategies and simple constant withdrawal/allocation
benchmarks. In particular, at comparable and economically relevant levels of lower-tail terminal-wealth CVaR, the tontine overlay delivers substantially higher expected cumulative real withdrawals, even after allowing for ``fees of the order of
50--100 basis points (bps) per year'' \cite[Section~1]{forsyth2024optimal}.

This efficiency gain, however, comes at the cost of forfeiting account wealth
upon death, which can limit the appeal of ``pure'' tontine overlays for retirees
with bequest or estate considerations \cite{forsyth2024optimal}. Reflecting this
practical constraint, recent industry innovation---particularly in the
Australian superannuation market---has introduced \emph{money-back guarantees}
(MBGs), sometimes described in product disclosures as ``money-back
protection'', for pooled retirement income products (e.g.\ QSuper and MyNorth
\cite{qsuper2025pds, MyNorth2025}).

Under such protection, if the member dies before cumulative
retirement-income payments reach the protected purchase price, the resulting
shortfall is paid to the member's beneficiaries or estate. This
return-of-purchase-price structure differs from the tontine-with-bequest
design in \cite{bernhardt2019modern}, which uses a separate bequest account.
Although \cite{forsyth2024optimal} notes that an MBG could be added as an
overlay and priced separately, it does not undertake that valuation.

From the perspective of the entity that ultimately bears the guarantee, whether
an insurer or the retirement-income pool, the MBG is a contingent,
right-skewed death-benefit liability: it is often zero, but can be large when
triggered. Accordingly, the economic cost of an MBG depends not only on
expected payouts, but also on tail risk and any prudential loading or
reserving rule applied to adverse outcomes. Crucially, the distribution of
guarantee payouts is endogenous to the decumulation policy: the
withdrawal and asset-allocation controls jointly determine the account and
withdrawal paths, while cumulative nominal withdrawals determine the guarantee
shortfall at death. At the portfolio level, contract-specific death-timing risk may
diversify across broadly homogeneous MBG contracts, so the per-contract
economic cost can differ from the stand-alone representative-contract
valuation.

This creates an actuarial pricing problem absent from the
pure-tontine analysis. The MBG must be valued under the decumulation policy it
accompanies, even though the death benefit is paid outside the member's account
and does not feed back into the withdrawal and rebalancing decisions under the
plan-to-live convention.

From a modelling perspective, two aspects are central in tontine decumulation: the specification of the asset market and the treatment of mortality risk.
While most individual account decumulation models assume a domestic stock--bond portfolio, in practice, for retirees, the investable universe is inherently richer and often includes international assets \cite{solnik1974not, de1997international}. International diversification is particularly relevant for investors in countries where the local equity market capitalization is small relative to the global market.
Extending the individual account decumulation setting to allow international diversification is economically important, since it can affect the risk--return trade-offs and, in turn, optimal withdrawal and asset allocation decisions.

Expanding the asset universe increases the dimensionality of both
the state and control spaces, making grid-based dynamic programming
computationally prohibitive.
Recent work in portfolio optimization and related control problems has shown that neural network (NN) parameterizations can approximate
high-dimensional, state-dependent controls without requiring a state-space
grid, thereby mitigating the dimensionality limitations of grid-based
methods \cite{li2019data, ni2022optimal, chen2023benchmark}.

A second modelling aspect is mortality at the pool level: deaths determine the redistribution to surviving members and hence the mortality credits. Following the discussion in \cite{forsyth2024optimal}, an important caveat is systematic mortality risk (e.g.\ unexpected improvements in life expectancy), which is typically ignored when mortality credits are computed from a fixed life table. To capture this source of uncertainty, stochastic mortality specifications such as Lee--Carter (LC) \cite{lee1992modeling} and Cairns--Blake--Dowd (CBD) \cite{cairns2006two} can be used to generate time-varying one-year death probabilities and hence stochastic mortality credits. If mortality improves unexpectedly, realized mortality credits can be lower than anticipated because lighter realized mortality implies fewer deaths and hence less redistribution to survivors.
This distinguishes idiosyncratic longevity risk, which is diversifiable within the pool, from systematic longevity risk, which is non-diversifiable.

Motivated by the above observations, this paper sets out to achieve three
primary objectives. First, building on the pure individual-account
tontine setting of \cite{forsyth2024optimal}, we add a
return-of-purchase-price MBG overlay and treat the resulting death benefit as a
contingent liability with a policy-dependent payout distribution. We
value this liability ex post under that policy. Because the
MBG may be funded through different operational mechanisms, we report its
economic cost as an implementation-agnostic equivalent up-front load rather
than specifying a particular fee schedule. The load combines the expected
payout under the physical measure with an explicit upper-tail CVaR prudential
buffer. {\dpurple{Using the representative-contract payout distribution, we
then develop a practical approximation to the effect of contract pooling on
per-contract tail risk and pricing.}}

Second, we extend the representative-member EW--CVaR control
framework of \cite{forsyth2024optimal} from a two-asset setting with
deterministic mortality to one with domestic and foreign assets and mortality
credits determined under either deterministic or stochastic mortality. We
compute the wealth- and time-dependent withdrawal and rebalancing controls using NN
parameterizations rather than the grid-based PIDE method used in
\cite{forsyth2024optimal}. The optimization follows a fixed-horizon plan-to-live convention; the 30-year horizon extends from age 65 to age 95, providing a prudent stress test of the sustainability of real (inflation-adjusted) retirement spending.

Third, using nearly a century of market data together with Australian
mortality data, we quantify how international diversification at the
individual-account level and systematic mortality at the pool level affect
EW--CVaR frontiers and MBG pricing. In particular, we measure the magnitude of
the frontier improvement relative to the domestic-only portfolio, examine the
state-dependent use of domestic and foreign assets, and determine how the
resulting controls and mortality specifications affect MBG payout distributions
and equivalent loads. We also assess the relative importance of expected
guarantee costs and upper-tail prudential buffers. In addition, we quantify how the approximate per-contract upper-tail risk and
equivalent load vary with MBG contract-book size.

Using an Australian-investor calibration with Australian and U.S.\
equity and government-bond indices, the numerical results show that expected
MBG payouts are modest, while the representative-contract payout has a severe
upper tail. Under the pooling approximation, upper-tail exposure is
substantially reduced on a per-contract basis, lowering the approximate pooled-contract load.
International diversification materially improves the EW--CVaR
retirement-income trade-off, with the largest frontier gains when combined
with the tontine overlay.
{\dpurple{Stochastic mortality has only a modest effect on the efficient
frontier, while both international diversification and stochastic mortality
have limited effects on representative-contract MBG pricing.}}

The remainder of the paper is organized as follows. Section~\ref{sc:mbg}
introduces the MBG overlay and its payoff and timing conventions, while
Section~\ref{sc:tontine_modeling} summarizes the representative-member tontine
mechanics and deterministic and stochastic mortality inputs.
Section~\ref{sc:modeling} presents the stochastic control framework for
individual-account decumulation, and Section~\ref{sc:formulation} formulates
the associated EW--CVaR optimization problem. Section~\ref{sec:NNs} develops
the NN-based computational approach. {\dpurple{Section~\ref{sc:mbg_pricing}
develops representative-contract MBG pricing, and
Section~\ref{sc:mbg_contract_pooling} introduces the contract-pooling
approximation. Section~\ref{sec:diverse_port} presents the market and mortality data and numerical results on efficient frontiers, MBG pricing, and contract pooling.}}
Section~\ref{sc:conclude} concludes and outlines directions for future
research. 
\section{Money-back guarantee overlay}
\label{sc:mbg}
\begin{center}
\begin{minipage}{0.8\textwidth}
``\emph{Your purchase price is always paid back as either income to you or a death benefit paid to your beneficiaries.}''
\end{minipage}

\vspace*{+0.25cm}
\hspace*{+5cm} QSuper's Lifetime Pension Product Disclosure Statement \cite{qsuper2025pds}
\end{center}
In a pure individual-account tontine, account wealth is forfeited at death and
redistributed within the pool. A money-back guarantee (MBG) changes the
actuarial accounting: early death may generate a death-contingent benefit
payable to the member's beneficiaries or estate. Thus, the problem is no longer
only how mortality credits support withdrawals while the member is alive, but
also how to value and express the cost of the death-contingent shortfall
determined by the member's cumulative withdrawals before death and the timing of
death. This guarantee overlay is the main product-design departure from the
pure tontine framework of \cite{forsyth2024optimal}.

The MBG is motivated by recent Australian pooled retirement income products,
where similar features are described as money-back protection; examples include
QSuper's Lifetime Pension and MyNorth Lifetime
\cite{qsuper2025pds, MyNorth2025}. We model this feature as an overlay on a
tontine retirement account, under which any shortfall between the purchase price
and cumulative withdrawals at death is paid as a death benefit to the member's
beneficiaries or estate.


\subsection{Guarantee payoff and timing}
\label{ssc:mbg_payoff_timing}
We define the MBG payout on the set of equally spaced decision times in $[0,T]$,
denoted by $\mathcal T$:
\begin{equation}
\label{eq: T_M}
\mathcal{T}=\{t_m \mid t_m=m\Delta t,\ m=0,\ldots,M\},
\qquad
\Delta t=T/M,
\end{equation}
where $t_0=0$ is the inception time and $T>0$ is the finite investment horizon.
Throughout, time is measured in years and we take annual decision times, so
$\Delta t=1$ and $T=M$. For notational convenience, define
$t_m^- = t_m-\varepsilon$ and $t_m^+ = t_m+\varepsilon$, with
$\varepsilon\to 0^+$, as the instants immediately before and after $t_m$,
respectively.

At inception, suppose a member invests an amount $L_0$ as the purchase price of
the retirement product with MBG. Throughout this paper, $L_0$ denotes a
nominal (unadjusted for inflation) dollar amount fixed at inception. At each
decision time $t_m$, $m=0,\ldots,M-1$, while the member is alive, the regular
tontine account mechanics apply: mortality credits are distributed, a real
(inflation-adjusted) withdrawal $q_m$ is made, and the portfolio is rebalanced.
Once cumulative withdrawals, converted to nominal (unadjusted for inflation)
terms, reach or exceed $L_0$, the MBG becomes inactive.

Let $m_{\tau}$ denote the interval index associated with death before the
horizon, so that $\tau\in(t_{m_{\tau}-1}^+,t_{m_\tau}^-]$, for some
$m_\tau\in\{1,\ldots,M\}$. If the member dies in this interval, no further
withdrawal occurs; in particular, the scheduled withdrawal at $t_{m_\tau}$ is
not made if such a withdrawal would otherwise be scheduled. If cumulative
withdrawals made up to and including $t_{m_{\tau}-1}$, converted to nominal
(unadjusted for inflation) terms, fall short of $L_0$, the MBG activates. If
the member survives to the horizon $T=t_M$, we set $m_\tau=M+1$ and the MBG
payout is zero. The nominal (unadjusted for inflation) MBG payout is
\begin{equation}
\label{eq:mbg}
\text{MBG-payout}
=
\begin{cases}
\displaystyle
\max\bigg( L_0 - \sum_{\ell = 0}^{m_{\tau}-1} q_\ell\,
\frac{\mathrm{CPI}_{\ell}}{\mathrm{CPI}_{0}}, \, 0 \bigg),
& m_\tau\le M,\\[10pt]
0,
& m_\tau=M+1.
\end{cases}
\end{equation}
Here, $\mathrm{CPI}_\ell$ is the consumer-price index at decision time $t_\ell$
(and $\mathrm{CPI}_{0}$ is the index level at inception). Because the guarantee
compares nominal cash flows, each real (inflation-adjusted) withdrawal $q_\ell$
is first expressed in nominal terms by multiplying by
$\mathrm{CPI}_{\ell}/\mathrm{CPI}_{0}$ before being summed in
\eqref{eq:mbg}.\footnote{When $m_\tau\le M$, this nominal shortfall is converted
to a real (inflation-adjusted) amount at inception by multiplying by
$\mathrm{CPI}_0/\mathrm{CPI}_{m_\tau}$ for valuation and reporting; see
Section~\ref{sc:mbg_pricing} (Eqn.~\eqref{eq:Z_g}).}
Under this accounting convention, all withdrawals made before death
are included in the cumulative nominal withdrawals in \eqref{eq:mbg}.

{\dpurple{The member's account balance at death is treated separately from the
MBG calculation: any outstanding negative balance is not deducted from the MBG
payout, while any positive balance is forfeited to the tontine pool and supports
mortality credits for surviving members. This differs from a variable-annuity
death-benefit guarantee, for which the deceased member's remaining investment
account is available to fund the benefit. The MBG death benefit must therefore
be financed separately from the deceased member's tontine account, whether
through the pool, external insurance, or another funding arrangement.}}
\begin{remark}[Timing convention]
\label{rm:mbg_convention}
In the discrete-time model, if a member dies at time
$\tau\in(t_{m_\tau-1}^{+},t_{m_\tau}^{-}]$, the MBG payout is made at the end of that interval, i.e.\ at the next decision time $t_{m_\tau}$,
to  the member's beneficiaries or estate. The scheduled withdrawal at $t_{m_\tau}$, if any,  is not made. This convention fixes only the death interval and the timing of the MBG
payout; the treatment of the deceased member's account within the tontine pool
is specified separately in Section~\ref{sc:tontine_modeling}.
\end{remark}
The MBG does not change the account dynamics or the controls chosen for the
individual tontine account while the member is alive. It is a contingent overlay
whose cost is computed under the withdrawal and rebalancing policy used for that
account. This separation is important: the guarantee does not feed back into the
member's withdrawal or rebalancing decisions in our formulation, but its payout
distribution is endogenous to those decisions.

In practice, the MBG may be subject to regulatory constraints, such as
Australia's Capital Access Schedule (CAS), which limits the refundable portion of
a pension's purchase price. We do not impose the CAS in the baseline model in
order to isolate the core MBG overlay and pricing mechanism. As a result, the
recoverable amount under \eqref{eq:mbg} may overstate the amount payable under a
regulated product when CAS limits bind.

\subsection{Equivalent MBG load}
\label{ssc:MBG_fee}
Product disclosures for pooled retirement income products typically emphasize
the presence of money-back protection, but the funding mechanism need not appear
as a stand-alone member-level charge. The cost may instead be reflected through
pool-level self-insurance, an external insurance arrangement, or adjustments to income rates.

Accordingly, we summarize the economic cost of the MBG using an equivalent
up-front load factor $f_g\ge 0$.
The quantity $f_gL_0$ is interpreted as a
one-time cost measure, or equivalently, a proportional reduction in a notional
starting income rate or benefit base, that represents the MBG cost under the pricing rule adopted in this paper. This translation device reports the guarantee
cost in a form that is comparable across operational implementations. Full
details of the pricing methodology and numerical implementation are provided in
Section~\ref{sc:mbg_pricing}.

\section{Tontine mechanics and mortality inputs}
\label{sc:tontine_modeling}
This section recalls the individual-account tontine framework of
\cite{forsyth2024optimal}, which provides the account dynamics underlying the
MBG overlay. In particular, we state the homogeneous, large-pool
representative-member mortality-credit rule from that framework. We then specify
the deterministic and stochastic mortality inputs needed for both the
mortality-credit amounts in the stochastic-control formulation and the death
times used in MBG valuation.

\subsection{Representative-member tontine gain rule}
\label{ssc:tontine}
The tontine-account mechanics use the decision-time grid
$\mathcal T=\{t_m\}_{m=0}^M$ defined in \eqref{eq: T_M}. For any
time-dependent quantity $f$, we use the shorthand
$f_{m^-}:=f(t_m^-)$ and $f_{m^+}:=f(t_m^+)$.

The tontine pool consists of {\dpurple{$L$}} members, indexed by {\dpurple{$\ell=1,\ldots,L$}}.
For a solvent account, the account update at each decision time
$t_m \in \mathcal{T}$ is ordered as
$t_m^{-}\;\longrightarrow\;t_m\;\longrightarrow\;t_m^{+}$.
\begin{itemize}[noitemsep, topsep=1pt, leftmargin=*]
  \item $t_m^{-}$, $m = 1, \ldots, M$: end-of-period portfolio balances are observed. Within the tontine account, balances of members who died in
$[t_{m-1}^+, t_m^-]$ are forfeited, and the corresponding mortality
credits are distributed to surviving members before withdrawals and rebalancing.

  \item $t_m$ and $t_m^+$, $m = 0, \ldots, M-1$: each surviving member withdraws an amount $q_m$ at $t_m$ and then rebalances their portfolio at $t_m^+$.
\end{itemize}
Between $t_{m-1}^{+}$ and $t_m^{-}$, $m = 1,\ldots,M$, asset returns and mortality evolve over the period. At $t_M=T$, the portfolio is liquidated with no withdrawal or rebalancing.

Let {\dpurple{$\bar{W}_{m^-}^{\ell}$}} denote the real (inflation-adjusted) account balance of member
{\dpurple{$\ell$}} immediately before any mortality-credit distribution at time $t_m^-$,
$m=0,\ldots,M$. Over each period $[t_{m-1}^{+},t_m^{-}]$, $m=1,\ldots,M$, mortality outcomes are realized: if
member {\dpurple{$\ell$}} dies during the period, their account balance is forfeited; if member
{\dpurple{$\ell$}} survives to $t_m^-$, they receive a mortality credit {\dpurple{$c_m^\ell$}} before the
scheduled withdrawal at $t_m$.

For $m=1,\ldots,M$ and {\dpurple{$\ell=1,\ldots,L$}}, let {\dpurple{$\delta_{m-1}^{\ell}$}} denote the
conditional one-period death probability of member {\dpurple{$\ell$}} over
$[t_{m-1}^{+},t_m^{-}]$, given survival to $t_{m-1}^{+}$:
\begin{equation}
\label{eq:prob_delta_m}
{\dpurple{
\delta_{m-1}^{\ell}
=
\mathbb P\!\left(
\text{member $\ell$ dies during }[t_{m-1}^+,t_m^-]
\,\middle|\,
\text{member $\ell$ is alive at }t_{m-1}^+
\right).
}}
\end{equation}
Thus, {\dpurple{$1-\delta_{m-1}^{\ell}$}} is the corresponding conditional probability that
member {\dpurple{$\ell$}} survives to $t_m^-$.

In \cite{forsyth2024optimal}, the mortality-credit rule is obtained by combining an individual fairness condition with a pool-level budget constraint. For finite heterogeneous pools, exact ex-post budget balance can be enforced through a group-gain adjustment; under the large-pool/small-bias approximation used below, this adjustment is set equal to one. Equivalently, no single member's expected mortality credit is material relative to aggregate expected forfeitures.
The large-pool mortality-credit rule used here is
\begin{equation}
{\dpurple{
  c_m^{\ell}= \frac{\delta_{m-1}^{\ell}}{1-\delta_{m-1}^{\ell}}\,\bar{W}_{m^-}^{\ell},
  \qquad m=1,\ldots,M,\quad \ell=1,\ldots,L.
}}
   \label{eq:tontine-gain-rule-j}
\end{equation}
The derivation and the finite-pool group-gain construction are given in \cite{forsyth2024optimal}; see also \cite{sabin2016analytics}. In the present paper, \eqref{eq:tontine-gain-rule-j} serves as the large-pool mortality-credit rule for the subsequent stochastic-control and MBG pricing formulation.

In the optimal control problem, we suppress the member superscript {\dpurple{$\ell$}} and work with a representative surviving member in a homogeneous pool. The mortality credit can then be written as
\begin{equation}
c_m = g_m\, \bar{W}_{m^-},
\quad \text{ where} \quad
   g_m = \frac{\delta_{m-1}}{1 - \delta_{m-1}}.
  \label{eq:tontine-gain}
\end{equation}
The tontine gain rate $g_m$ is the proportional mortality-credit uplift received
by a surviving member over $[t_{m-1}^+,t_m^-]$.
Thus the mortality-credit component of the subsequent wealth recursion is
determined by the annual death probabilities
$\{\delta_{m-1}\}_{m=1}^M$ and the
associated gain rates $\{g_m\}_{m=1}^M$.

\subsection{Mortality inputs}
\label{ssc:mortality_models}
The annual death-probability sequence $\{\delta_{m-1}\}_{m=1}^M$ is the
mortality input for both the representative-member mortality-credit rule and the
death-time simulation used in MBG valuation. We now describe how this sequence
is specified under deterministic mortality and generated under stochastic
mortality.

%
%

\subsubsection{Deterministic mortality}
\label{ssc:determ_mort}

In the deterministic case, we work with a fixed period life table, such as the
Canadian Pensioner Mortality Tables or Australian mortality from the Human
Mortality Database (HMD) \cite{HMD}. Let $q_{a,y_0}$ denote the one--year
death probability for an individual aged $a$ in the selected period table,
where $y_0$ denotes its fixed reference year. Let the retiree be aged $a_0$ at
retirement. In our tontine framework, the decision times are measured in years
since retirement, so $t_m=m$ for $m=0,\ldots,M$, with $t_0=0$. At decision
time $t_m$, the retiree is aged $a_0+m$.

Holding the period-table year fixed at $y_0$, we define the corresponding
one--year conditional death probabilities by
\begin{equation}
\label{eq:delta_m}
  \delta_{m-1}
  \;:=\;
  q_{a_0+(m-1),\,y_0},
  \qquad m=1,\ldots,M.
\end{equation}

The deterministic sequence $\{\delta_{m-1}\}_{m=1}^M$ then determines the
corresponding tontine gain rates
$\{g_m\}_{m=1}^M$, where
$g_m=\delta_{m-1}/(1-\delta_{m-1})$.

\subsubsection{Stochastic mortality}
\label{ssc:stoch_mort}
To incorporate systematic longevity risk, we allow the mortality surface to be
generated by a stochastic model in the generalized age--period--cohort (GAPC)
family \cite{villegas2018stmomo}. Let $D_{a,y}$ denote the number of
deaths at age $a$ in calendar year $y$ and $E^0_{a,y}$ the corresponding initial
exposure. We assume
$D_{a,y}\sim\mathrm{Binomial}\bigl(E^0_{a,y},q_{a,y}\bigr)$, where $q_{a,y}$
is the one--year death probability. Its systematic component is captured by a
linear predictor
\begin{equation}
  \eta_{a,y}
  =
  \alpha_a
  + \sum_{i=1}^{\dpurple{G}} \beta_a^{(i)}\,\kappa_y^{(i)}
  + \beta_a^{(0)}\,\gamma_{y-a},
  \label{eq:gapc-eta}
\end{equation}
where $\alpha_a$ describes the age profile, $\kappa_y^{(i)}$ are period factors
and $\gamma_{y-a}$ is a cohort effect{\dpurple{; $G$ denotes the number of GAPC period factors}}. For the numerical implementation
used in this paper, the linear predictor is related directly to the one--year
death probability through the logit link $\mathrm{logit}\,q_{a,y}=\eta_{a,y}$,
with $\mathrm{logit}(u)=\log\bigl(u/(1-u)\bigr)$ for $u\in(0,1)$.
In this general GAPC formulation the cohort term is optional; the specific
Lee--Carter \cite{lee1992modeling} and Cairns--Blake--Dowd
\cite{cairns2006two} models used below are special cases that omit the cohort
component.

\subsubsection*{Lee--Carter (LC) model}
In the Lee--Carter specification, the predictor has a single age--period term,
\begin{equation}
  \eta_{a,y} = \alpha_a + \beta_a\,\kappa_y,
  \label{eq:lc-eta}
\end{equation}
and we set $\mathrm{logit}\,q_{a,y}=\eta_{a,y}$. The period index
$\kappa_y$ captures the overall mortality level and is commonly modelled as a
random walk with drift,
\begin{equation}
  \kappa_y = d_\kappa + \kappa_{y-1} + \xi_y,
  \qquad
    \xi_y \overset{\mathrm{i.i.d.}}{\sim} {\dpurple{\mathcal{N}}}(0,\sigma_\kappa^2).
  \label{eq:lc-rw}
\end{equation}

\subsubsection*{Cairns--Blake--Dowd (CBD) model}
The Cairns--Blake--Dowd model \cite{cairns2006two} describes mortality as
approximately linear in age around a reference age $\bar a$:
\begin{equation}
  \eta_{a,y}
  = \kappa_y^{(1)} + (a-\bar a)\,\kappa_y^{(2)}.
  \label{eq:cbd-eta}
\end{equation}
As in the LC specification, we set
$\mathrm{logit}\,q_{a,y}=\eta_{a,y}$.
The bivariate period factor
$\boldsymbol{\kappa}_y=(\kappa_y^{(1)},\kappa_y^{(2)})^\top$
is usually specified as a random walk with drift,
\begin{equation}
  \boldsymbol{\kappa}_y
  =
  \boldsymbol{d}_\kappa
  + \boldsymbol{\kappa}_{y-1}
  + \boldsymbol{\xi}_y,
  \qquad
   \boldsymbol{\xi}_y
  \overset{\mathrm{i.i.d.}}{\sim}
  {\dpurple{\mathcal{N}}}(\mathbf 0,\Sigma_\kappa).
  \label{eq:cbd-rw}
\end{equation}
For both the LC and CBD specifications, model estimation, including the required
normalizations and any model-specific identifiability constraints, follows the
standard \textsf{StMoMo} implementation \cite{villegas2018stmomo}.

\paragraph{Death-probability inputs.}
Once an LC or CBD model has been calibrated to historical deaths and
initial exposures, its fitted and projected period factors determine a
surface of one--year death probabilities $\{q_{a,y}\}$.
Let $y_{\mathrm{proj}}$ denote the first projected calendar year. For a
representative retiree aged $a_0$ at the start of year
$y_{\mathrm{proj}}$, with annual decision times $t_m=m$, we define
\begin{equation}
\label{eq:delta_m_sto}
  \delta_{m-1}
  \;:=\;
  q_{a_0+(m-1),\,y_{\mathrm{proj}}+(m-1)},
  \qquad m=1,\ldots,M.
\end{equation}
In the homogeneous pool considered here,  we set
{\dpurple{$\delta_{m-1}^{\ell}\equiv\delta_{m-1}$}} for all members~{\dpurple{$\ell$}}. Unlike the fixed-period
deterministic specification, the stochastic sequence advances in both age and
projected calendar year.
These probabilities feed directly into the tontine gain rate
$g_m=\delta_{m-1}/(1-\delta_{m-1})$ in \eqref{eq:tontine-gain}. Thus the
tontine mechanics developed above apply unchanged; only the sequence
$\{\delta_{m-1}\}$ differs between deterministic life--table mortality and
stochastic GAPC-based mortality.

\begin{remark}[Mortality inputs in simulation]
\label{rm:stochastic_mortality}
The same one--year conditional death probabilities also underpin our simulation
framework. Let $n$ index a simulated path. In the deterministic case, the
sequence $\{\delta_{m-1}\}_{m=1}^M$ is fixed, so
$\delta_{m-1}^{(n)}\equiv\delta_{m-1}$ for all $n$. Under stochastic LC or CBD
mortality, each simulated mortality surface $\{q_{a,y}^{(n)}\}$ yields
\begin{equation}
\label{eq:delta_mk}
  \delta_{m-1}^{(n)}
  \;:=\;
  q^{(n)}_{a_0+(m-1),\,y_{\mathrm{proj}}+(m-1)},
  \qquad m=1,\ldots,M.
\end{equation}
These probabilities determine the pathwise tontine gain rates through
\eqref{eq:mod_Tg}; when the account is solvent, the corresponding gain rate is
$\delta_{m-1}^{(n)}/(1-\delta_{m-1}^{(n)})$. They enter the wealth recursion
in the NN training (Section~\ref{sec:NNs}) and are also used to generate death
times and payouts in the MBG valuation (Section~\ref{sc:mbg_pricing}).
\end{remark}

\section{Stochastic control framework}
\label{sc:modeling}
{\apurple{We now turn to the stochastic-control framework for the representative member in the homogeneous large-pool tontine. This member is modelled as a domestic
investor with access to four asset classes:}}
(i) a domestic stock index fund,
(ii) a domestic bond index fund,
(iii) a foreign stock index fund, converted to domestic currency, and
(iv) a foreign bond index fund, converted to domestic currency.
{\apurple{Throughout the stochastic-control formulation, unless otherwise stated, account and portfolio quantities are measured in real (inflation-adjusted) terms.}}
This setup allows us to examine both an internationally diversified portfolio, which includes all four asset classes, and a non-diversified alternative that is restricted to domestic assets only.
The construction of these indices, the inflation adjustment, and the data sources used for calibration are detailed in Sections~\ref{sec:validation_model_data} and \ref{sec:diverse_port}.

\subsection{Index dynamics}
\label{ssc:wealth_dynamics}
For simplicity and clarity, we establish the following notational conventions. A subscript $\iota \in \{d,f\}$ is used to distinguish quantities related to the domestic stock or bond ($\iota = d$) from those corresponding to the foreign counterparts ($\iota = f$). Additionally, a superscript ``$s$'' denotes quantities associated with stock indices, while a superscript ``$b$'' identifies those related to their bond index counterparts.

Let $S^{\dom}(t)$, $B^{\dom}(t)$, $S^{\f}(t)$, and $B^{\f}(t)$ denote the real (inflation-adjusted) \emph{amounts} invested in the stock and bond indices of the domestic and foreign markets, respectively, at time $t \in [0, T]$.
To avoid notational clutter, we occasionally use the shorthand notation:
$S_t^{\dom} \equiv S^{\dom}(t)$,
$B_t^{\dom} \equiv B^{\dom}(t)$,
$S_t^{\f} \equiv S^{\f}(t)$, and
$B_t^{\f} \equiv B^{\f}(t)$.
We denote by $\{X_t\}_{0 \le t \le T}$ (resp.\ by $x$) the controlled underlying index process (resp.\ a generic state of the system), where
\begin{equation}
\label{eq:Xt_x}
X_t \quad \left(\text{resp.\ } x\right) =
\begin{cases}
\bigl(S_t^{\dom},\, B_t^{\dom} \bigr) \quad \left(\text{resp.\ } (s^{\dom},\, b^{\dom})\right), & \text{domestic-only}, \\[4pt]
\bigl(S_t^{\dom},\, B_t^{\dom},\, S_t^{\f},\, B_t^{\f} \bigr) \quad \left(\text{resp.\ } (s^{\dom},\, b^{\dom},\, s^{\f},\, b^{\f})\right), & \text{internationally diversified}.
\end{cases}
\end{equation}
Recall the set of decision times $\mathcal{T}$ defined in \eqref{eq: T_M}.
Between decision times $t_m\in\mathcal T$, the process evolves passively, driven solely by index-return dynamics.
For any $t\in [t_m^+,\, t_{m+1}^-]$,  we write
\begin{equation}
\label{eq:dynamics_generic}
   X_t \;=\; \mathcal M_{t_m,t}\bigl(X_{t_m^{+}},\,\varepsilon_{m+1}\bigr),
   \qquad t\in [t_m^+,\, t_{m+1}^-],
\end{equation}
where $\mathcal M_{t_m,t}$ is a transition operator and
$\varepsilon_{m+1}$ collects all exogenous drivers acting over the interval $[t_m^+,\, t_{m+1}^-]$. These drivers may include Brownian or jump shocks (in parametric models) or resampled blocks of historical returns (in bootstrapped settings).
The operator $\mathcal M_{t_m,t}$ maps the post-decision state $X_{t_m^{+}}$ and the
exogenous drivers on $[t_m^+, t]$ to the state $X_t$.

In this paper, we consider two specifications for the driver set $\varepsilon_{m+1}$ as follows.
\begin{itemize}[noitemsep, topsep=1pt, leftmargin=*]
\item Parametric (jump–diffusion) model:
For benchmarking against PIDE-based methods, we implement a two-asset domestic-only case
in which the domestic stock and bond indices $\{S_t^{d}\}$ and $\{B_t^{d}\}$ follow Kou‐type jump–diffusion dynamics with asymmetric double-exponential jumps, capturing empirically observed heavy tails \cite{kou01}. The full SDE specification appears in Subsection~\ref{ssc:asset_dynamics}

\item Historical block bootstrap:
In our main experiments, both domestic-only and internationally diversified, we simulate asset paths nonparametrically via the stationary block bootstrap \cite{politis1994stationary,patton2009correction,politis2004automatic,dichtl2016testing}.
Implementation details of the bootstrapping techniques are provided in Subsection~\ref{ssc:data}.
\end{itemize}
In both cases, mortality-credit realizations (when applicable) are simulated independently and applied only at $t_m^-$. The resulting dynamics $\{X_t\}_{0 \le t \le T}$ from \eqref{eq:dynamics_generic} capture the passive evolution of the index values between decision times. Active decisions, namely mortality updates, withdrawals, and rebalancing, are applied only at $\{t_m\}_{m=0}^{M-1}$, as detailed in the next subsection.

\subsection{Mortality-credit and fee updates}
{\red{We define the investor's portfolio wealth at time $t$ by}}
\begin{equation}
\label{eq:wealth}
{\red{\bar W_t}} =
\begin{cases}
S_t^{\dom} + B_t^{\dom}, & \text{domestic-only}, \\[4pt]
S_t^{\dom} + B_t^{\dom} + S_t^{\f} + B_t^{\f},
& \text{internationally diversified}.
\end{cases}
\end{equation}
{\red{At a decision time $t_m^-$, $\bar W_{m^-}$ denotes the investor's
pre-update wealth, immediately before the mortality-credit and
tontine-management-fee updates. The account is solvent at $t_m^-$ when
$\bar W_{m^-}>0$.}}

At each $t_m\in\mathcal T$, the sequence of actions at
$t_m^{-}\!\rightarrow t_m\!\rightarrow t_m^{+}$ described in
Subsection~\ref{ssc:tontine} applies.
{\red{To simplify bookkeeping, we use a uniform event structure across all
$t_m\in\mathcal T$, with the following initial- and terminal-time
conventions:}}
\begin{itemize}[noitemsep, topsep=1pt, leftmargin=*]
\item
{\red{At each time $t_m^-$, $m=0,1,\ldots,M$, the mortality credit $c_m$
defined in \eqref{eq:tontine-gain} is applied when $m\ge1$ and
$\bar W_{m^-}>0$. No mortality credit is applied at $t_0$, since no
forfeitures have yet occurred, or after the account becomes insolvent.}}

\item
At each time $t_m$, $m = 0, \ldots, M-1$, the investor withdraws an
amount $q_m$, followed by portfolio rebalancing at $t_m^+$. At $t_M$, the
portfolio is liquidated (i.e.\ no rebalancing or withdrawal occurs), and
terminal wealth $W_T$ is realized. For notational completeness, this is
enforced by setting $q_M = 0$.
\end{itemize}
Accordingly, the tontine gain rate in \eqref{eq:tontine-gain} is defined
by
\EQ
\label{eq:mod_Tg}
g_m=
\begin{cases}
\displaystyle\frac{\delta_{m-1}}{1-\delta_{m-1}},
& {\red{m=1,\ldots,M,\quad \bar W_{m^-}>0}},\\[6pt]
0,
& {\red{\text{otherwise}.}}
\end{cases}
\EN
Let
\begin{equation}
\label{eq:Wtilde_def}
\widetilde W_{m^-}
=
{\red{(1+g_m)\bar W_{m^-}}}.
\end{equation}
That is, $\widetilde W_{m^-}$ is the portfolio wealth immediately
after the mortality credit distribution at $t_m^{-}$ but before any
fee, withdrawal, or rebalancing.

We introduce a baseline tontine management fee that is deducted once per year,
at each decision time $t_m$, $m=1,\ldots,M$. In line with Australian
superannuation practice, and in particular the administration fee structure in
QSuper's Lifetime Pension PDS \cite{qsuper2025pds}, we model this as a
proportional charge on the account balance. Let $\varrho\in(0,1)$ denote the
yearly fee rate (e.g.\ $\varrho=0.11\%$ for an 11 bp charge).

{\red{To ensure that the tontine fee is applied only when the account is
solvent and is omitted at $t_0$, define}}
\EQ
\label{eq:delta_t_m}
{\red{
\varrho_m
=
\varrho\,
\mathbf 1_{\{\,m\ge1,\;\bar W_{m^-}>0\,\}},
\qquad m=0,\ldots,M.
}}
\EN
With the conventions established in
\eqref{eq:mod_Tg}--\eqref{eq:delta_t_m}, the investor's
pre-withdrawal wealth, i.e. wealth immediately before withdrawal,
after applying the mortality credit and deducting the tontine
fee, is
\EQ
\label{eq:Wm_post_tontine}
W_{m^-}
=
\bigl(1-\varrho_m\bigr)\widetilde W_{m^-},
\qquad
\text{$\widetilde W_{m^-}$ given by \eqref{eq:Wtilde_def}.}
\EN
Unless otherwise stated, we use $W_{m^-}$ to denote this pre-withdrawal wealth.

\subsection{Withdrawal and rebalancing controls}
At each decision time $t_m\in\mathcal T$, the withdrawal $q_m$ reduces
$W_{m^-}$ to the post-withdrawal wealth
\EQ
\label{eq:Wm_post_withdrawal}
W_{m^+}
=
W_{m^-}-q_m,
\qquad t_m\in\mathcal T,
\quad
\text{$W_{m^-}$ given by \eqref{eq:Wm_post_tontine}.}
\EN
We model $q_m$, for $m=0,\ldots,M$, as a withdrawal control that depends on
{\red{pre-withdrawal wealth}} $W_{m^-}$ and time $t_m$. Recalling the convention
that $q_M=0$, we define the withdrawal control function as
\[
q_m(\cdot):(W_{m^-},t_m)\mapsto
q_m=q(W_{m^-},t_m),
\]
where $W_{m^-}$ is given by \eqref{eq:Wm_post_tontine}.

At each rebalancing time $t_m$, for $m=0,\ldots,M-1$, we denote the
rebalancing control by $\boldsymbol p_m(\cdot)$, the vector of proportions of
{\red{post-withdrawal wealth}} allocated to the asset indices. This control
depends on $t_m$ and on the wealth $W_{m^+}$ after the cash
withdrawal $q_m$ in \eqref{eq:Wm_post_withdrawal}. Formally,
\[
\boldsymbol p_m(\cdot):(W_{m^+},t_m)
\mapsto
\boldsymbol p_m=\boldsymbol p(W_{m^+},t_m),
\]
where
\begin{equation}
\label{eq:reb_control}
\boldsymbol{p}_m =
   \begin{cases}
      \bigl(p^{\dom}_{s,m}\bigr), & \text{domestic-only case}, \\[4pt]
      \bigl(p^{\dom}_{s,m},\, p^{\dom}_{b,m},\, p^{\f}_{s,m}\bigr),
      & \text{internationally diversified case}.
   \end{cases}
\end{equation}
We denote by $X_{m^+}$ the state of the system immediately after applying the
rebalancing control $\boldsymbol p_m$, where
\begin{equation}
\label{eq:Xmplus}
X_{m^+}  =
\begin{cases}
\bigl( S_{m^+}^{\dom},\, B_{m^+}^{\dom} \bigr),
& \text{domestic-only},
\\[4pt]
\quad \text{with }
S_{m^+}^{\dom}
=
p_{s,m}^{\dom}W_{m^+},
\qquad
B_{m^+}^{\dom}
=
p_{b,m}^{\dom}W_{m^+},
\\[6pt]
\bigl(S_{m^+}^{\dom},\,B_{m^+}^{\dom},
      \,S_{m^+}^{\f},\,B_{m^+}^{\f}\bigr),
& \text{internationally diversified},
\\[4pt]
\quad \text{with }
\begin{cases}
S_{m^+}^{\dom}
=
p_{s,m}^{\dom}W_{m^+},
\qquad
B_{m^+}^{\dom}
=
p_{b,m}^{\dom}W_{m^+},
\\[4pt]
S_{m^+}^{\f}
=
p_{s,m}^{\f}W_{m^+},
\qquad
B_{m^+}^{\f}
=
p_{b,m}^{\f}W_{m^+}.
\end{cases}
\end{cases}
\end{equation}
Here, in the domestic-only case, we define
$p_{b,m}^{\dom}=1-p_{s,m}^{\dom}$, while in the internationally diversified
case, $p_{b,m}^{\f}:=
1-p_{s,m}^{\dom}-p_{b,m}^{\dom}-p_{s,m}^{\f}$.

{\red{
\subsection{Insolvency and debt convention}
\label{ssc:insolvency}
The account becomes insolvent at $t_m^+$ when a withdrawal results in
$W_{m^+}\le 0$. In the event of insolvency, the portfolio is liquidated and
trading stops. The negative balance is carried in the domestic bond position
and accumulates at the borrowing rate (a spread above the bond rate), and no further mortality credits or tontine management fees are applied. Minimum withdrawals nevertheless continue, so once the account is exhausted, they are funded by borrowing and contribute to the accumulated debt.}}


\subsection{Admissible control sets}
We denote by $\mathcal Z_q(W_{m^-},t_m)$ and
$\mathcal Z_p(W_{m^+},t_m)$ the wealth- and time-dependent admissible sets for
the withdrawal and rebalancing controls, respectively, and by
$\mathcal Z(W_{m^-},W_{m^+},t_m)$ the admissible set for the control pair
$\bigl(q_m(\cdot),\boldsymbol p_m(\cdot)\bigr)$. We discuss these admissible
sets below.

Contrary to the common perception that retiree spending is largely
inflexible from year to year, survey evidence suggests that retirees adjust
their lifestyle in response to cash-flow changes, including in categories often
perceived as ``fixed'' expenses \cite{banerjee2021decoding}. This supports
modelling withdrawals as a flexible control within bounds, rather than as a
rigid rule.

To reflect this evidence, and consistent with industry practice, we impose
lower and upper bounds, $q_{\min}$ and $q_{\max}$, on withdrawals. Formally, for
$t_m\in\mathcal T$, with $W_{m^-}$ given by
\eqref{eq:Wm_post_tontine}, we define the admissible withdrawal control set as
\begin{equation}
\label{eq:Zq}
\mathcal{Z}_{q}\bigl(W_{m^-},t_m\bigr)=
\begin{cases}
\bigl[q_{\min},\,q_{\max}\bigr],
    & t_m\neq t_M,\; W_{m^-}\ge q_{\max},\\[4pt]
\bigl[q_{\min},\,\max\{q_{\min},\,W_{m^-}\}\bigr],
    & t_m\neq t_M,\; W_{m^-}<q_{\max},\\[4pt]
\{0\}, & t_m=t_M.
\end{cases}
\end{equation}
While somewhat complicated, the piecewise structure in \eqref{eq:Zq} captures
the intuition that the retiree aims to avoid insolvency as much as possible by
tightening the upper withdrawal limit when wealth declines. If wealth falls
below $q_{\min}$, the retiree can still withdraw the minimum amount, accepting
that this may lead to insolvency (i.e.\ $W_{m^+}<0$). At the terminal time
$t_M$, no withdrawal occurs, as the portfolio is liquidated.

This constraint also admits a natural economic interpretation: $q_{\max}$
represents the retiree's desired annual level of real (inflation-adjusted)
spending, while $q_{\min}$ is a contingency floor the retiree is willing to
adopt, potentially temporarily, to reduce the risk of depletion. The optimal
control, therefore, uses the flexibility in \eqref{eq:Zq} to preserve higher
withdrawals when the account is well funded, but cuts withdrawals toward
$q_{\min}$ in adverse market/wealth states to mitigate the long-run
consequences of drawing at the maximum rate during a downturn.

This state-contingent spending cut mechanism is closely related to practitioner
rules such as the \emph{Canasta} strategy, which advocates reducing withdrawals
following poor market performance.\footnote{
\url{https://www.theglobeandmail.com/report-on-business/math-prof-tests-investing-formulas-strategies/article22397218/};
\url{https://cs.uwaterloo.ca/~paforsyt/Canasta.html}.}
Time-varying bounds (e.g.\ imposing $q_{\min}=q_{\max}$ over an initial period
to force high early spending) can also be accommodated, but would mechanically
increase the likelihood of early depletion when adverse returns occur early.

\begin{remark}[Minimum withdrawals under insolvency]
\label{rm:insolvency_qmin}
At first sight it might seem natural to cease withdrawals once the account is
depleted. However, the lower bound $q_{\min}$ is intended to represent a
required minimum level of real (inflation-adjusted) spending. Accordingly, when
$W_{m^-}<q_{\min}$ we still enforce the minimum withdrawal by allowing
$W_{m^+}=W_{m^-}-q_m<0$, which is equivalent to borrowing the shortfall.
If a withdrawal renders the account insolvent, the insolvency convention in
Subsection~\ref{ssc:insolvency} applies.
This treatment ensures
that tail-risk measures of terminal wealth penalize insolvency and, in
particular, penalize early depletion more than late depletion, since debt has
more time to accumulate.

This convention also admits a practical interpretation in which minimum
spending can be funded from assets outside the managed account (e.g.\ housing
equity monetized via a reverse mortgage) once financial wealth is exhausted
\cite{pfeiffer2013increasing}. This ``hedge of last resort'' interpretation is
also consistent with mental accounting views in which housing wealth is
treated as a separate bucket that is tapped only in extreme circumstances, and
otherwise can form part of the retiree's bequest
\cite{shefrin1988behavioral}.
\end{remark}

We enforce no leverage and no shortselling when the portfolio is solvent, and once the account becomes insolvent, we halt trading entirely and direct all wealth to the domestic bond. To define the set of admissible rebalancing controls, we introduce
\[
\Delta^{(k)}
:=
\Bigl\{
(p_1,\ldots,p_k)\in[0,1]^k
\,\Bigm|\,
\sum_{i=1}^{k}p_i\le1
\Bigr\},
\qquad
\mathbf 0^{(1)}:=(0),
\qquad
\mathbf b^{(3)}:=(0,1,0),
\]
where $\Delta^{(k)}$ denotes the $k$-dimensional simplex of explicitly
parameterized asset proportions. We set $k=1$ for the domestic-only portfolio
and $k=3$ for the internationally diversified case.

Under these conventions, the admissible rebalancing control set at time
$t_m\in\mathcal T$ is
\begin{equation}
\label{eq:Zp}
{\red{
\mathcal Z_p(W_{m^+},t_m)
=
\begin{cases}
\Delta^{(k)},
& t_m\neq t_M,\quad W_{m^+}>0,
\\[4pt]
\{\mathbf 0^{(1)}\},
& \text{domestic-only and }
  \bigl(t_m=t_M\text{ or }W_{m^+}\le0\bigr),
\\[4pt]
\{\mathbf b^{(3)}\},
& \text{internationally diversified and }
  \bigl(t_m=t_M\text{ or }W_{m^+}\le0\bigr),
\end{cases}
}}
\end{equation}
where $W_{m^+}$ is the post-withdrawal wealth defined in
\eqref{eq:Wm_post_withdrawal}. {\red{In the
domestic-only case, $\mathbf 0^{(1)}$ assigns the entire balance to the domestic
bond. In the internationally diversified case,
$\mathbf b^{(3)}=(0,1,0)$ assigns the entire balance to the domestic bond and
zero to the other assets. When the balance is negative, this fixed allocation
is the bookkeeping convention under which debt accrues at the borrowing rate. At $t_M$, these singleton controls are notational conventions only, since the
portfolio is liquidated and no rebalancing occurs.}}

The admissible control set for the pair
$\bigl(q_m(\cdot),\boldsymbol p_m(\cdot)\bigr)$,
$t_m\in\mathcal T$, can then be written as
\begin{equation}
\label{eq:Z}
(q_m,\boldsymbol p_m)
\in
\mathcal Z(W_{m^-},W_{m^+},t_m)
=
\mathcal Z_q(W_{m^-},t_m)
\times
\mathcal Z_p(W_{m^+},t_m),
\end{equation}
where $\mathcal Z_q(W_{m^-},t_m)$ and
$\mathcal Z_p(W_{m^+},t_m)$ are defined in \eqref{eq:Zq} and
\eqref{eq:Zp}, respectively.

Let $\mathcal A$ be the set of admissible controls:
\begin{equation}
\label{eq:A}
\mathcal A
=
\left\{
\mathcal U
=
\left(q_m,\boldsymbol p_m\right)_{0\le m\le M}
\,\Bigm|\,
\left(q_m,\boldsymbol p_m\right)
\in
\mathcal Z(W_{m^-},W_{m^+},t_m)
\right\}.
\end{equation}
For any $t_m$, we define the subset of controls applicable from $t_m$ onward
by
\begin{equation}
\label{eq:U_m}
\mathcal U_m
=
\left\{
\left(q_\ell,\boldsymbol p_\ell\right)_{m\le\ell\le M}
\,\Bigm|\,
\left(q_\ell,\boldsymbol p_\ell\right)
\in
\mathcal Z(W_{\ell^-},W_{\ell^+},t_\ell)
\right\}.
\end{equation}

\section{EW--CVaR portfolio formulation}
\label{sc:formulation}
We use CVaR to quantify downside risk in terminal wealth.
For subsequent use, we denote by
$\mathbb{E}_{\mathcal{U}_0}^{X_{0^+},t_0^+}\left[W_T\right]$
the expectation of terminal wealth $W_T$ under the real-world measure,
conditional on the system being in state $X_{0^+}$ at time $t_0^+$, and using
the control $\mathcal{U}_0$ over $[t_0,T]$. Let $x_0=X_{0^-}$ be the initial
state.

{\apurple{For a tail probability $\alpha\in(0,1)$, typically
$\alpha=0.01$ or $\alpha=0.05$, we define the lower-tail CVaR of terminal
wealth directly through the optimization representation of \cite{RT2000}:}}
\begin{eqnarray}
\mathrm{CVaR}_{\alpha,\,\mathcal{U}_0}^{x_0,t_0}
&=&
\sup_{W\in\mathbb R}
\Ebb_{\mathcal{U}_0}^{X_{0^+},t_0^+}
\!\Bigl[
W+\tfrac{1}{\alpha}\min\!\bigl(W_T-W,0\bigr)
\,\big|\,
X_{0^-}=x_0
\Bigr].
\label{eq: CVaR_optimization}
\end{eqnarray}
{\apurple{Here, $W$ is a candidate threshold that partitions the distribution
of $W_T$ into its lower tail and the remainder. Intuitively, $\mathrm{CVaR}_{\alpha}(W_T)$ represents the average terminal
wealth in the worst $\alpha$ fraction of outcomes \cite{rock2014}.}}

As a measure of reward, we consider the expected total withdrawals.
For any admissible control $\mathcal{U}_0 \in \mathcal{A}$, we define expected withdrawals (EW)
\begin{equation}
\label{eq:ew}
\operatorname{EW}_{\mathcal{U}_0}^{x_0,t_0}=\Ebb_{\mathcal{U}_0}^{X_{0^+}, t_0^+}\left[\sum_{m=0}^M q_m \right]
\end{equation}
where $q_m = q(W_{m^-}, t_m)$ and $W_{m^-}$ is defined in \eqref{eq:Wm_post_tontine} (i.e.\ after mortality credit distribution).  {\apurple{Here, we assume the investor survives the full decumulation period, consistent with the convention underlying \cite{bengen1994}.}}
\begin{remark}[Discounting and mortality weighting]
\label{rm:discounting_mortality_weighting}
We remark that \eqref{eq:ew} applies no time discounting.
With all quantities expressed in real (inflation-adjusted) dollars, this is
equivalent to assuming a zero real discount rate.
Over long horizons, short-term real rates have often been modest on average in
advanced economies, so adopting a zero real discount rate provides a reasonable
and conservative baseline. 

In addition, we do not mortality-weight withdrawals by survival probabilities.
Such weighting is natural when valuing cash flows averaged across a population
(e.g.\ in annuity pricing), but our objective represents the strategy of an
individual retiree: conditional on being alive, the retiree requires the full
spending amount, not its survival-probability-weighted value. For example, if the
probability of surviving to a later age were 50\%, mortality-weighting would
mechanically halve the corresponding cash flows (ignoring discounting), even
though the retiree would still require the full amount if alive. Hence, a
conservative approach is to condition on survival to a fixed horizon (age 95 in
our numerical illustration), which is consistent with the Bengen spending-rule
scenario that has proved popular with retirees \cite{bengen1994}. Accordingly,
we condition on survival to the horizon, consistent with the practitioner
``plan-to-live, not to die'' convention; see \cite{pfau2018overview}.
\end{remark}
The goal is to simultaneously maximize expected withdrawals (EW) and the CVaR of terminal wealth---two objectives that are inherently in conflict. Solving this trade-off requires solving a bi-objective optimization problem. To obtain Pareto-optimal solutions,
we apply a scalarization approach, which converts the bi-objective problem into a single-objective one by combining CVaR and EW using a scalarization parameter $\gamma > 0$.
Formally, for a given $\gamma>0$, we seek a control \mbox{$\mathcal{U}_0$ that solves}
\begin{equation}
\label{eq:scalarization}
 \sup_{\mathcal{U}_0 \in \mathcal{A}} \left\{\operatorname{EW}_{\mathcal{U}_0}^{x_0,t_0} + \gamma\, \text{CVaR}_{\alpha,\, \mathcal{U}_0}^{x_0,t_0}\right\}.
\end{equation}
\begin{remark}[Pre-commitment and induced time-consistent interpretation]
\label{rm:precommitment_time_consistency}
The scalarized objective in \eqref{eq:scalarization} is evaluated at $t_0$ and thus
defines a \emph{pre-commitment} control: the retiree selects the policy at inception
and then follows it over the horizon. Because $\mathrm{CVaR}_\alpha$ is generally not
time-consistent in multi-period settings, the EW--CVaR optimization is formally
time-inconsistent and is often viewed as non-implementable, since at time $t>0$ the
retiree would typically have an incentive to deviate from the $t_0$ pre-commitment
strategy.  Nonetheless, we use the pre-commitment formulation to determine the joint optimizer $W^\ast$ in \eqref{eq: CVaR_optimization}. Fixing $W^\ast$ thereafter yields the equivalent expected-value control problem
\[
\sup_{\mathcal{U}_0\in\mathcal{A}}
\Ebb_{\mathcal{U}_0}^{X_{0^+},t_0^+}\!\bigg[\sum_{m=0}^M q_m
+\frac{\gamma}{\alpha}\min(W_T-W^\ast,0)\bigg].
\]
{\apurple{This equivalent control problem satisfies the Bellman principle and is therefore time-consistent.}} Accordingly, we interpret the computed control $\mathcal{U}_0^\ast(\gamma)$ as the
associated \emph{induced time-consistent} strategy determined by $(\alpha,\gamma,W^\ast)$
(see, e.g.\ \cite{PA2020a, strub2019, dang2026multi}).
\end{remark}
Using \eqref{eq: CVaR_optimization}--\eqref{eq:ew}, we recast \eqref{eq:scalarization} as a control problem involving both system dynamics, mortality credit distributions, withdrawals, and rebalancing actions. For a fixed scalarization parameter $\gamma > 0$, the pre-commitment EW--CVaR problem, denoted $PCEC_{t_0}(\alpha, \gamma)$, \mbox{is defined via the value function $V(x_0, t_0^-)$ as}
\begin{equation}
\label{eq:PCEC}
\begin{aligned}
\left(PCEC_{t_0}(\alpha, \gamma)\right): V\left(x_0, t_0^{-}\right)=
 \sup_{\mathcal{U}_0 \in \mathcal{A}}\, \sup_{W}\bigg\{\Ebb_{\mathcal{U}_{0}}^{ X_{0^+}, t_{0 }^{+}} \bigg[\sum_{m=0}^M q_m &+ \gamma\left(W+\frac{1}{\alpha} \min \left(W_T-W, 0\right)\right)
\\
& \qquad \qquad +\epsilon\, W_T \bigg{|} X_{0^-}=x_0\bigg]\bigg\}
\end{aligned}
\end{equation}
\begin{equation}
\text{subject to:}
\begin{cases}
~~\text{(dynamics)} \quad~~\{X_t\} \text{ evolves via } 
\eqref{eq:dynamics_generic}
& t \notin \mathcal{T}, \\[4pt]
\left.\begin{array}{ll}
\text{(tontine)}  &W_{m^-}  \text{ given by } \eqref{eq:Wm_post_tontine}, \\
\text{(withdrawal)} &W_{m^+} = W_{m^-} - q_m \text{ as in } \eqref{eq:Wm_post_withdrawal}, \\
\text{(rebalancing)} &X_{m^+} \text{ as defined in  \eqref{eq:Xmplus}, using } \boldsymbol{p}_m(\cdot), \\
&(q_m(\cdot),\, \boldsymbol{p}_m(\cdot)) \in \mathcal{Z}(W_{m^-}, W_{m^+}, t_m),
\quad m = 0,\dots, M,
\end{array}\right\} & t_m \in \mathcal{T}.
    \end{cases}\label{eq:PCEC_constraints}
\end{equation}
We denote by $\mathcal{U}_0^\ast = \mathcal{U}_0^\ast(\gamma)$ the control that attains the supremum in \eqref{eq:PCEC}-\eqref{eq:PCEC_constraints}, i.e.\ the optimal policy for a given scalarization parameter $\gamma$.

Note that we have added the extra term
$\Ebb_{\mathcal{U}_0}^{X_{0^+},t_0^+}\!\left[\epsilon\, W_T\right]$
to \eqref{eq:PCEC}. If we have a maximum withdrawal constraint, and if
$W_t \gg W$ as $t \rightarrow T$, the investment controls
{\apurple{may become non-unique}}. {\apurple{In such well-funded states}},
we can break investment policy ties either by setting $\epsilon<0$, which
{\apurple{selects}} investments in bonds, or by setting $\epsilon>0$, which
{\apurple{selects}} investments in stocks. Choosing $|\epsilon| \ll 1$ ensures
that this term only has an effect if $W_t \gg W$ and $t \rightarrow T$.
See \cite{forsyth2022stochastic} for more discussion of this.

We interchange the supremum operators in \eqref{eq:PCEC} to express the value function in a more computationally tractable form:
\begin{equation}
\label{eq:PCEC_interchange}
V(x_0, t_0^{-})=
  \sup_{W}\, \sup _{\mathcal{U}_0 \in \mathcal{A}}\bigg\{\Ebb_{\mathcal{U}_{0}}^{ X_{0^+}, t_{0 }^{+}} \bigg[\sum_{m=0}^M q_m + \gamma\big(W+\frac{1}{\alpha} \min \left(W_T-W, 0\right)\big) +\epsilon\,  W_T \bigg{|} X_{0^-}=x_0\bigg]\bigg\}.
\end{equation}
The inner optimization yields a continuous function of $W$, and the optimal threshold \mbox{$W^*(x_0)$ is defined as}
\begin{equation}
\label{eq:W_opt}
W^*(x_0)=  \underset{W}{\arg \max }\, \sup _{\mathcal{U}_0 \in \mathcal{A}}\bigg\{\Ebb_{\mathcal{U}_{0}}^{ X_{0^+}, t_{0 }^{+}} \bigg[\sum_{m=0}^M q_m + \gamma\big(W+\frac{1}{\alpha} \min \left(W_T-W, 0\right)\big) +\epsilon\,  W_T \bigg{|} X_{0^-}=x_0\bigg]\bigg\}.
\end{equation}
Finally, the scalarization parameter $\gamma$ reflects the investor's level of risk aversion. For a given {\apurple{$\alpha\in (0,1)$}},  the EW--CVaR efficient frontier is defined as the following set of points in $\mathbb{R}^2$:
\begin{eqnarray}
   \mathcal{S}(\alpha): \quad \left\{ \left( \text{CVaR}_{\alpha,\,  \mathcal{U}_{0}^{\ast}}^{x_{0}, t_0}\left[ W_{T} \right], \  \text{EW}_{\mathcal{U}_{0}^{\ast}}^{x_{0}, t_0}\right) : \gamma >0 \right\},
    \label{eq:PCEC_set}
\end{eqnarray}
traced out by solving \eqref{eq:PCEC} for each $\gamma > 0$. In other words, for any fixed level of risk aversion $\gamma$, the corresponding point in \eqref{eq:PCEC_set} cannot be improved in the EW--CVaR sense by any other admissible strategy in $\mathcal{A}$.

\section{Neural network formulation}
\label{sec:NNs}
{\apurple{We numerically solve the EW--CVaR stochastic-control problem for the
representative member by parameterizing the withdrawal and rebalancing controls
with NNs.}}
Our approach builds on the growing literature that uses NNs to approximate the
optimal control directly, avoiding dynamic programming methods
\cite{buehler2019deep, li2019data, reppen2023deep, reppen2023deepMF,
van2024global}.
These methods have been termed ``global-in-time'' machine-learning approaches to stochastic control \cite{hu2024recent}. This contrasts with the stacked NN approach, in which a separate NN is used to approximate the control at each rebalancing step \cite{tsang2020deep, han2016deep}. More generally,
these techniques are special cases of ``policy function approximation'' for optimal stochastic control \cite{powell2023universal}.
{\apurple{Related NN-based approximation methods have also been developed for
certain classes of mean-field games and mean-field control problems
\cite{carmona2021deepmfg, phamwarin2023meanfieldnn}.}}

For any admissible control $\mathcal{U}_0 \in \mathcal{A}$ and any CVaR threshold $W$, we define the objective
\begin{equation}
\label{eq:Vhat}
\begin{aligned}
V(x_0, t_0^{-};\, \mathcal{U}_0, W) &=
    \Ebb_{\mathcal{U}_0}^{X_{0^+}, t_0^+} \bigg[
        \sum_{m=0}^M q_m
        + \gamma\left(W + \frac{1}{\alpha} \min \left(W_T - W,\, 0\right)\right)
        + \epsilon W_T
    \,\bigg|\, X_{0^-} = x_0
    \bigg],
\\[-2pt]
&\text{ subject to the dynamics and constraints specified in \eqref{eq:PCEC_constraints}}.
\end{aligned}
\end{equation}
The value function $V(x_0, t_0^-)$ of the $PCEC_{t_0}(\alpha, \gamma)$ problem
can be written as
 \begin{equation}
 \label{eq:V_Vhat}
V(x_0, t_0^-) = \sup_{W} \, \sup_{\mathcal{U}_0 \in \mathcal{A}} \, V(x_0, t_0^-;\, \mathcal{U}_0, W).
\end{equation}

\subsection{{\apurple{NN parameterization of admissible controls}}}
The essence of our NN approach is to directly {\apurple{parameterize}} the
withdrawal and rebalancing components of an admissible control policy
$\mathcal{U}_0$ in \eqref{eq:V_Vhat} using feedforward, fully connected neural
networks. More specifically, given NN parameters (weights and biases)
$\boldsymbol{\theta}_{q}$ and $\boldsymbol{\theta}_{p}$, we denote by
$\widehat{q}(W_{m^-}, t_m;\boldsymbol{\theta}_{q})$ and
$\widehat{\boldsymbol{p}}(W_{m^+}, t_m; \boldsymbol{\theta}_{p})$ the
{\apurple{NN parameterizations}} of the withdrawal control $q_m(\cdot)$ and the
rebalancing control $\boldsymbol{p}_m(\cdot)$, respectively. Formally, we write
\begin{eqnarray}
\label{eq:NN_control}
    \widehat{q}_m(\cdot)
    := \widehat{q}\left( W_{m^-}, t_m; \boldsymbol{\theta}_{q} \right)
    &\simeq&
    q_m(W_{m^-}, t_m), \quad m = 0, 1, \ldots, M, \nonumber \\
    \widehat{\boldsymbol{p}}_m(\cdot)
    := \widehat{\boldsymbol{p}}\left( W_{m^+}, t_m; \boldsymbol{\theta}_{p} \right)
    &\simeq&
    \boldsymbol{p}_m(W_{m^+}, t_m), \quad m = 0, 1, \ldots, M, \\
    \widehat{\mathcal U}_0
    := \left\{ (\widehat{q}_m(\cdot), \widehat{\boldsymbol{p}}_m(\cdot))
    \,\big|\, m = 0, \ldots, M \right\}
    &\simeq&
    \mathcal{U}_0. \nonumber
\end{eqnarray}
{\apurple{Thus,  $\widehat{\mathcal U}_0$ denotes the NN-parameterized control
policy. The collection of all such policies satisfying the admissibility
conditions defining $\mathcal A$ is denoted by
\[
\widehat{\mathcal A}
:=
\big\{
\widehat{\mathcal U}_0\in\mathcal A
~\big|~
\widehat{\mathcal U}_0 \text{ has the form \eqref{eq:NN_control}}
\big\}.
\]
}}
{\apurple{We approximate the original policy search by restricting attention to
$\widehat{\mathcal A}$. Specifically, for fixed
$\widehat{\mathcal U}_0 \in \widehat{\mathcal A}$ and fixed $W$, define the
NN-induced objective $V_{NN}(x_0,t_0^{-};\,\widehat{\mathcal U}_0,W)$ by}}
 \begin{equation}
V_{NN}(x_0, t_0^{-};\, \widehat{\mathcal U}_0, W) = \Ebb_{\widehat{\mathcal{U}}_0}^{X_{0^+}, t_0^+} \bigg[
        \sum_{m=0}^M \widehat{q}_m
        + \gamma\left(W + \frac{1}{\alpha} \min \left(W_T - W,\, 0\right)\right)
        + \epsilon W_T
    \,\bigg|\, X_{0^-} = x_0
    \bigg],\label{eq: PCEC NN}
\end{equation}
subject to the system evolution under the NN-parameterized policy given by:
\begin{equation}
\begin{cases}
~~\text{(dynamics)} \quad~~ \{X_t\} \text{ evolves via } 
\eqref{eq:dynamics_generic},
& t \notin \mathcal{T}, \\[4pt]
\left.\begin{array}{ll}
\text{(tontine)}  &W_{m^-}  \text{ computed via } \eqref{eq:Wm_post_tontine}, \\
\text{(withdrawal)} &W_{m^+} = W_{m^-} - \widehat{q}_m(\cdot) \text{ as in } \eqref{eq:Wm_post_withdrawal}, \\
\text{(rebalancing)} &\text{$X_{m^+}$ computed via \eqref{eq:Xmplus} using } \widehat{\boldsymbol{p}}_m(\cdot), \\
&(\widehat{q}_m(\cdot),\, \widehat{\boldsymbol{p}}_m(\cdot)) \in \mathcal{Z}\left(W_{m^-}, W_{m^+}, t_m\right),
\end{array}\right\} & t_m \in \mathcal{T}.
    \end{cases}\label{eq:PCEC_constraints_NN}
\end{equation}
The value function $V(x_0,t_0^-)$, defined in \eqref{eq:V_Vhat}, is approximated
by the NN-induced value function $V_{NN}(x_0,t_0^-)$, obtained by restricting
the policy search to $\widehat{\mathcal A}$:
\begin{equation}
 \label{eq:VNN}
  V(x_0,t_0^-)
  \simeq
  V_{NN}(x_0,t_0^-)
  :=
  \sup_{W}
  \sup_{\widehat{\mathcal U}_0 \in \widehat{\mathcal A}}
  V_{NN}(x_0,t_0^{-};\,\widehat{\mathcal U}_0,W).
\end{equation}
\subsection{Network architecture for controls}
\label{ssc:NN_architecture}
We use two fully connected feedforward neural networks: one for withdrawals $\widehat{q}(\cdot)$, and one for rebalancing $\widehat{\boldsymbol{p}}(\cdot)$, parameterized by $\boldsymbol{\theta}_{q}\in \mathbb{R}^{\nu_q}$ and $\boldsymbol{\theta}_{p} \in \mathbb{R}^{\nu_p}$, respectively. These networks take inputs of the form $(W_{m^\pm}, t_m)$, but at different stages of investor actions:
(i) $\widehat{q}(\cdot)$ uses pre-withdrawal wealth after mortality credits have been applied, i.e.\ $(W_{m^-}, t_m)$; and (ii) $\widehat{\boldsymbol{p}}(\cdot)$ uses post-withdrawal wealth, i.e.\ $(W_{m^+}, t_m)$.

{\apurple{To enforce the control constraints in \eqref{eq:Zq} and \eqref{eq:Zp}
within the NN-parameterized policy, we use output activation functions together
with deterministic terminal and insolvency branches.}}
\begin{itemize}[noitemsep, topsep=1pt, leftmargin=*]
  \item Withdrawal control: Let $z_q \in \mathbb{R}$ denote the
  pre-activation output of the withdrawal network's final layer.
  {\apurple{For non-terminal decision times, we apply a sigmoid transformation scaled by a wealth-dependent range;   at the terminal date, we impose the deterministic convention $q_M=0$}}:
\begin{equation}
\label{eq:hatq_theta}
{\apurple{
\widehat{q}_m(W_{m^-}, t_m;\boldsymbol{\theta}_{q})
=
\begin{cases}
q_{\min}
+
\max\left( \min(q_{\max}, W_{m^-}) - q_{\min},\, 0 \right)
\displaystyle \frac{1}{1+e^{-z_q}},
& m=0,\ldots,M-1,\\[8pt]
0,
& m=M .
\end{cases}
}}
\end{equation}
{\apurple{Since the sigmoid function $1/(1+e^{-z_q})$ maps into $(0,1)$, the
first branch enforces the non-terminal withdrawal bounds in
$\mathcal{Z}_q(W_{m^-},t_m)$. The second branch enforces the terminal
convention $q_M=0$. Hence,
$\widehat{q}_m(\cdot)\in\mathcal{Z}_q(W_{m^-},t_m)$ for all $m=0,\ldots,M$,
without introducing optimization constraints on the trainable parameters.}}

  \item Rebalancing control:
  {\apurple{Let $(z_1,\ldots,z_{k+1})\in\mathbb{R}^{k+1}$ denote the logits
  output by the rebalancing network's final layer, where $k=1$ for the
  domestic-only portfolio and $k=3$ for the internationally diversified
  portfolio. Define the auxiliary $(k+1)$-dimensional softmax vector
  $\widetilde{\boldsymbol p}_m
  =(\widetilde p_m^{\, (1)},\ldots,\widetilde p_m^{\, (k+1)})$ by}}
 \[
{\apurple{
\widetilde p_m^{\, (i)}
=
\frac{e^{z_i}}{\sum_{\ell=1}^{k+1} e^{z_\ell}},
\qquad i=1,\ldots,k+1 .
}}
\]
{\apurple{For solvent non-terminal decision times, the first $k$ softmax
components are used as the rebalancing control; at the terminal date or after
insolvency, the relevant deterministic convention is imposed:}}
\begin{equation}
\label{eq:hatp_theta}
{\apurple{
\widehat{\boldsymbol p}_m(W_{m^+},t_m;\boldsymbol{\theta}_{p})
=
\begin{cases}
\left(\widetilde p_m^{\, (1)},\ldots,\widetilde p_m^{\, (k)}\right),
& m=0,\ldots,M-1,\quad W_{m^+}>0,\\[4pt]
\mathbf 0^{(1)},
& \text{domestic-only, if } m=M \text{ or } W_{m^+}\le 0,\\[4pt]
\mathbf b^{(3)},
& \text{internationally diversified, if } m=M \text{ or } W_{m^+}\le 0 .
\end{cases}
}}
\end{equation}
{\apurple{Therefore,
$\widehat{\boldsymbol p}_m(\cdot)\in\mathcal{Z}_p(W_{m^+},t_m)$ for all
$m=0,\ldots,M$. These deterministic branches are part of the NN-parameterized
policy map, not post-processing steps in the optimization.}}
\end{itemize}

{\apurple{With these output transformations in place,  an  NN-parameterized control policy $\widehat{\mathcal{U}}_0$ is fully determined by its trainable network parameters $(\boldsymbol{\theta}_{q}, \boldsymbol{\theta}_{p})$. Accordingly,  we write $V_{NN}(x_0, t_0^{-};\, \widehat{\mathcal{U}}_0, W)$ as $V_{NN}(x_0, t_0^{-};\, \boldsymbol{\theta}_{q}, \boldsymbol{\theta}_{p}, W)$ to make this dependency
explicit.}}

Then, the control optimization problem \eqref{eq:VNN} becomes an unconstrained optimization over the network parameters $(\boldsymbol{\theta}_{q}$, $\boldsymbol{\theta}_{p})$, and the CVaR threshold $W$:
\begin{align}
 \label{eq:VNN_theta}
V_{NN}\left(x_0, \,   t_0^{-}\right) & \, =\, \sup _{W} \,
\sup _{\boldsymbol{\theta}_{q} \,  \in \, \mathbb{R}^{\nu_q }}
\,
\sup _{\boldsymbol{\theta}_{p} \,  \in \,\mathbb{R}^{\nu_p }}
V_{NN}\left(x_0,\,  t_0^-; \boldsymbol{\theta}_{q},\,  \boldsymbol{\theta}_{p},\,  W\right)
\nonumber
\\
& \, =\, \sup_{\left(W, \,  \boldsymbol{\theta}_{q}, \, \boldsymbol{\theta}_{p}\right)\,  \in \, \mathbb{R}^{\nu_q+\nu_p+1}}
V_{NN}\left(x_0,\,  t_0^-; \boldsymbol{\theta}_{q},\,  \boldsymbol{\theta}_{p},\,  W\right).
\end{align}
We denote by $(\boldsymbol{\theta}_{q}^*$, $\boldsymbol{\theta}_{p}^*)$, and $W^*$ the optimal network parameters and threshold.

We emphasize that, while the original control problem is constrained via the admissible set $\mathcal{A}$ (see \eqref{eq:A}), the reformulated {\apurple{NN-induced}} objective \eqref{eq:VNN_theta} is unconstrained in terms of the training parameters. This allows the use of standard gradient-based methods for optimization. In our numerical implementation (Sections~\ref{sec:validation_model_data}
and \ref{sec:diverse_port}), we use Adam stochastic gradient descent to train the networks and determine the optimal parameters.
\subsection{Approximation of the NN-induced objective}
\label{ssc:approx_NN}
To evaluate the NN-induced objective in \eqref{eq:VNN_theta}, we approximate its expectation using a finite set of $N$ independent sample paths, each representing a full set of exogenous drivers for the asset indices (generated by \eqref{eq:dynamics_generic}). Mortality enters only through the pathwise tontine gain rates (see Remark~\ref{rm:stochastic_mortality}).
These sample paths are indexed by $n = 1,\ldots,N$, and all path-dependent quantities carry the superscript~``$(n)$''.  For example, ${X_t^{(n)}}_{t\in[0,T]}$ is the simulated state trajectory for the $n$-th path, and $W_T^{(n)}$ is the corresponding terminal wealth.

Specifically, given any set of NN parameters $(\boldsymbol{\theta}_{q}, \boldsymbol{\theta}_{p})$ and threshold $W$,
the NN-induced objective
$V_{NN}\left(x_0,\,  t_0^-; \boldsymbol{\theta}_{q},\,  \boldsymbol{\theta}_{p},\,  W\right)$
is approximated by
\begin{equation}
\label{eq:vnn_vnnhat}
V_{NN}\left(x_0,\,  t_0^-; \boldsymbol{\theta}_{q},\,  \boldsymbol{\theta}_{p},\,  W\right)
\approx
\widehat{V}_{NN}\left(x_0,\, t_0^-,\, \boldsymbol{\theta}_{q}, \boldsymbol{\theta}_{p}, W \right)
\end{equation}
where $\widehat{V}_{NN}\left(x_0,\, t_0^-,\, \boldsymbol{\theta}_{q}, \boldsymbol{\theta}_{p}, W \right) := \ldots$
\begin{equation}
 \label{eq:VNN_sim}
\ldots :=  \frac{1}{N} \sum_{n=1}^N \bigg[
\sum_{m=0}^M \widehat{q}_m\big(W_{m^-}^{(n)}, t_m;\, \boldsymbol{\theta}_{q}\big)
+ \gamma \big(W + \tfrac{1}{\alpha} \min(W_T^{(n)} - W,\, 0)\big)
+ \epsilon W_T^{(n)}\bigg],
\end{equation}
\begin{equation}
\text{s.t.~ }
\begin{cases}
~~\text{(dynamics)} \quad~~ \{X_t^{(n)}\} \text{ drawn from the $n$-th sample path of index returns},
\\
\qquad\qquad\qquad\quad~\text{(generated by \eqref{eq:dynamics_generic})}
& t \notin \mathcal{T}, \\[4pt]
\left.\begin{array}{ll}
\text{(tontine)}  &W_{m^-}^{(n)}  \text{ computed via \eqref{eq:Wm_post_tontine}},
\\
& {\myblue{\text{using the pathwise $g_m^{(n)}$ given by \eqref{eq:mod_Tg}}}}
 {\myblue{ \text{(see Remark~\ref{rm:stochastic_mortality})}}},
\\
\text{(withdrawal)} &W_{m^+}^{(n)} = W_{m^-}^{(n)} - \widehat{q}\big(W_{m^-}^{(n)}, t_m; \boldsymbol{\theta}_{q}\big), \\
\text{(rebalancing)} &\text{$X_{m^+}^{(n)}$ computed from  \eqref{eq:Xmplus}, using } \widehat{\boldsymbol{p}}\big(W_{m^+}^{(n)}, t_m; \boldsymbol{\theta}_{p}\big),
\\
&\big(\widehat{q}_m(\cdot),\, \widehat{\boldsymbol{p}}_m(\cdot)\big) \in \mathcal{Z}\left(W_{m^-}^{(n)}, W_{m^+}^{(n)}, t_m\right),
\end{array}\right\} & t_m \in \mathcal{T}.
    \end{cases}\label{eq:PCEC_constraints_NN_sim}
\end{equation}
Given the sample-based objective
$\widehat{V}_{NN}\!\bigl(x_{0},t_{0}^{-};\,
\boldsymbol{\theta}_{q},\boldsymbol{\theta}_{p},W\bigr)$
in \eqref{eq:VNN_sim}–\eqref{eq:PCEC_constraints_NN_sim}, we train the networks by solving the empirical maximization problem\footnote{Equivalently, we minimize the empirical loss function
$\mathcal{L}\bigl(\boldsymbol{\theta}_{q},\boldsymbol{\theta}_{p},W\bigr)
:= -\widehat{V}_{NN}\!\bigl(x_{0},t_{0}^{-};
          \boldsymbol{\theta}_{q},\boldsymbol{\theta}_{p},W\bigr).$}
\begin{equation}
\label{eq:loss}
(\boldsymbol{\theta}_{q}^{*},
 \boldsymbol{\theta}_{p}^{*},
 W^{*})
\;:=\;
\underset{\boldsymbol{\theta}_{q},\,
          \boldsymbol{\theta}_{p},\,
          W}{\arg\max}\;
          \widehat{V}_{NN}\!\bigl(x_{0},t_{0}^{-};
          \boldsymbol{\theta}_{q},\boldsymbol{\theta}_{p},W\bigr)
\end{equation}
with a gradient-based optimizer, such as Adam stochastic gradient descent.
The resulting parameters define the learned controls
\[
\widehat{q}_m^*(\cdot) := \widehat{q}(\cdot;\, \boldsymbol{\theta}_{q}^*), \qquad
\widehat{\boldsymbol{p}}_m^*(\cdot) := \widehat{\boldsymbol{p}}(\cdot;\, \boldsymbol{\theta}_{p}^*), \quad m = 0, \ldots, M.
\]
{\apurple{The learned NN-parameterized control policy is given by }}
\[
\widehat{\mathcal{U}}_0^* := \bigl\{\, (\widehat{q}_m^*(\cdot),\, \widehat{\boldsymbol{p}}_m^*(\cdot)) \;\big|\; m = 0, \ldots, M \,\bigr\}.
\]
{\apurple{The NN architecture, training parameters, sample sizes, and
sequential transfer-learning procedure over the ordered $\gamma$-grid are
reported in Appendix~\ref{app:nn_reproducibility}.}}

We highlight that the learned policy may be affected by
approximation, sampling, and optimization errors, and that gradient-based
training is not guaranteed to attain the global optimum of the empirical NN
problem. Under the assumption that this global optimum is attained,
\cite{chang2026convergence} establishes convergence in probability of the
sample-based NN value $\widehat V_{NN}$ to the population value $V$ as both the
number of training samples and the NN capacity increase.
{\dpurple{The benchmark comparison with the PIDE-based frontiers of
\cite{forsyth2024optimal} in Appendix~\ref{app:benchmark_validation} provides numerical validation of the
learned policies, but does not establish their optimality for the original
EW--CVaR stochastic-control problem.}}

\section{Representative-contract MBG pricing}
\label{sc:mbg_pricing}
{\dpurple{This section first develops the stand-alone valuation of the MBG for
a representative contract. Section~\ref{sc:mbg_contract_pooling} then uses the
resulting payout distribution to construct an approximate pooled per-contract
valuation. The MBG is valued ex post}} under the learned NN-parameterized
control policy $\widehat{\mathcal U}_0^*$ obtained in
Section~\ref{sec:NNs}. This policy is obtained by optimizing the NN-induced
approximation of the pre-commitment EW--CVaR problem
$PCEC_{t_0}(\alpha,\gamma)$ in \eqref{eq:PCEC}, under the plan-to-live
objective used for the withdrawal and rebalancing problem. Death-contingent
MBG benefits are therefore not fed back into the member's account balance,
withdrawal decisions, or rebalancing decisions.

The MBG overlay is then valued under this fixed policy. If the member dies
before cumulative nominal withdrawals reach the initial purchase price $L_0$,
the overlay pays the shortfall in \eqref{eq:mbg} to the member's beneficiaries
or estate. Because the payout occurs only at death and is not credited to the
member's account, valuation is based on the payout distribution induced by
$\widehat{\mathcal U}_0^*$.

Under the end-of-interval convention in Remark~\ref{rm:mbg_convention}, when
$m_\tau\le M$ the MBG payout is made at time $t_{m_\tau}$. If the member
survives to the horizon, then $m_\tau=M+1$ and the MBG payout is zero. We
therefore define the random real (inflation-adjusted) MBG payout at inception by
\EQ
\label{eq:Z_g}
Z_g :=
\begin{cases}
\displaystyle
\frac{\mathrm{CPI}_{0}}{\mathrm{CPI}_{m_\tau}}\,
\max\bigl(
      L_{0}-\sum_{\ell=0}^{m_\tau-1}
             \widehat q^{*}_{\ell}(\cdot)\,
             \frac{\mathrm{CPI}_{\ell}}{\mathrm{CPI}_{0}},
      0
\bigr),
& m_\tau\le M,\\[12pt]
0,
& m_\tau=M+1.
\end{cases}
\EN
In the first branch of \eqref{eq:Z_g}, the ratio
$\mathrm{CPI}_\ell/\mathrm{CPI}_0$ converts each real (inflation-adjusted)
withdrawal $\widehat q_\ell^*(\cdot)$ into the nominal amount counted by the
guarantee, while $\mathrm{CPI}_0/\mathrm{CPI}_{m_\tau}$ converts the resulting
nominal payout made at $t_{m_\tau}$ into real (inflation-adjusted) dollars at
inception. Consistent with Remark~\ref{rm:discounting_mortality_weighting}, MBG
payouts are valued using a zero real discount rate. Hence, the real discount
factor is one and no additional discount term appears in \eqref{eq:Z_g}.

\subsection{Actuarial pricing formula}
\label{ssc:MBG_pricing_load}
To report the economic cost of the MBG in a transparent and
implementation-agnostic way, we summarize it using an equivalent up-front load
factor $f_g \ge 0$ applied to the initial contribution $L_0$.
All quantities in the pricing rule below are expressed in real
(inflation-adjusted) dollars at $t_0$. We choose $f_g$ so that the real value of
the load equals the physical-measure expected MBG payout plus an explicit
prudential buffer based on an upper-tail risk measure (a standard
risk-measure/premium-principle approach in actuarial pricing; see, e.g.\
\cite{Wang2000Distortion,Denuit2005DependentRisks}).

Formally,
\begin{equation}
\label{eq:actuarial}
f_g\,L_{0}
= \mathbb{E}^{\,x_{0},t_{0}}_{\widehat{\mathcal U}_{0}^{*}}
           \bigl[Z_g\bigr]
\;+\;
\lambda\,
\mathrm{CVaR}_{\alpha_g}^{+,\,x_{0},t_{0}}
    \bigl[Z_g\bigr],
\quad
\lambda\ge 0,\quad \alpha_g\in(0,1).
\end{equation}
\begin{itemize}[noitemsep, topsep=0.5pt, leftmargin=*]
\item $\mathbb{E}^{\,x_{0},t_{0}}_{\widehat{\mathcal U}_{0}^{*}}
           \bigl[Z_g\bigr]$ is the physical-measure expectation of the MBG
           payout $Z_g$ (in real (inflation-adjusted) $t_0$ dollars),
           evaluated under the models for index dynamics, mortality, and CPI,
           with the wealth process generated by the learned NN-parameterized
           policy $\widehat{\mathcal U}_0^*$.

\item $\mathrm{CVaR}_{\alpha_g}^{+,\,x_{0},t_{0}}
    \bigl[Z_g\bigr]$ is the upper-tail CVaR of the same real
      (inflation-adjusted) dollar payout at tail level $\alpha_g$,
      computed under the same models and learned policy.
      The tail level $\alpha_g$ is a
      prudential-buffer choice and need not coincide with the lower-tail
      probability $\alpha$
      used in the retiree's EW--CVaR objective.\footnote{The MBG payout is a
      low-frequency, high-severity liability with a right-skewed distribution,
      so upper-tail measures such as CVaR provide a natural basis for a
      prudential buffer.}

\item $\lambda$ is the prudential-buffer coefficient: $\lambda=0$ corresponds
      to an actuarially fair (expected-cost) equivalent load, while
      $\lambda>0$ adds an explicit buffer for adverse outcomes (e.g.\ to
      reflect internal risk limits or capital-adequacy considerations).
\end{itemize}
For fixed $(\lambda,\alpha_g)$, \eqref{eq:actuarial} defines the equivalent
MBG load $f_g$ from the payout distribution induced by
$\widehat{\mathcal U}_0^*$. Because $f_g$ is computed ex post and does not alter
the initial state $x_0$ or enter the EW--CVaR control problem, no load--policy
fixed point is required.

The valuation functionals in \eqref{eq:actuarial} are written explicitly as
follows, with both quantities expressed in real (inflation-adjusted) dollars at
time~$t_0$:
\begin{equation}
\label{eq:mbg_theoretical}
\begin{aligned}
\mathbb{E}^{\,x_{0},t_{0}}_{\widehat{\mathcal U}_{0}^{*}}
           \bigl[Z_g\bigr]
&=
\mathbb{E}_{\widehat{\mathcal U}_0^{*}}^{X_{0^+},t_0^+}
\bigl[Z_g \;\big|\;X_{0^-}=x_0 \bigr],
\\[3pt]
\mathrm{CVaR}_{\alpha_g}^{+,\,x_{0},t_{0}}
           \bigl[Z_g\bigr]
&:=
\inf_{\eta\in\mathbb R}
\left\{
\eta+
\frac{1}{\alpha_g}
\mathbb{E}_{\widehat{\mathcal U}_0^{*}}^{X_{0^+},t_0^+}
\!\left[
(Z_g-\eta)^+\;\big|\;X_{0^-}=x_0
\right]
\right\}.
\end{aligned}
\end{equation}
For compactness, and under the convention that $m_\tau=M+1$ when the member
survives to $T$, set $\bar m_\tau:=\min(m_\tau,M)$. Thus
$\bar m_\tau=m_\tau$ when death occurs before the horizon, while
$\bar m_\tau=M$ when the member survives to $T$. The valuation functionals in
\eqref{eq:mbg_theoretical} are evaluated under the system evolution generated
by the learned NN-parameterized policy $\widehat{\mathcal U}_0^*$ up to this
capped interval index:
\begin{equation}
\text{subject to~ }
\begin{cases}
~~\text{(dynamics)} \quad~~ \{X_t\} \text{ evolves via }\eqref{eq:dynamics_generic},
& t \notin \mathcal{T},\; 0\le t < t_{\bar m_\tau} \\[4pt]
\left.\begin{array}{ll}
\text{(tontine)}  &W_{m^-}  \text{ computed via } \eqref{eq:Wm_post_tontine}, \\
\text{(withdrawal)} &W_{m^+} = W_{m^-} - \widehat q_m^*(\cdot) \text{ as in } \eqref{eq:Wm_post_withdrawal}, \\
\text{(rebalancing)} &\text{$X_{m^+}$ computed via \eqref{eq:Xmplus} using } \widehat{\boldsymbol p}_m^*(\cdot),
\end{array}\right\} & m = 0, \ldots, \bar m_{\tau}-1.
    \end{cases}\label{eq:MBG_constraints}
\end{equation}
\begin{remark}[Interpreting $f_g$ in starting--rate terms]
\label{rm:beta0_translation}
Equation~\eqref{eq:actuarial} defines an equivalent up-front load
$f_g\ge0$ on the initial contribution $L_0$ (i.e.\ a one-time cost
measure expressed as a proportion of $L_0$). For ease of interpretation in the
numerical results, we also translate $f_g$ into an implied reduction in a
notional starting payment rate.

Fix a notional starting income rate $\beta_0$ (per dollar of contribution) for
the same tontine design \emph{without} the MBG. Under the reporting convention
that the MBG cost is expressed as an equivalent up-front adjustment to the
initial contribution (or benefit base), for $0\le f_g<1$, the
implied post--load starting rate is
$\beta_g=(1-f_g)\beta_0$, i.e.\ a reduction of $f_g\beta_0$ (or
$10^4f_g\beta_0$ basis points) from the reference rate.

If $f_g=1$, the equivalent load exhausts the initial contribution
and the translated starting rate is zero. If $f_g>1$, the required load exceeds
the initial contribution and therefore cannot be funded solely through an
up-front deduction from $L_0$. Accordingly, the starting-rate translation is
economically meaningful only for $0\le f_g<1$.
We emphasize that $\beta_0$ is used only as a translation device for reporting
and need not correspond to any specific operational funding mechanism.
\end{remark}

\subsection{Simulation-based numerical methods}
\label{ssc:sim_method}
{\dpurple{The valuation functionals in \eqref{eq:mbg_theoretical} are
approximated from $K$ independent sample paths, each representing a full set of
exogenous drivers for index dynamics, mortality inputs, and CPI. The resulting
Monte Carlo estimators of the exact expectation and upper-tail CVaR are denoted
by $\widehat{\mathbb{E}}[Z_g]$ and
$\widehat{\mathrm{CVaR}}^{+}_{\alpha_g}(Z_g)$, respectively.}}
These sample paths are indexed by $k=1,\ldots,K$, and all path-dependent
quantities carry the superscript~$(k)$.

The pricing procedure uses the learned NN-parameterized controls obtained from
Section~\ref{sec:NNs}. These controls are held fixed during the MBG valuation
and applied across all sample paths.
For each path $k = 1, \ldots, K$, the wealth evolution follows the dynamics
generated by the learned NN-parameterized policy
$\widehat{\mathcal U}_{0}^{*}$.
We write $\{X_t^{(k)}\}_{t\in[0,T]}$ for the resulting state trajectory on path
$k$; all withdrawals and allocations at the rebalancing times
$t_m\in\mathcal T$ are the learned network outputs
$\widehat q_m^{*}(W_{m^-}^{(k)},t_m)$ and
$\widehat{\boldsymbol p}_m^{*}(W_{m^+}^{(k)},t_m)$
evaluated with the path-specific inputs.

As summarized in Remark~\ref{rm:stochastic_mortality}, each simulation path
$k$ is equipped with a sequence of one--year conditional death probabilities
$\{\delta_{m-1}^{(k)}\}_{m=1}^M$. These probabilities are converted to
tontine gain rates
$g_m^{(k)} = \delta_{m-1}^{(k)}/\bigl(1-\delta_{m-1}^{(k)}\bigr)$
when computing mortality credits via
\eqref{eq:mod_Tg}--\eqref{eq:Wm_post_tontine}. In addition, they drive the
simulation of death times and MBG payouts along each path as outlined in
Algorithm~\ref{alg:mbg_pricing} below. Let
$\{\mathrm{CPI}_m^{(k)}\}_{m=0}^{M}$ denote the CPI values used on path $k$ at
the decision times. The mortality inputs determine the timing of the MBG
payout, while the CPI values enter the nominal withdrawal accumulation and the
nominal--real conversion used to compute the pathwise real
(inflation-adjusted) MBG payout.

The simulation-based pricing details are described in
Algorithm~\ref{alg:mbg_pricing}. For each path $k = 1,\ldots,K$, we denote by
$Z_g^{(k)}$ the simulated real (inflation-adjusted) dollar MBG payout, obtained
by applying \eqref{eq:Z_g} with the path-specific withdrawal history and death
time.
\begin{algorithm}[!htb]
\caption{Simulation-based MBG pricing under learned NN-parameterized controls}
\label{alg:mbg_pricing}
\begin{algorithmic}[1]
\STATE initialize the active index list $\mathcal K \leftarrow \{1,\dots,K\}$;
\STATE initialize MBG payouts $Z_g^{(k)} \leftarrow 0$, $k = 1,\ldots,K$;
\STATE initialize cumulative nominal withdrawals $\mathcal{C}^{(k)}\leftarrow 0$,
$k = 1, \ldots, K$;
\STATE initialize $X_{0^-}^{(k)}=x_0$ and compute $W_{0^-}^{(k)}$ for all
$k\in\mathcal K$;

\FOR{$m = 0$ \textbf{to} $M-1$}
    \FOR{\textbf{each} $k \in \mathcal K$}
        \STATE compute pathwise $g_{m}^{(k)}$ via \eqref{eq:mod_Tg}
         (see Remark~\ref{rm:stochastic_mortality})
         and $W_{m^-}^{(k)}$ via \eqref{eq:Wm_post_tontine};
        \hfill \COMMENT{mortality credit}

        \STATE $q_m^{(k)} \leftarrow
        \widehat q_m^{*}\!\bigl(W_{m^-}^{(k)},t_m\bigr)$;
        \hfill\COMMENT{real (inflation-adjusted) dollars}
        \STATE $\mathcal{C}^{(k)} \leftarrow \mathcal{C}^{(k)} +
                q_m^{(k)}\,\dfrac{\mathrm{CPI}_{m}^{(k)}}{\mathrm{CPI}_{0}}$;
                \hfill\COMMENT{nominal withdrawal accumulation}
                \label{line:one}
        \STATE $W_{m^+}^{(k)} \leftarrow W_{m^-}^{(k)} - q_m^{(k)}$;
        \STATE obtain $X_{m^+}^{(k)}$ via \eqref{eq:Xmplus} using
               $\widehat{\boldsymbol p}_m^{*}\!\bigl(W_{m^+}^{(k)},t_m\bigr)$;
               \hfill\COMMENT{rebalancing}
        \STATE compute $X_{(m+1)^-}^{(k)}$ via \eqref{eq:dynamics_generic};
        \hfill\COMMENT{simulation over $[t_m^+,t_{m+1}^-]$}
        \STATE draw $B_m^{(k)}\sim\mathrm{Bernoulli}(\delta_m^{(k)})$;
        \hfill\COMMENT{death-time}
        \label{ln:death_bernoulli}
        \IF{$B_m^{(k)}=1$}
            \STATE
            \label{ln:mbg}
              $\displaystyle Z_g^{(k)} \leftarrow
              \frac{\mathrm{CPI}_{0}}{{{\mathrm{CPI}_{m+1}^{(k)}}}}\,
              \max\Bigl(L_{0}-\mathcal{C}^{(k)},\,0\Bigr)$;
              \label{line:two}
            \STATE remove path $k$ from the active list $\mathcal K$;
        \ENDIF
    \ENDFOR
    \IF{$\mathcal K=\emptyset$}
    \STATE break; \hfill \COMMENT{all deaths processed}
    \ENDIF
\ENDFOR

\STATE compute the sample mean:
       ${\displaystyle
       \widehat{\mathbb{E}}[Z_g]
       \;=\;
       \frac{1}{K}\sum_{k=1}^{K}Z_g^{(k)}}$;

\STATE sort the simulated payouts in descending order to obtain
       $Z_{(1)}\ge Z_{(2)}\ge\cdots\ge Z_{(K)}$;

\STATE compute the empirical upper-tail CVaR estimator:
       ${\displaystyle
         \widehat{\mathrm{CVaR}}^{+}_{\alpha_g}(Z_g)
         \;=\;
         \frac{1}{\lceil\alpha_g K\rceil}
         \sum_{j=1}^{\lceil\alpha_g K\rceil} Z_{(j)}}$;

\STATE return
${\bigl(\widehat{\mathbb{E}}[Z_g],
\widehat{\mathrm{CVaR}}^{+}_{\alpha_g}(Z_g)\bigr)}$;

\end{algorithmic}
\end{algorithm}
Algorithm~\ref{alg:mbg_pricing} simulates the account under the learned
NN-parameterized policy until death or the terminal horizon.\footnote{The
estimator of $\mathrm{CVaR}^{+}_{\alpha_g}(Z_g)$ in
Algorithm~\ref{alg:mbg_pricing} is an order-statistic upper-tail average,
computed after sorting the simulated payouts in descending order. When
$\alpha_g K$ is an integer, it coincides with the empirical CVaR associated
with the infimum representation in \eqref{eq:mbg_theoretical}; otherwise, it is
implemented as the average of the largest $\lceil \alpha_g K\rceil$ simulated
MBG payouts.}
On Line~\ref{ln:death_bernoulli}, the death event over
$(t_m^+,t_{m+1}^-]$ is determined by a Bernoulli experiment based on the
pathwise death probability $\delta_m^{(k)}$. If death occurs, the algorithm
computes the real (inflation-adjusted) MBG payout on Line~\ref{ln:mbg}, records
it as $Z_g^{(k)}$, and removes the path from the active list. Paths with no
death before the horizon retain $Z_g^{(k)}=0$.

With $\widehat{\mathbb{E}}[Z_g]$ and
$\widehat{\mathrm{CVaR}}^{+}_{\alpha_g}(Z_g)$ computed by
Algorithm~\ref{alg:mbg_pricing}, the actuarial load factor $f_g$ defined by
\eqref{eq:actuarial} is approximated by
\EQ
\label{eq:widehat_g}
\widehat{f}_g
\;:=\;
(L_0)^{-1}\left(\widehat{\mathbb{E}}[Z_g]
      +\lambda\,\widehat{\mathrm{CVaR}}^{+}_{\alpha_g}(Z_g)\right).
\EN
In the numerical results, we report $\widehat{f}_g$ (and the associated
up-front deduction $\widehat{f}_g L_0$).

\begin{remark}[Inflation treatment in MBG pricing]
\label{rm:cpi_pathwise}
In the numerical experiments, wealth is simulated in real
(inflation-adjusted) units using bootstrapped real (inflation-adjusted) asset
returns, so CPI does not enter the wealth recursion. CPI is used only in
Algorithm~\ref{alg:mbg_pricing} to implement the money-back test in nominal
dollars at the death time and to express the resulting MBG payout back in real
(inflation-adjusted) $t_0$ dollars.

To obtain pathwise inflation adjustments consistent with the bootstrap, CPI is
treated as an additional series in the resampling procedure: monthly CPI
changes are bootstrapped jointly with the asset-return blocks, a CPI index path
is reconstructed along each simulated path, and the corresponding pathwise CPI
ratios are used in the nominal--real conversion step of
Algorithm~\ref{alg:mbg_pricing}. See $\mathrm{CPI}_{m}^{(k)}$ on
Line~\ref{line:one} and $\mathrm{CPI}_{m+1}^{(k)}$ on
Line~\ref{line:two}.
\end{remark}

\section{Contract pooling and tail risk}
\label{sc:mbg_contract_pooling}
The preceding section evaluates the representative contract on a
stand-alone basis. In practice, an issuer would hold a portfolio of broadly
homogeneous MBG contracts, so part of the payout risk can diversify across
contracts. Let $J$ denote the number of contracts over which MBG costs are
aggregated; this is distinct from the tontine-pool size underlying the
homogeneous large-pool mortality-credit approximation. Let
$Z_{g,1},\ldots,Z_{g,J}$ denote MBG payouts on these contracts, each with the
same marginal distribution as the representative-contract payout $Z_g$. Define
the aggregate and pooled per-contract payouts by
\begin{equation}
\label{eq:pooled_mbg_per_contract}
S_{g,J}
:=
\sum_{j=1}^{J}Z_{g,j},
\qquad
\overline Z_{g,J}
:=
\frac{S_{g,J}}{J}.
\end{equation}
Linearity gives
$\mathbb E[\overline Z_{g,J}]=\mathbb E[Z_g]$. More importantly, subadditivity
and positive homogeneity of CVaR give the diversification inequality
\begin{equation}
\label{eq:pooled_mbg_cvar_bound}
\mathrm{CVaR}_{\alpha_g}^{+}(\overline Z_{g,J})
=
\frac{1}{J}\,
\mathrm{CVaR}_{\alpha_g}^{+}(S_{g,J})
\le
\frac{1}{J}\sum_{j=1}^{J}
\mathrm{CVaR}_{\alpha_g}^{+}(Z_{g,j})
=
\mathrm{CVaR}_{\alpha_g}^{+}(Z_g).
\end{equation}
Thus, contract pooling cannot increase the upper-tail CVaR per
contract.

To obtain a practical analytical approximation, decompose the
representative-contract CVaR as
\begin{equation}
\label{eq:mbg_excess_tail_decomposition}
\mathrm{CVaR}_{\alpha_g}^{+}(Z_g)
=
\mathbb E[Z_g]
+
\underbrace{\left(
\mathrm{CVaR}_{\alpha_g}^{+}(Z_g)-\mathbb E[Z_g]
\right)}_{\text{excess-tail risk}}.
\end{equation}
Suppose that contract-specific mortality risk is independent across
contracts and is the dominant source of diversifiable excess-tail variation,
while common-factor dependence is abstracted from in this approximation. For
sufficiently large $J$, a normal approximation suggests that the excess-tail
component of the aggregate CVaR scales approximately as $\sqrt{J}$, and hence
as $J^{-1/2}$ per contract. Approximating the coefficient of this
$J^{-1/2}$ term by the representative-contract excess-tail risk in
\eqref{eq:mbg_excess_tail_decomposition} gives the heuristic
\begin{equation}
\label{eq:pooled_mbg_cvar_approx}
\mathrm{CVaR}_{\alpha_g}^{+}(S_{g,J})
\approx
J\,\mathbb E[Z_g]
+
\sqrt{J}\left(
\mathrm{CVaR}_{\alpha_g}^{+}(Z_g)-\mathbb E[Z_g]
\right),
\end{equation}
and hence
\begin{equation}
\label{eq:pooled_mbg_cvar_per_contract_approx}
\mathrm{CVaR}_{\alpha_g}^{+}(\overline Z_{g,J})
\approx
\mathbb E[Z_g]
+
J^{-1/2}\bigl(
\mathrm{CVaR}_{\alpha_g}^{+}(Z_g)-\mathbb E[Z_g]
\bigr).
\end{equation}
Under this approximation,
$\mathrm{CVaR}_{\alpha_g}^{+}(\overline Z_{g,J})
\to\mathbb E[Z_g]$ as $J\to\infty$.

Using \eqref{eq:pooled_mbg_cvar_per_contract_approx}, define the corresponding
approximate pooled-contract load $f_{g,J}^{\mathrm{a}}$ by
\begin{equation}
\label{eq:pooled_mbg_load_approx}
f_{g,J}^{\mathrm{a}}
:=
(L_0)^{-1}\,\left(
\mathbb E[Z_g]
+
\lambda\left[
\mathbb E[Z_g]
+
J^{-1/2}\bigl(
\mathrm{CVaR}_{\alpha_g}^{+}(Z_g)-\mathbb E[Z_g]
\bigr)
\right]
\right).
\end{equation}
Consequently, as $J\to\infty$,
$f_{g,J}^{\mathrm{a}}
\longrightarrow
\frac{(1+\lambda)\mathbb E[Z_g]}{L_0}$.

For numerical implementation, the exact expectation and upper-tail
CVaR in
\eqref{eq:pooled_mbg_cvar_per_contract_approx}--\eqref{eq:pooled_mbg_load_approx}
are replaced by the Monte Carlo estimators
$\widehat{\mathbb E}[Z_g]$ and
$\widehat{\mathrm{CVaR}}_{\alpha_g}^{+}(Z_g)$ from
Algorithm~\ref{alg:mbg_pricing}. The resulting estimated approximate quantities
are defined by
\begin{equation}
\label{eq:pooled_mbg_estimates}
\begin{aligned}
\widehat{\mathrm{CVaR}}_{\alpha_g}^{+,\mathrm{a}}
    (\overline Z_{g,J})
&:=
\widehat{\mathbb E}[Z_g]
+
J^{-1/2}
\left(
\widehat{\mathrm{CVaR}}_{\alpha_g}^{+}(Z_g)
-
\widehat{\mathbb E}[Z_g]
\right),
\\[3pt]
\widehat f_{g,J}^{\mathrm{a}}
&:=
(L_0)^{-1}\left(
\widehat{\mathbb E}[Z_g]
+
\lambda\,
\widehat{\mathrm{CVaR}}_{\alpha_g}^{+,\mathrm{a}}
    (\overline Z_{g,J})
\right).
\end{aligned}
\end{equation}
These are the quantities evaluated in the contract-pooling numerical
results.

The approximation abstracts from common market, inflation, and
systematic-mortality dependence, which may leave residual tail risk. A direct
portfolio-level CVaR calculation preserving these dependencies would require
joint simulation of MBG liabilities across contracts and is left for future
work.

\section{Numerical results: retirement outcomes and MBG pricing}
\label{sec:diverse_port}
\subsection{Asset return data and preprocessing}
\label{ssc:data}
For the internationally diversified experiments, we take the perspective of a domestic (Australian) investor and work with four indices: Australian equities and government bonds (domestic), and U.S.\ equities and government bonds (foreign) converted into
AUD. Unless otherwise noted, all return
series in this section are measured at a monthly frequency in real AUD terms,
where ``real'' means inflation-adjusted. Monetary quantities in this section are reported in thousands of real AUD.

Our empirical sample runs from 1935:1 to 2022:12. Where necessary, individual series are backcast and truncated so that all four indices and the Australian CPI are jointly available over a common horizon. This long historical panel is used both to describe the ``historical market'' and as input to the bootstrap simulations described below. Detailed construction of the asset-return, AUD/USD exchange rate, and CPI series is reported in Appendix~\ref{app:diversified_data}.

\vspace*{-0.25cm}
\paragraph{Historical market and bootstrap sampling.}
Future index paths in the internationally diversified experiments are generated nonparametrically using the stationary block bootstrap \cite{politis1994stationary,politis2004automatic,patton2009correction,dichtl2016testing}. We apply geometrically distributed block lengths and sample blocks of four--asset returns jointly to preserve both cross--sectional correlations and serial dependence. The expected block length is chosen following the data--driven procedure of \cite{patton2009correction} applied to each series, and then harmonised across assets to obtain a common effective block size of the order of a few years. All bootstrap sampling is carried out on the preprocessed monthly real AUD return series described above.
\vspace*{-0.25cm}
\paragraph{Return characteristics and dependence structure.}
Tables~\ref{tab:return_stats_summary} and~\ref{tab:return_stats_corr}
summarize the preprocessed monthly real AUD returns
across all four assets.
\begin{table}[!htbp]\centering
\vspace*{-0.25cm}
\caption{Summary statistics of monthly real returns (1935:1--2022:12): annualized mean,
annualized geometric mean, and annualized volatility. $\mathrm{VaR}_{0.05}$ (mo.) and $\mathrm{CVaR}_{0.05}$ (mo.) report the $5\%$ Value-at-Risk and CVaR of monthly returns.}
\label{tab:return_stats_summary}
\begin{tabular}{lccccc}
\hline
Asset & Mean (ann.) & Geo. mean (ann.) & Vol (ann.) & $\mathrm{VaR}_{0.05}$ (mo.) & $\mathrm{CVaR}_{0.05}$ (mo.) \\
\hline
U.S.\ 30--day T--bill  & 0.003 & -0.001 & 0.095 & -0.041 & -0.056 \\
U.S.\ equity index     & 0.082 & 0.071  & 0.165 & -0.064 & -0.096 \\
AU 10--year bond & 0.012 & 0.012 & 0.036 & -0.015 & -0.025 \\
AU equity index & 0.069 & 0.059 & 0.151 & -0.065 & -0.100 \\
\hline
\end{tabular}
\vspace*{-0.25cm}
\end{table}

\begin{table}[!htbp]\centering
\vspace*{-0.25cm}
\caption{Correlation matrix of monthly real returns (1935:1--2022:12).}
\label{tab:return_stats_corr}
\begin{tabular}{lcccc}
\hline
\noalign{\vspace{2mm}}
 & \shortstack{U.S.\ 30-day\\T--bill} & \shortstack{U.S.\\equity} & \shortstack{AU 10-year \\bond} & \shortstack{AU\\equity}\\
 \noalign{\vspace{1mm}}
 \hline
U.S.\ 30-day T--bill & 1.00 & 0.34 & 0.17 & -0.22
\\[+3pt]
U.S.\ equity           & 0.34 & 1.00 & 0.10 & 0.33
\\
AU 10-year bond     & 0.17 & 0.10 & 1.00 & 0.11
\\
AU equity            & -0.22 & 0.33 & 0.11 & 1.00
\\
\hline
\end{tabular}
\end{table}

As indicated in Tables~\ref{tab:return_stats_summary}--\ref{tab:return_stats_corr}, U.S.\ equity has the highest
historical average real return in this sample and is
only weakly correlated with Australian equity (correlation $\approx 0.33$, well below 1)\footnote{Over 1990:1--2022:12, the correlation
is 0.48 (vs 0.33 in the full sample), indicating higher dependence in the recent period.}. This comparatively low cross-country equity correlation suggests
meaningful scope for international diversification, even among developed equity markets, a point we revisit later.
By contrast, once expressed in real AUD, the U.S.\ 30--day T--bill has a near--zero average real return and substantial volatility inherited from exchange--rate fluctuations, making short--maturity foreign fixed income a less attractive defensive asset than domestic government bonds. These empirical patterns are revisited when interpreting the learned controls below.

\subsection{Mortality assumptions}
\label{ssc:diverse_mortality}
In the internationally diversified experiments, we model the lifetime of a
65--year--old Australian male and consider both deterministic and stochastic
mortality, consistent with the general framework in
Subsection~\ref{ssc:mortality_models} and
Remark~\ref{rm:stochastic_mortality}.

For the deterministic specification, we use the 2021 column of the Australian
male $1\times1$ period life table from the Human Mortality Database
(HMD, \cite{HMD}); the full file \texttt{mltper\_1x1.txt} covers calendar
years 1921--2021.%
\footnote{Available at
\url{https://www.mortality.org/File/GetDocument/hmd.v6/AUS/STATS/mltper_1x1.txt}.}
We hold the period-table year fixed at $y_0=2021$ over the retirement
horizon. Thus, with $a_0=65$, the deterministic death-probability sequence is
$\delta_{m-1}=q_{65+(m-1),\,2021}$, $m=1,\ldots,M$, consistently with
\eqref{eq:delta_m}.
In a homogeneous pool we set
{\dpurple{$\delta_{m-1}^{\ell}\equiv\delta_{m-1}$ for all members~$\ell$}}, and the associated
tontine gain rates $g_m$ entering the wealth recursion follow from
\eqref{eq:tontine-gain}.

For stochastic mortality, annual Australian male deaths and central
exposures for ages $55$--$95$ and calendar years $1980$--$2021$ are used.
The central exposures are converted to initial exposures, and both the LC and
CBD models are fitted using a Binomial likelihood with a logit link in
\textsf{R} using \textsf{StMoMo} \cite{villegas2018stmomo}. We simulate
$256{,}000$ mortality sequences and extract each sequence diagonally from the
projected mortality surface, beginning with age $65$ in
$y_{\mathrm{proj}}=2022$. The data construction, calibration, and simulation
details are reported in
Appendix~\ref{app:diversified_mortality_calibration}.

%
%
%

\subsection{Experimental setup and scenarios}
\label{ssc:diverse_setup}
For the internationally diversified experiments, we retain the same decumulation framework used in the benchmark validation (Table~\ref{tab:scenario}), but now from the perspective of a 65--year--old Australian male retiree investing in the four indices described in Subsection~\ref{ssc:data}. The mortality assumptions follow Subsection~\ref{ssc:diverse_mortality}.

More specifically, the retiree starts at time $t_0 = 0$ with initial wealth $W_0 = 1000$, corresponding to an initial purchase price $L_0 = 1000$ (thousand AUD at inception) for the
tontine product with the embedded MBG overlay described in Section~\ref{sc:mbg}.
The planning horizon is $T = 30$ years with annual decision times $t_m = m$, $m = 0,\ldots,M$, where $M=30$.
The minimum and maximum annual withdrawals are fixed at $q_{\min} = 40$ and $q_{\max} = 80$. In addition, we also enforce the no--shorting and no--leverage constraints
as in Section~\ref{sc:modeling}. In these experiments, following QSuper's Lifetime Pension PDS~\cite{qsuper2025pds}, we use an annual tontine fee of \mbox{$\varrho = 0.11\%$ (11~bps).}

Risk and reward are measured by the same EW--CVaR criterion as in Section~\ref{sc:formulation}:
expected total real withdrawals over $[0,T]$ and the lower-tail $\mathrm{CVaR}_{0.05}(W_T)$.
For a given scalarization parameter $\gamma>0$, we first train the NNs to obtain the learned policy $\widehat{\mathcal U}_0^*(\gamma)$ and
compute the corresponding point on the EW--CVaR efficient frontier
under either deterministic table mortality or stochastic GAPC--based mortality. The MBG overlay is then priced ex post, holding $\widehat{\mathcal U}_0^*(\gamma)$ fixed, using the actuarial pricing rule \eqref{eq:actuarial} and the simulation procedure in Section~\ref{sc:mbg_pricing}.

\subsection{EW--CVaR frontiers: deterministic mortality}
\label{ssc:diverse_frontier}
We now discuss the EW--CVaR efficient frontiers for both cases:
two--asset (domestic--only) and four--asset (internationally diversified).

As in \cite{forsyth2024optimal}, we include a constant--real--withdrawal, constant--weight benchmark as a reference point on the EW--CVaR frontiers.
In line with the 4\% rule of \cite{bengen1994},
the benchmark always
withdraws $40$ (thousand real AUD) per year and maintains fixed portfolio
weights over time. For each case,
we select the benchmark by grid search over asset weights in 10\% increments,
choosing the allocation that maximizes
$\mathrm{CVaR}_{0.05}$ of terminal wealth.


In the two--asset case, the search is over equity fractions
$p^{\dom}_s \in \{0,0.1,\ldots,1\}$ with $p^{\dom}_b = 1 - p^{\dom}_s$; the best
constant--weight benchmark has a 50\% domestic equity allocation
and $\mathrm{CVaR}_{0.05}(W_T) \approx -617.1$. In the four--asset case, we
search over all weight vectors on the 10\% grid whose components sum to one; the
best benchmark has allocation
$(p^{\dom}_s, p^{\dom}_b, p^{\f}_s, p^{\f}_b) = (0.1, 0.3, 0.2, 0.4)$
with $\mathrm{CVaR}_{0.05}(W_T) \approx -446.1$.
These two constant--strategy benchmarks appear as single
points in the efficient--frontier figures for the two--asset and
four--asset experiments, respectively.

\begin{figure}[hbt!]
\centering

\begin{minipage}[t]{0.5\textwidth}
\vspace{0pt}
\centering
\includegraphics[
    width=0.9\textwidth,
]{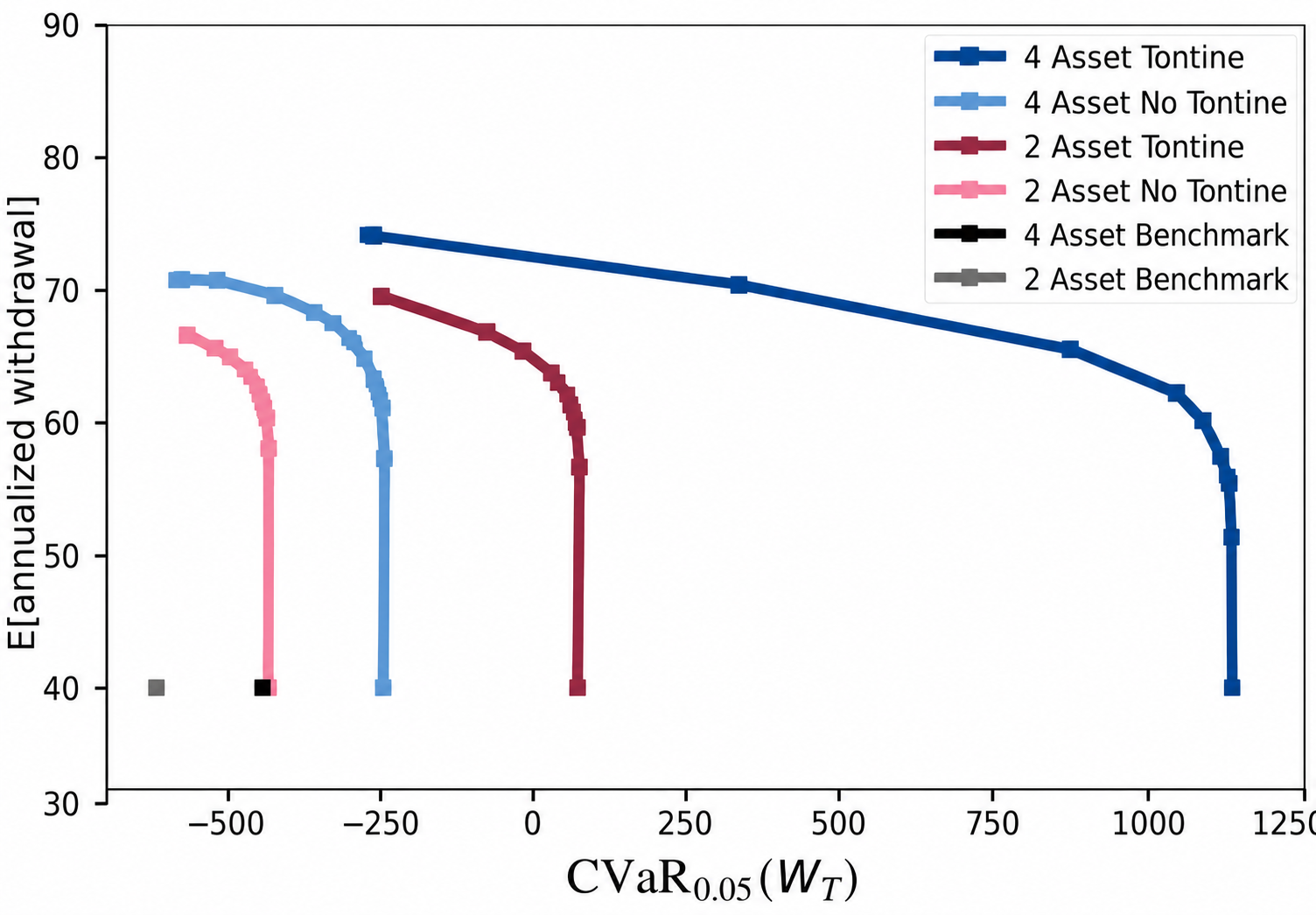}
\end{minipage}
\hfill
\begin{minipage}[t]{0.45\textwidth}
\vspace{0pt}
\raggedright
\begingroup
\setlength{\abovecaptionskip}{0pt}
\captionof{figure}{EW--CVaR efficient frontiers for domestic--only
(two--asset) and internationally diversified (four--asset) portfolios,
with and without a tontine overlay. Constant--strategy benchmarks are
shown as single points.}
\label{fig:diversified_frontier}
\endgroup
\end{minipage}

\vspace*{-0.5cm}
\end{figure}
\noindent Figure~\ref{fig:diversified_frontier} shows the EW--CVaR efficient frontiers
traced out by the learned policies in the two--asset
(domestic--only) and four--asset (internationally diversified) cases, both
with and without a tontine overlay, together with the corresponding
constant--strategy benchmarks.

\noindent We make the following key observations about Figure~\ref{fig:diversified_frontier}:
\begin{itemize}[noitemsep, topsep=2pt, leftmargin=*]
\item
The internationally diversified (four--asset)
constant--strategy benchmark lies essentially on the two--asset
no--tontine frontier. Thus, at the 4\%--rule withdrawal level, the
internationally diversified constant--strategy benchmark achieves
an EW--CVaR trade--off comparable to that generated by the
domestic--only learned EW--CVaR policy without a tontine
overlay. Moreover, relative to the two--asset constant--strategy
benchmark, international diversification produces a substantial improvement
in CVaR at the same withdrawal level. These
comparisons provide evidence that international diversification
can materially improve the risk--reward trade--off even before longevity
pooling is introduced.
\\
\item
In both the domestic--only and internationally diversified settings,
the efficient frontiers traced out by the learned policies lie
well above and to the right of their corresponding
constant--strategy benchmark points.
Within the calibrated setting, this shows that 4\%--rule--style
strategies with constant real withdrawals and constant portfolio weights are
substantially less efficient than the learned NN dynamic policies obtained
from our EW--CVaR optimization.
For example, even at
$\mathrm{CVaR}_{0.05}(W_T)\approx 500$ (thousand real AUD) at age 95,
corresponding to a substantial lower--tail terminal--wealth buffer,
the four--asset tontine frontier supports expected annual withdrawals of about
$70$ (thousand real AUD), or roughly $7\%$ of initial wealth, compared with the
fixed $40$ (thousand real AUD, $4\%$) annual withdrawal under the
constant--strategy benchmark.

\item
Comparing the two ``no tontine'' curves, the four--asset frontier is generally shifted
upwards and to the right relative to the two--asset frontier over the
practically relevant range of $\mathrm{CVaR}_{0.05}(W_T)$.
Allowing the retiree to invest in both Australian and U.S.\ assets therefore supports higher expected annualized withdrawals
for a comparable, or even improved, level of downside risk relative to a purely domestic portfolio.

\smallskip
This is consistent with Tables~\ref{tab:return_stats_summary}--\ref{tab:return_stats_corr}:
in this sample, U.S.\ equity has a higher average real return than Australian equity, and the cross--country equity correlation is well below one. Thus, adding foreign equity provides potential to improve the EW--CVaR trade--off even when used selectively rather than through uniformly higher equity exposure.

\item
For both asset settings, the tontine frontier dominates the corresponding
no--tontine frontier, with especially pronounced gains in the four--asset
case. Across a broad range of scalarization parameters $\gamma$,
the four--asset tontine strategies deliver both higher expected annual
withdrawals and higher $\mathrm{CVaR}_{0.05}(W_T)$ than the
corresponding four--asset no--tontine strategies.
Together with the preceding comparison, this illustrates the
substantial combined benefit of international diversification and longevity
pooling.
\end{itemize}
In the next subsection, we examine how these frontiers shift when mortality is modelled stochastically.

\subsection{EW--CVaR frontiers: effects of stochastic mortality}
\label{ssc:effect_mortality}
\label{ssc:mortality_model_robustness}
We now examine how introducing stochastic mortality affects the EW--CVaR
efficient frontier in the four--asset tontine setting. Throughout this
subsection, we keep the assets, fee structure, control constraints, and
scalarization setup identical to Subsection~\ref{ssc:diverse_frontier}; only
the mortality specification changes.

\begin{figure}[hbt!]
\centering
\begin{minipage}[t]{0.5\textwidth}
\vspace{0pt}
\centering
\includegraphics[
    width=0.9\textwidth,
]{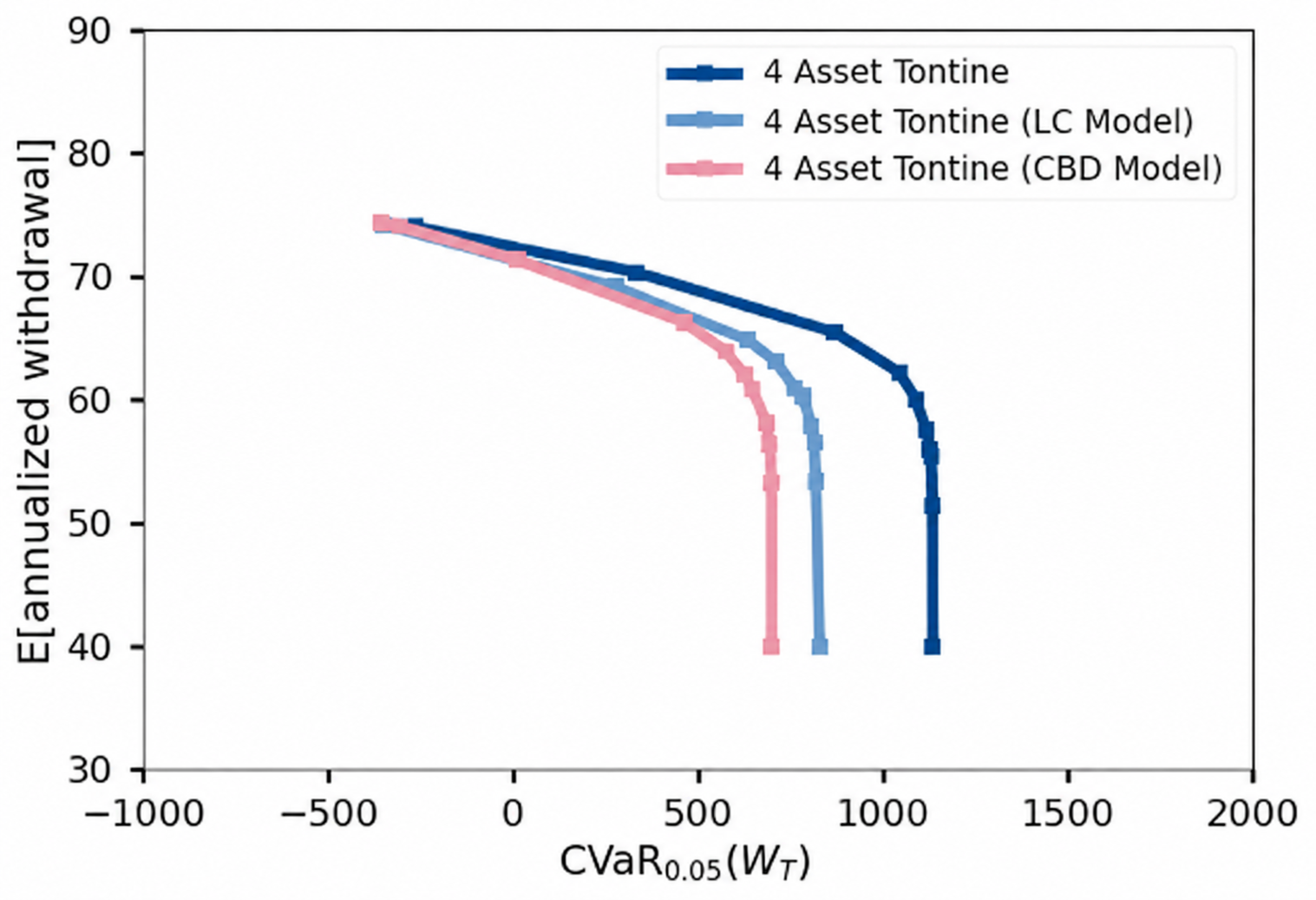}
\end{minipage}
\hfill
\begin{minipage}[t]{0.48\textwidth}
\vspace{0pt}
\raggedright
\begingroup
\setlength{\abovecaptionskip}{0pt}
\captionof{figure}{EW--CVaR efficient frontiers for the four--asset tontine
portfolio under deterministic table mortality and stochastic mortality (LC and CBD models).}
\label{fig:stoch_mort_frontier}
\endgroup
\end{minipage}
\end{figure}
Figure~\ref{fig:stoch_mort_frontier} compares the EW--CVaR efficient frontiers
for the four--asset tontine portfolio under three mortality specifications:
(i) deterministic table mortality as in
Subsection~\ref{ssc:diverse_mortality},
(ii) stochastic LC mortality, and
(iii) stochastic CBD mortality.

A key point for interpretation is that the horizon corresponds to a very
advanced age (age 95 in the baseline calibration with a 30--year horizon).
Consequently, frontier points with very high terminal--wealth CVaR,
for example, $\mathrm{CVaR}_{0.05}(W_T)\approx 1000$ (thousand real AUD),
correspond to policies that place substantial weight on preserving residual
wealth at age 95. Values around
$\mathrm{CVaR}_{0.05}(W_T)\approx 500$ already represent a sizeable
terminal--wealth buffer relative to the annual withdrawal amounts considered.
Therefore, for economic interpretation, we focus on the region of
Figure~\ref{fig:stoch_mort_frontier} at or below roughly
$\mathrm{CVaR}_{0.05}(W_T)=500$, where the
controls correspond to higher expected withdrawals rather than
preserving large terminal balances.

\medskip
\noindent
Over this region, the impact of stochastic mortality is visible but
modest: the stochastic frontiers track the deterministic table frontier
closely, with only a relatively small
leftward and downward shift.
This shift is consistent with systematic longevity improvement in
the stochastic mortality models: when death probabilities fall relative to the
deterministic baseline table, survivors receive smaller mortality credits,
which lowers terminal wealth in the adverse outcomes that determine
$\mathrm{CVaR}_{0.05}(W_T)$.
Although systematic longevity improvement under the LC and CBD
models reduces the mortality-credit support available to survivors relative to the deterministic baseline, the four--asset tontine continues to
offer an attractive combination of high expected withdrawals and improved
$\mathrm{CVaR}_{0.05}(W_T)$ relative to the domestic--only and
constant--weight strategies considered in
Subsection~\ref{ssc:diverse_frontier}
(Figure~\ref{fig:diversified_frontier}).

A further observation is that the LC and CBD frontiers are close to each other,
with the CBD curve generally lying slightly to the left of the LC curve over
this region. Conditional on modelling systematic longevity risk, the specific
choice between the LC and CBD specifications therefore has only a relatively
minor effect on the EW--CVaR trade--off.

{\dpurple{As a further robustness check, we evaluate the learned controls
trained under LC mortality on out-of-sample CBD mortality realizations, without
retraining, and compare them with controls trained and evaluated under CBD
mortality. The resulting frontiers are nearly coincident; for $\gamma=1.5$ and
$\lambda=0.05$, the corresponding representative-contract equivalent MBG loads
are $\widehat f_g=0.114$ and $\widehat f_g=0.113$, respectively. Full results
are reported in Appendix~\ref{app:mortality_model_robustness}.}}

The learned controls provide a wealth- and time-dependent interpretation of the frontier gains from international diversification.
The complete control heatmaps and the comparison with the domestic-only
allocation are reported in Appendix~\ref{ssc:heatmaps}.

\subsection{Representative-contract MBG pricing}
\label{ssc:mbg_pricing_results}
\label{sssc:mbg_basecase}
We now report simulation--based representative-contract MBG pricing
results using the actuarial pricing formula~\eqref{eq:actuarial} and the Monte
Carlo procedure in Algorithm~\ref{alg:mbg_pricing}.
As noted in Section~\ref{sc:mbg}, the deceased member's tontine account is
forfeited to the pool and supports mortality credits for surviving members; it
is therefore unavailable to finance the MBG death benefit, which must be funded
separately. {\dpurple{The effect of aggregating MBG costs across contracts is discussed in
Subsection~\ref{ssc:mbg_contract_pooling}.}} In all cases, the MBG is priced
\emph{ex post} under the fixed learned policy as discussed previously.

\noindent
\begin{minipage}[t]{0.62\textwidth}
Unless otherwise stated, the MBG pricing experiments use the common base--case
parameter set reported in Table~\ref{tab:mbg_basecase_parameters}. The choice
$\gamma=1.5$ selects a representative point on the EW--CVaR frontier, $L_0$ is
measured in thousands of AUD at inception, and $\beta_0$ is the reference
payment rate device described in
Remark~\ref{rm:beta0_translation}.
\end{minipage}
\hfill
\begin{minipage}[t]{0.38\textwidth}
\vspace*{-0.5cm}
\centering
\captionof{table}{Base--case parameters for the MBG pricing experiments.}
\label{tab:mbg_basecase_parameters}
\begin{tabular}{|c|c|c|c|c|}
\hline
$\gamma$
& $L_0$
& $\alpha_g$
& $\lambda$
& $\beta_0$ \\
\hline
$1.5$
& $1000$
& $5\%$
& $0.05$
& $0.05$ \\
\hline
\end{tabular}
\end{minipage}

\vspace*{0.25cm}
\noindent The prudential-buffer coefficient $\lambda=0.05$ is an illustrative
loading choice.
Sensitivity to the prudential-buffer coefficient $\lambda$, the
upper-tail probability $\alpha_g$, and the retiree's scalarization parameter
$\gamma$, is reported in
Appendix~\ref{app:mbg_pricing_supplement_sensitivity}.

Under this base case, Table~\ref{tab:mbg_basecase_compare} compares MBG pricing
across four settings: two--asset (domestic--only) and four--asset
(internationally diversified) portfolios, each under deterministic table
mortality and stochastic LC mortality. The final column reports the implied
post--load notional starting payment rate
$(1-\widehat{f}_g)\beta_0$ (see Remark~\ref{rm:beta0_translation}).
Monte Carlo standard errors are reported in
Appendix~\ref{app:mbg_pricing_supplement_error}
(Table~\ref{tab:mbg_mc_se_basecase}).

\begin{table}[!htb]
\vspace*{-0.25cm}
\centering
\caption{MBG pricing across asset and mortality settings under the base--case
parameters in
Table~\ref{tab:mbg_basecase_parameters}.}
\label{tab:mbg_basecase_compare}
\begin{tabular}{|l|c|c|c|c|}
\hline
\textbf{Setting}
& $\widehat{\mathbb{E}}[Z_g]$
& $\widehat{\mathrm{CVaR}}^{+}_{0.05}(Z_g)$
& $\widehat{f}_g$
& $(1-\widehat{f}_g)\beta_0$ \\
\hline
2 assets, table mortality
& $65.63$
& $736.62$
& $0.102$
& $4.49\%$ \\

2 assets, LC mortality
& $66.68$
& $755.48$
& $0.104$
& $4.48\%$ \\

4 assets, table mortality
& $69.33$
& $736.82$
& $0.106$
& $4.47\%$ \\

4 assets, LC mortality
& $70.69$
& $758.28$
& $0.109$
& $4.46\%$ \\
\hline
\end{tabular}
\end{table}
The central finding from Table~\ref{tab:mbg_basecase_compare} is that the
material improvement in the EW--CVaR frontier from international diversification
translates into only a modest change in MBG pricing. Across the four settings,
the expected payouts and upper-tail CVaR estimates remain close, yielding
equivalent loads between $0.102$ and $0.109$.

To examine this pricing stability further, we use the fact that the MBG payout
$Z_g\ge0$ to write
$\widehat{\mathbb E}[Z_g]=
\widehat{\mathbb P}(Z_g>0)
\widehat{\mathbb E}[Z_g\mid Z_g>0]$,
which decomposes the expected MBG payout into the frequency and conditional
severity of positive payouts. Table~\ref{tab:mbg_frequency_severity} reports
these components across the four settings.

\begin{table}[h]
\vspace*{-0.25cm}
\centering
\caption{Decomposition of $\widehat{\mathbb E}[Z_g]$
from Table~\ref{tab:mbg_basecase_compare}.}
\label{tab:mbg_frequency_severity}
\begin{tabular}{|l|c|c|}
\hline
\textbf{Setting}
& $\widehat{\mathbb P}(Z_g>0)$
& $\widehat{\mathbb E}[Z_g\mid Z_g>0]$
\\
\hline
2 assets, table mortality
& $17.59\%$
& $373.2$ \\

2 assets, LC mortality
& $16.78\%$
& $397.3$ \\

4 assets, table mortality
& $19.30\%$
& $359.2$ \\

4 assets, LC mortality
& $18.33\%$
& $385.7$ \\
\hline
\end{tabular}
\vspace*{-0.25cm}
\end{table}
As shown in Table~\ref{tab:mbg_frequency_severity}, moving from two
to four assets slightly raises the probability of a positive MBG payout but
lowers its conditional mean, while replacing table mortality with LC mortality
has the opposite effect. These offsetting frequency and severity effects keep
expected payouts close across the four settings. The corresponding CVaR
estimates in Table~\ref{tab:mbg_basecase_compare} also vary only modestly.

While this decomposition explains the modest variation across the asset
and mortality specifications, the large upper tail remains
economically important for the equivalent MBG load. The actuarially fair
expected--cost component $\widehat{\mathbb{E}}[Z_g]$ is modest (about
$6.6\%$--$7.1\%$ of $L_0$). Thus, under actuarially fair pricing
($\lambda=0$), the implied equivalent load would be only
$\widehat{f}_g\approx\widehat{\mathbb{E}}[Z_g]/L_0
\approx6\%$--$7\%$, which translates, for $\beta_0=5\%$, into an
illustrative starting-rate reduction of roughly
$33$--$35$~bps. By contrast, the tail measure
$\widehat{\mathrm{CVaR}}^{+}_{0.05}(Z_g)$ is large (about
$0.74$--$0.76$ of $L_0$, roughly an order of magnitude larger than
$\widehat{\mathbb{E}}[Z_g]$). With $\lambda=0.05$, the resulting equivalent
load is $\widehat{f}_g\approx0.102$--$0.109$, corresponding to a post--load
notional starting payment rate of about $4.46\%$--$4.49\%$.

\begin{figure}[h]
  \centering
  \begin{minipage}{0.49\textwidth}
    \centering
    \includegraphics[width=0.9\textwidth]{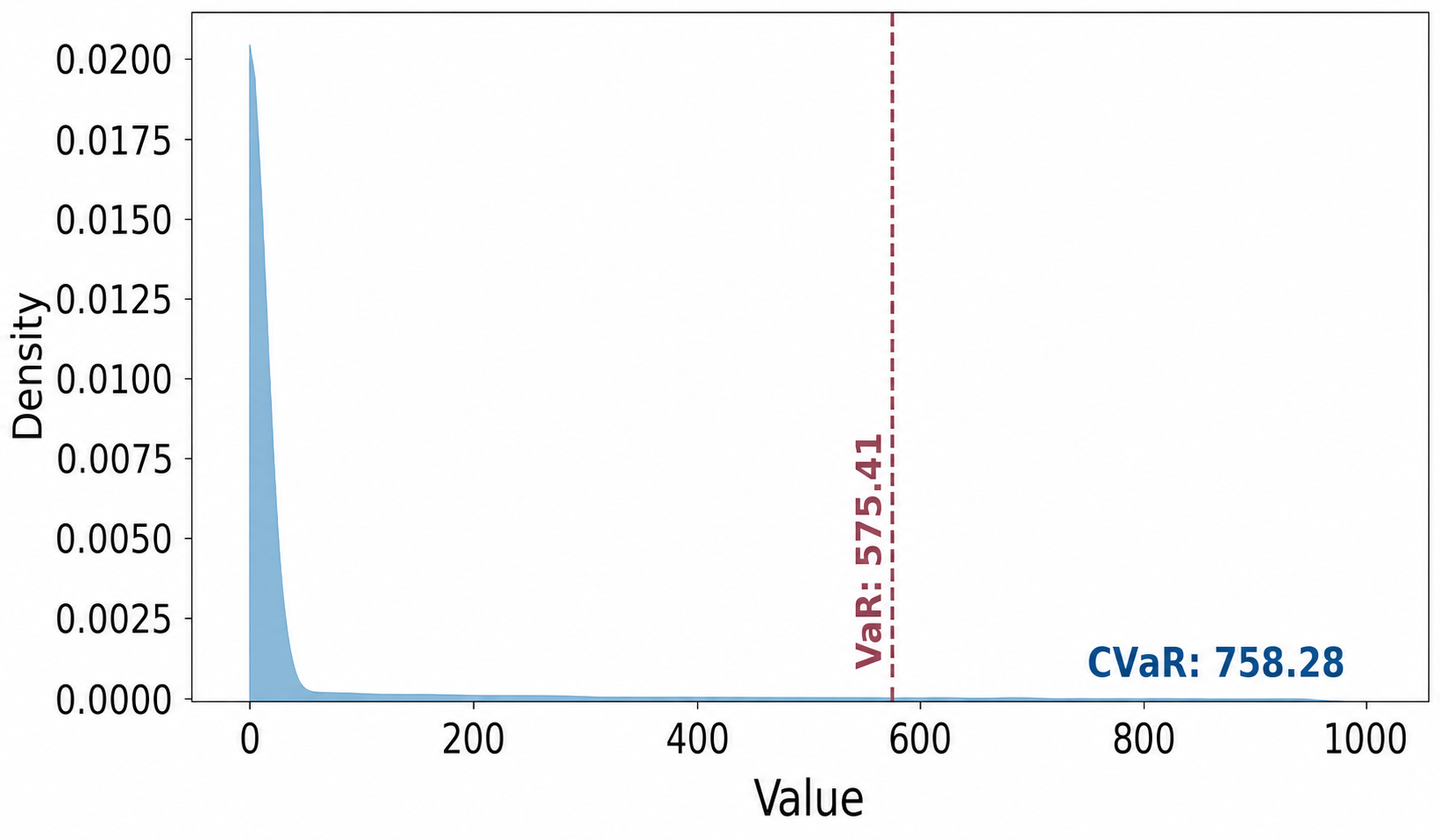}
    \caption*{(a) Unconditional empirical distribution of $Z_g$}
  \end{minipage}
  \hfill
  \begin{minipage}{0.5\textwidth}
    \centering
    \includegraphics[width=0.9\textwidth]{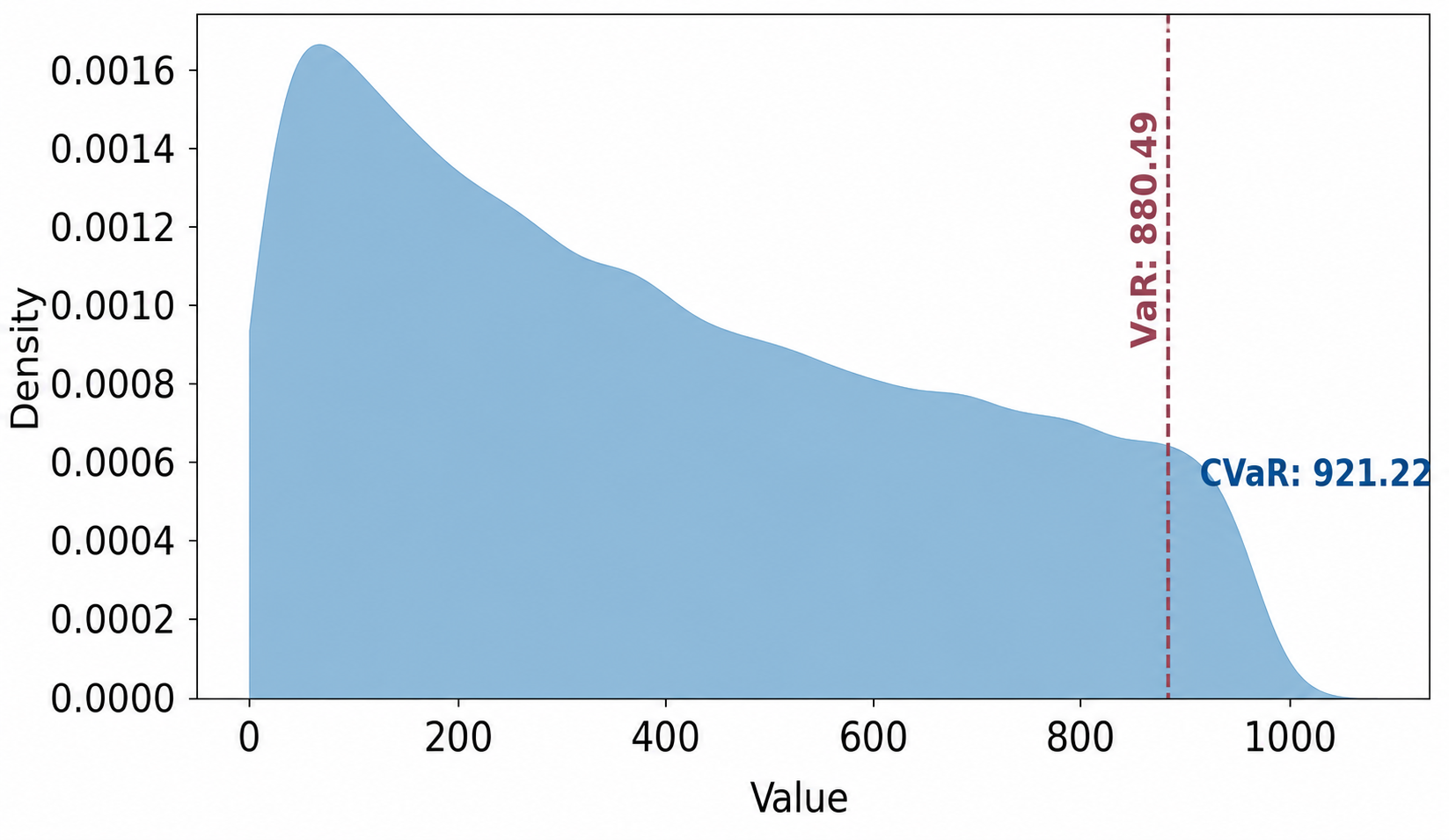}
    \caption*{(b) Empirical distribution conditional on $Z_g>0$}
  \end{minipage}
  \caption{Empirical distributions of the representative-contract MBG payout
  $Z_g$ (in real dollars at inception): the four--asset tontine with stochastic
  mortality (LC) under the base--case parameters from
  Table~\ref{tab:mbg_basecase_parameters}.}
  \label{fig:mbg_density}
\end{figure}
Figure~\ref{fig:mbg_density} illustrates the distributional source of
this large tail contribution. It plots the empirical distribution of the MBG
payout $Z_g$ for the four--asset (internationally diversified) tontine with
stochastic mortality (LC) and $\gamma=1.5$; the upper-tail probability is
$\alpha_g=5\%$.
Figure~\ref{fig:mbg_density}~(a) shows the unconditional empirical
distribution of $Z_g$, which has a substantial point mass at zero (many
scenarios in which the guarantee is out of the money) and a long right tail.
Figure~\ref{fig:mbg_density}~(b) shows the empirical distribution
conditional on $Z_g>0$, highlighting that when the MBG is triggered the
payout can still be large and widely spread. In
Figure~\ref{fig:mbg_density}~(a), the
5\% upper-tail VaR threshold is approximately $575$
(thousand real AUD) and the corresponding upper-tail mean is
$\widehat{\mathrm{CVaR}}^{+}_{0.05}(Z_g)\approx758$, consistent
with Table~\ref{tab:mbg_basecase_compare}.

Taken together, the representative-contract results show that
international diversification materially improves retirement outcomes while
changing the MBG load only modestly; the large upper-tail exposure nevertheless
materially raises the load above its expected-cost level.

\subsection{Contract pooling results}
\label{ssc:mbg_contract_pooling}
Section~\ref{sc:mbg_contract_pooling} develops the contract-book
approximation and defines the estimated approximate quantities in
\eqref{eq:pooled_mbg_estimates}. We now evaluate these quantities for the
four--asset LC base case. Table~\ref{tab:mbg_basecase_compare} reports the
representative-contract estimates for this case:
$\widehat{\mathbb E}[Z_g]=70.69$,
$\widehat{\mathrm{CVaR}}_{0.05}^{+}(Z_g)=758.28$, and
$\widehat f_g=0.109$ for $\lambda=0.05$.
Table~\ref{tab:mbg_pooling_approx} reports the resulting estimated approximate
pooled per-contract CVaR, {\dpurple{equivalent load, and implied post--load
notional starting payment rate
$(1-\widehat f_{g,J}^{\mathrm{a}})\beta_0$}}, with Monte Carlo standard errors
reported in Appendix~\ref{app:mbg_pooling_uncertainty}.
\begin{table}[!htb]
\vspace*{-0.25cm}
\centering
\caption{Approximate contract-pooling results: the four--asset
tontine with stochastic mortality (LC), based on the estimators in
\eqref{eq:pooled_mbg_estimates} with $\lambda=0.05$
{\dpurple{and $\beta_0=0.05$. }}}
\label{tab:mbg_pooling_approx}
\begin{tabular}{|c|c|c|c|}
\hline
$J$
& $\widehat{\mathrm{CVaR}}_{0.05}^{+,\mathrm{a}}(\overline Z_{g,J})$
& {\dpurple{$\widehat f_{g,J}^{\mathrm{a}}$}}
& {\dpurple{$(1-\widehat f_{g,J}^{\mathrm{a}})\beta_0$}}
\\
\hline
$1$
& $758.28$
& $0.109$
& {\dpurple{$4.46\%$}} \\

$10$
& $288.13$
& $0.085$
& {\dpurple{$4.57\%$}} \\

$50$
& $167.93$
& $0.079$
& {\dpurple{$4.60\%$}} \\

$100$
& $139.45$
& $0.078$
& {\dpurple{$4.61\%$}} \\

$500$
& $101.44$
& $0.076$
& {\dpurple{$4.62\%$}} \\

$1000$
& $92.44$
& $0.075$
& {\dpurple{$4.62\%$}} \\

$10{,}000$
& $77.57$
& $0.075$
& {\dpurple{$4.63\%$}} \\

$20{,}000$
& $75.56$
& $0.074$
& {\dpurple{$4.63\%$}} \\

$J\to\infty$
& $70.69$
& $0.074$
& {\dpurple{$4.63\%$}} \\
\hline
\end{tabular}
\vspace*{-0.25cm}
\end{table}

Table~\ref{tab:mbg_pooling_approx} shows that the estimated
approximate pooled per-contract upper-tail CVaR
$\widehat{\mathrm{CVaR}}_{0.05}^{+,\mathrm{a}}
(\overline Z_{g,J})$ declines substantially as $J$ increases.
Specifically, relative to the representative-contract
$\widehat{\mathrm{CVaR}}_{0.05}^{+}(Z_g)=758.28$ in
Table~\ref{tab:mbg_basecase_compare}, the estimated approximate pooled
per-contract counterpart
$\widehat{\mathrm{CVaR}}_{0.05}^{+,\mathrm{a}}
(\overline Z_{g,J})$ falls to $92.44$ at $J=1000$ and to $75.56$ at
$J=20{,}000$.

{\dpurple{At $\lambda=0.05$, Table~\ref{tab:mbg_pooling_approx} shows that the
estimated approximate equivalent load $\widehat f_{g,J}^{\mathrm{a}}$ falls
from $0.109$ at $J=1$ to $0.075$ at $J=1000$ and $0.074$ at $J=20{,}000$,
while the corresponding implied post--load starting rate
$(1-\widehat f_{g,J}^{\mathrm{a}})\beta_0$ rises from $4.46\%$ to $4.62\%$
and $4.63\%$, respectively. As $J\to\infty$, evaluating the limit implied by
\eqref{eq:pooled_mbg_load_approx} at the representative-contract estimates
gives $\widehat f_{g,J}^{\mathrm{a}}\to0.074$, with a corresponding implied
post--load starting rate of $4.63\%$.}}
This limiting load is close to the representative-contract expected-cost load
$\widehat{\mathbb E}[Z_g]/L_0=70.69/1000=0.071$, and materially below the
representative-contract load $\widehat f_g=0.109$ reported in
Table~\ref{tab:mbg_basecase_compare}.
The limiting load remains slightly above the expected-cost load because the
prudential buffer is applied to the entire pooled per-contract CVaR. As
$J\to\infty$, the approximate pooled CVaR converges to $\mathbb E[Z_g]$, so the
buffer converges to $\lambda\mathbb E[Z_g]$ rather than to zero.

\section{Conclusion and future work}
\label{sc:conclude}
This paper develops a pricing framework for a
return-of-purchase-price MBG overlay on an individual-account tontine.
The representative-member EW--CVaR optimization problem incorporates dynamic
withdrawals and rebalancing across domestic and foreign assets, with mortality
credits determined under either deterministic or stochastic mortality.
We use NN parameterizations to compute the wealth- and time-dependent withdrawal and rebalancing controls.
Under the fixed-horizon plan-to-live convention, the decumulation policy is
determined first and the MBG is then valued ex post through an
implementation-agnostic equivalent up-front load combining the expected payout
under the physical measure with an upper-tail CVaR prudential buffer.
{\dpurple{The resulting representative-contract payout distribution is then
used to approximate the effect of contract pooling on per-contract tail risk
and pricing.}}

{\dpurple{The central MBG result is that expected payouts are modest, while the
representative-contract payout has a severe upper tail. Under the pooling
approximation, diversification of contract-specific death-timing risk
substantially reduces the excess-tail component and lowers 
the approximate pooled-contract load. International diversification materially improves the EW--CVaR retirement-income trade-off, with the largest frontier gains when
combined with the tontine overlay. Stochastic mortality has only a modest
effect on the efficient frontier, while both international diversification and
stochastic mortality have limited effects on representative-contract MBG
pricing.}}

Natural extensions include modelling death as a stopping time with
explicit bequest preferences; allowing the MBG terms and funding mechanism to
feed back into the decumulation problem; {\dpurple{developing a direct
portfolio-level CVaR calculation accounting for cross-contract dependence;}}
incorporating regulatory constraints and heterogeneous pool members; and
testing broader asset, currency-hedging, return, and mortality specifications.

\section*{Appendices}
\appendix
\section{NN implementation and transfer learning}
\label{app:nn_reproducibility}
Separate fully connected feedforward networks parameterize the
withdrawal and rebalancing controls. The networks use the same hidden-layer
architecture throughout the numerical experiments; only the dimension of the
rebalancing output changes with the number of assets. The principal
implementation settings are summarized in
Table~\ref{tab:NN_training_architecture}.

\begin{table}[!htb]
\vspace*{-0.45cm}
\centering
\caption{Common NN architecture and principal training parameters.}
\label{tab:NN_training_architecture}
\small
\setlength{\tabcolsep}{5pt}
\renewcommand{\arraystretch}{1.02}
\begin{tabular}{@{}ll@{}}
\hline
{\apurple{\textbf{Specification}}}
&
{\apurple{\textbf{Value}}}
\\
\hline
{\apurple{Inputs}}
&
{\apurple{Standardized wealth and time-to-go}}
\\
{\apurple{Hidden architecture}}
&
{\apurple{2 layers, 10 sigmoid units per layer}}
\\
{\apurple{Output maps}}
&
{\apurple{Scaled sigmoid (withdrawal); softmax (rebalancing)}}
\\
{\apurple{Optimization}}
&
{\apurple{Adam}}
\\
{\apurple{Mini-batch size}}
&
{\apurple{$1000$}}
\\
{\apurple{Initial learning rates}}
&
{\apurple{$0.04$--$0.05$ for the NN parameters and CVaR threshold $W$}}
\\
{\apurple{Multistep schedule}}
&
{\apurple{Yes}}
\\
{\apurple{Maximum iterations}}
&
{\apurple{$30{,}000$}}
\\
{\apurple{Retained iterate}}
&
{\apurple{Best empirical objective attained during training}}
\\
{\apurple{Training / independent test / pricing paths}}
&
{\apurple{$N=256{,}000$ / $256{,}000$ / $K=256{,}000$}}
\\
\hline
\end{tabular}
\vspace*{-0.25cm}
\end{table}

{\apurple{\paragraph{Transfer learning across the $\gamma$-grid.}
The efficient frontiers are computed over the ordered grid
$\gamma\in\{0.05,\allowbreak 0.10,\allowbreak 0.15,\allowbreak 0.18,\allowbreak
0.20,\allowbreak 0.50,\allowbreak 0.80,\allowbreak 1,\allowbreak 1.5,\allowbreak
2,\allowbreak 5,\allowbreak 10,\allowbreak 50,\allowbreak 10^{11}\}$.
At smaller values of $\gamma$, the objective places relatively greater weight
on expected withdrawals. As $\gamma$ increases, greater weight is placed on
lower-tail CVaR, whose empirical contribution is determined by only the lowest
$\alpha$ fraction of terminal-wealth outcomes. Training the more tail-focused
problems directly from a cold start can therefore be less stable.}}

{\apurple{Following the sequential transfer-learning procedure in
\cite{chen2023benchmark}, the initial low-$\gamma$ grid points are trained from
cold starts. Thereafter, the trained weights and biases at one value of
$\gamma$ are used to initialize the optimization at the next value. In the
reported four--asset experiments, the first four grid points use cold starts,
and transfer learning is applied subsequently. At every grid point, all NN
parameters and the CVaR threshold $W$ are re-optimized; transfer learning
therefore provides only the initialization and does not constrain the solution
at the new value of $\gamma$. The independent out-of-sample test set is
generated using a different random seed from the training sample.}}

\section{\dpurple{Benchmark validation}}
\label{app:benchmark_validation}
\label{sec:validation_model_data}
This {\dpurple{appendix}} provides a benchmark validation of the NN implementation by
reproducing the synthetic-market efficient frontiers of
\cite{forsyth2024optimal}. That paper solves the two-asset,
deterministic-mortality EW--CVaR decumulation problem, with and without a
tontine overlay, using a dynamic-programming/PIDE method. The purpose of this
benchmark exercise is to verify that the NN-parameterized policy approach in
Section~\ref{sec:NNs} reproduces the reference frontiers in a setting where a
PIDE-based benchmark is available.

Accordingly, we adopt the benchmark specification of
\cite{forsyth2024optimal}: a broad equity index and a short-term government
bond index, modelled by correlated double-exponential jump-diffusion processes,
together with deterministic mortality from the 2014 Canadian Pensioner
Mortality Table (CPM2014). In the notation of
Section~\ref{ssc:mortality_models}, CPM2014 supplies a fixed deterministic
sequence of age-specific one-year conditional death probabilities
$\{\delta_{m-1}\}_{m=1}^M$; hence
$\delta_{m-1}^{(n)}\equiv\delta_{m-1}$ on every training path. The full
synthetic-market dynamics and calibration are reported {\dpurple{below}} for
reproducibility.

\subsection{Benchmark scenario}
\label{subsec:scenario}
The base-case retirement scenario mirrors the specification in Section~11 and
Table~3 of \cite{forsyth2024optimal}. Unless otherwise stated, monetary
quantities in this {\dpurple{appendix}} are expressed in thousands of real
dollars, where ``real'' means inflation-adjusted. A concise summary is given in
Table~\ref{tab:scenario}.
\begin{table}[!htb]
  \centering
  \caption{Base-case retirement scenario used for benchmark validation, matching
  the specification in \cite{forsyth2024optimal}. Monetary units are thousands
  of real dollars.}
  \label{tab:scenario}
  \begin{tabular}{ll}
    \hline
    Item & Value \\
    \hline
    Retiree & 65-year-old Canadian male \\
    Mortality table & CPM2014, deterministic life table \\
    Investment horizon $T$ & 30 years \\
    Equity index & Real CRSP capitalization-weighted total return index \\
    Bond index & Real US 30-day T-bill index \\
    Initial wealth $W_0$ & 1000 \\
    Rebalancing / withdrawal times & $t = 0,1,\ldots,29$ (annual) \\
    Minimum annual withdrawal $q_{\min}$ & 40 \\
    Maximum annual withdrawal $q_{\max}$ & 80 \\
    Equity fraction range & $p_m \in [0,1]$ \\
    Borrowing spread $\mu_b^c$ & 0.02 \\
    Mortality-credit rule &
    Homogeneous large-pool representative-member rule\\
    & in \eqref{eq:tontine-gain-rule-j}--\eqref{eq:tontine-gain}
    (Section~\ref{ssc:tontine}) \\
    Tontine fee $\varrho$ &
    $1-e^{-0.005}\approx0.00499$ (49.9 bps/year)\footnote{This matches the
    continuously charged $0.5\%$ per annum tontine fee used in
    \cite{forsyth2024optimal} under annual decision times.} \\
    Risk tail level $\alpha$ & 0.05 \\
    Stabilization parameter $\epsilon$ & $-10^{-4}$ \\
    Market parameters & {\dpurple{Table~\ref{tab:calibration}}} \\
    \hline
  \end{tabular}
\vspace*{-0.5cm}
\end{table}

In this setting, the investor controls annual withdrawals
$q_m\in[q_{\min},q_{\max}]$ and the equity fraction $p_m\in[0,1]$. If required
withdrawals exhaust the account, the portfolio is liquidated and subsequent
withdrawals are financed as debt growing at the bond rate plus the borrowing
spread $\mu_b^c$. Mortality affects the account only through deterministic
mortality credits based on CPM2014, so all synthetic-market training paths share
the same mortality profile.

\subsection{Synthetic-market asset dynamics}
\label{ssc:asset_dynamics}
We restrict attention to the two domestic assets
used in \cite{forsyth2024optimal}: a broad domestic equity index and a
constant-maturity domestic short-term government bond index. The benchmark uses
the same correlated double-exponential jump-diffusion model as
\cite[Section~4]{forsyth2024optimal}.

Let $\{S_t\}_{0\le t\le T}$ denote the real (inflation-adjusted) value invested
in the domestic stock index at time $t$, and let $\{B_t\}_{0\le t\le T}$ denote
the real (inflation-adjusted) value invested in the domestic bond index. In the
absence of investor actions, both assets evolve according to correlated
double-exponential jump-diffusion processes.

Let $\xi_s$ be the stock-index jump multiplier, so that a jump at time $t$
changes the index value according to $S_t=\xi_s S_{t^-}$. We assume
$\ln(\xi_s)$ has the asymmetric double-exponential density
\begin{equation}
\label{eq:density_S}
\varphi_s(y) = \zeta_s \,\eta_{s,1} \, e^{-\eta_{s,1} y} \, \mathbf{1}_{y \ge 0}
+ (1 - \zeta_s) \,\eta_{s,2} \, e^{\eta_{s,2} y} \, \mathbf{1}_{y < 0},
\quad
\zeta_s \in [0,1],\quad \eta_{s,1} > 1,\quad \eta_{s,2} > 0.
\end{equation}
Between rebalancing times, and in the absence of active control, the stock index
process evolves as
\begin{equation}
\label{eq:dynamics_S}
\frac{dS_t}{S_{t^-}} =
\left(\mu_s - \lambda_s\, \kappa_s \right) dt
+ \sigma_s\, dZ_s(t)
+ d\left( \sum_{i=1}^{\pi_s(t)} (\xi_{s,i} - 1) \right),
\quad t \in [t_{m-1}^+,\, t_m^-],\quad t_{m-1} \in \mathcal{T}.
\end{equation}
Here, $\mu_s$ and $\sigma_s$ are the real (inflation-adjusted) drift and
instantaneous volatility, $\{Z_s(t)\}_{t\in[0,T]}$ is a standard Brownian motion,
and $\{\pi_s(t)\}_{0\le t\le T}$ is a Poisson process with intensity
$\lambda_s>0$. The jump amplitudes $\{\xi_{s,i}\}_{i=1}^{\infty}$ are i.i.d.\
with the same distribution as $\xi_s$. The compensated drift adjustment is
\begin{equation}
\label{eq:kappa_S}
\kappa_s = \mathbb{E}[\xi_s - 1]
= \frac{\zeta_s \, \eta_{s,1}}{\eta_{s,1} - 1}
+ \frac{(1 - \zeta_s)\, \eta_{s,2}}{\eta_{s,2} + 1} - 1.
\end{equation}
The Brownian motion, Poisson process, and jump multipliers are mutually
independent.

Similarly, between rebalancing times, the bond index process evolves as
\begin{equation}
\label{eq:dynamics_B}
\frac{dB_t}{B_{t^-}} =
\big( \mu_b - \lambda_b \kappa_b
+ \mu_b^c \mathbf{1}_{\{B_{t^-} < 0\}} \big)\, dt
+ \sigma_b \, dZ_b(t)
+ d\bigg( \sum_{i=1}^{\pi_b(t)} (\xi_{b,i} - 1) \bigg),
\quad t \in [t_{m-1}^+,\, t_m^-],\quad t_{m-1} \in \mathcal{T}.
\end{equation}
The parameters in \eqref{eq:dynamics_B} are defined analogously to those in
\eqref{eq:dynamics_S}. The term
$\mu_b^c\mathbf{1}_{\{B_{t^-}<0\}}$ applies the borrowing spread
$\mu_b^c\ge0$ when the bond amount is negative. The bond jump multiplier
$\xi_b$ satisfies
\begin{equation}
\label{eq:density_Ab}
\varphi_b(y) = \zeta_b \,\eta_{b,1} \, e^{-\eta_{b,1} y} \, \mathbf{1}_{y \ge 0}
+ (1 - \zeta_b) \, \eta_{b,2} \, e^{\eta_{b,2} y} \, \mathbf{1}_{y < 0},
\quad
\zeta_b \in [0, 1],\quad \eta_{b,1} > 1,\quad \eta_{b,2} > 0.
\end{equation}
The diffusion components are correlated, $dZ_s(t)\,dZ_b(t)=\rho_{sb}\,dt$,
while the jump processes and jump sizes are mutually independent and independent
of the Brownian motions.

\subsection{Synthetic-market calibration}
\label{subsec:synthetic_calibration}
Following \cite{forsyth2024optimal}, the synthetic market is calibrated to
monthly real (inflation-adjusted) total-return series for the CRSP
value-weighted equity index and the CRSP US 30-day T-bill index over
1926:1--2020:12, with both series deflated by the US CPI. We do not re-estimate
these parameters. Instead, the benchmark validation uses the annualized real
(inflation-adjusted) parameter values reported in Table~1 of
\cite{forsyth2024optimal}, reproduced in Table~\ref{tab:calibration}.
\vspace*{-0.25cm}
\begin{table}[!htb]
  \centering
  \caption{Annualized real (inflation-adjusted) parameter values for the
  double-exponential jump-diffusion model, taken from
  \cite{forsyth2024optimal}. The stock asset is the CRSP value-weighted total
  return index; the bond asset is the 30-day US T-bill index, both deflated by
  CPI.}
  \label{tab:calibration}
  \begin{tabular}{lccccccc}
  \hline
  Stock (CRSP) & $\mu_s$ & $\sigma_s$ & $\lambda_s$ & $\zeta_s$
  & $\eta_{s, 1}$ & $\eta_{s, 2}$ & $\rho_{sb}$ \\
  & 0.08912 & 0.1460 & 0.3263 & 0.2258 & 4.3625 & 5.5335 & 0.08420 \\
  \hline
  30-day T-bill & $\mu_b$ & $\sigma_b$ & $\lambda_b$ & $\zeta_b$
  & $\eta_{b, 1}$ & $\eta_{b, 2}$ & $\rho_{sb}$ \\
  & 0.00460 & 0.0130 & 0.5053 & 0.3958 & 65.801 & 57.793 & 0.08420 \\
  \hline
  \end{tabular}
\end{table}

\subsection{Validation results}
\label{ssc:valid}
We validate our NN implementation by reproducing the synthetic-market
EW--CVaR efficient frontiers reported in \cite{forsyth2024optimal}, both with
and without a tontine overlay. Figure~\ref{fig:validation_frontier} compares
the PIDE-based reference frontiers from \cite{forsyth2024optimal} with the NN
frontiers obtained from our training procedure, together with the
constant-withdrawal/constant-allocation benchmark.

For $\alpha=0.05$, the horizontal axis reports the lower-tail
$\mathrm{CVaR}_{0.05}(W_T)$, while the vertical axis reports the expected
annualized withdrawal
$\mathbb{E}[\sum_{m=0}^{M}q_m]/T$; larger values on both axes are preferred.
The constant-policy benchmark uses an annual withdrawal of $40$ and a constant
equity allocation of $10\%$, with the remaining wealth invested in bonds and
with no tontine overlay.

\vspace*{-0.5cm}
\begin{minipage}[t]{0.5\textwidth}
\strut\vspace*{-\baselineskip}\newline
\flushleft
    \centering
    \includegraphics[width=0.85\textwidth
    ]{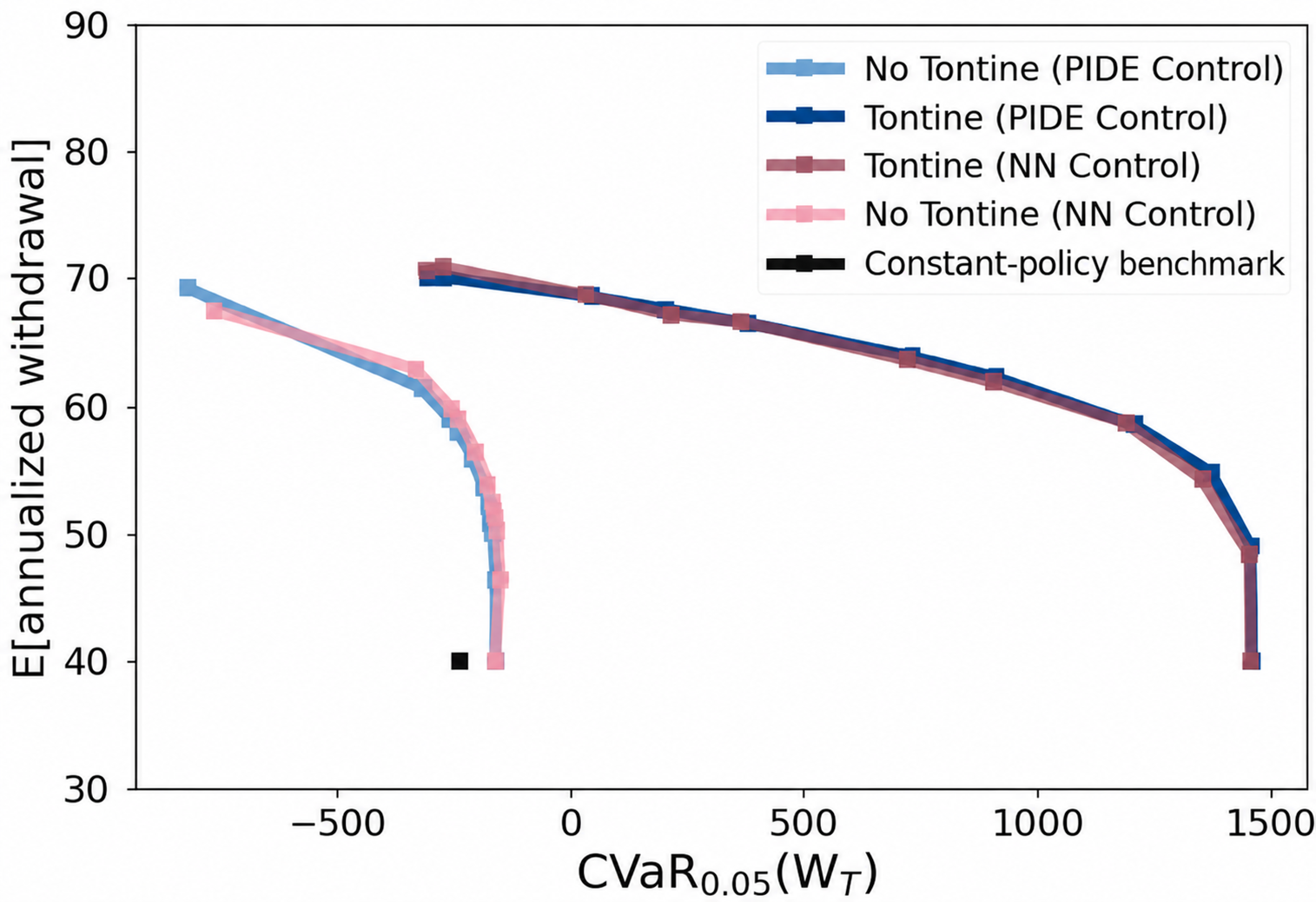}
\end{minipage}
~~
\begin{minipage}[t]{0.45\textwidth}
\strut\vspace*{-\baselineskip}\newline
\flushright
    \captionof{figure}{Validation of the EW--CVaR efficient frontiers in the
    synthetic market. Reference frontiers with and without a tontine overlay,
    labelled respectively as ``No Tontine (PIDE Control)'' and
    ``Tontine (PIDE Control)'', are compared with the corresponding NN
    frontiers, together with the constant withdrawal/constant allocation
    benchmark. Units: thousands of real dollars.}
    \label{fig:validation_frontier}
\end{minipage}

\vspace*{+0.1cm}
Overall, the NN approach produces EW--CVaR efficient frontiers in close
agreement with the PIDE-based reference frontiers and preserves their
qualitative structure. Across the range of scalarization parameters $\gamma$
used to trace the risk--reward trade-off, the NN frontiers closely track the
reference curves in both the tontine and no-tontine cases.

Beyond pointwise agreement, the NN frontiers preserve the expected qualitative
behavior as $\gamma$ varies: as risk aversion increases, the expected annualized
withdrawal decreases, while the lower-tail
$\mathrm{CVaR}_{0.05}(W_T)$ improves. The substantial improvement in the
risk--reward trade-off generated by the tontine overlay is likewise preserved.

\section{Subsections~\ref{ssc:data} and \ref{ssc:diverse_mortality}:
supplementary details}
\label{app:diversified_experiments}
This appendix provides supplementary details on data construction,
mortality calibration, and NN control comparisons for the internationally
diversified portfolio experiments in Section~\ref{sec:diverse_port}.

\subsection{Subsection~\ref{ssc:data}: Construction of the asset-return and CPI series}
\label{app:diversified_data}
\paragraph{Domestic equity.}
As a proxy for the Australian equity market, we construct a capitalization--weighted total return series for the All Ordinaries index. Monthly price levels are taken from the Bloomberg All Ordinaries Price Index (AS30) and combined with trailing dividend yield data from \cite{Mathews2019AustEquities} and the Reserve Bank of Australia (RBA) to impute reinvested dividends prior to the start of the Bloomberg All Ordinaries Accumulation Index (XAOA) in 1999. From 1999 onward, the constructed series is aligned to XAOA, so that the resulting index is a consistent long--horizon Australian equity total return series.

\vspace*{-0.5cm}
\paragraph{Domestic bonds.}
The domestic bond asset is represented by a 10--year Commonwealth Government bond total return index constructed from two sources. From December 1935 to December 2011 we use the annual Australian government bond total return series of \cite{jorda2019rate}, which is based on micro--level Commonwealth bond data targeting a 10--year maturity \cite{chen2022documentation}. We convert this annual total return index to monthly frequency by interpolating the level of the total return index and then computing month--on--month returns. From May 2011 onward we splice in the S\&P/ASX Australian Government Bond 7--10 Year Total Return Index (Bloomberg), available at a monthly frequency, to obtain a single long--horizon nominal total return bond index for the Australian market.

\vspace*{-0.35cm}
\paragraph{U.S.\ equity and bonds.}
For the U.S.\ market,  we start from the same CRSP nominal total return series as in Subsection~\ref{subsec:synthetic_calibration}, namely the value--weighted equity index and the 30--day Treasury bill (T--bill) index, both in USD. These nominal USD returns are first converted into AUD using the end--of--month AUD/USD exchange rate, constructed from Federal Reserve Banking and Monetary Statistics archives (January 1935--December 1968) and the RBA historical exchange rate series thereafter, spliced to form a continuous monthly AUD/USD series, and are then placed on the same real AUD basis as the domestic assets via CPI deflation (see below).

\vspace*{-0.35cm}
\paragraph{Inflation and real returns.}
Inflation is measured by the Australian Consumer Price Index published by the RBA. The CPI series is obtained entirely from the RBA: from September 2017 onward it is available monthly, while before that date it is published quarterly. We treat the quarterly CPI as the benchmark series and linearly interpolate it to monthly frequency to match the financial data. For each asset we form a nominal total return index, convert U.S.\ series to AUD where appropriate, and then obtain real returns by deflating with the interpolated Australian CPI. In this way all four assets are expressed in real AUD terms, consistent with the objective of funding real retirement spending for an Australian investor.

\subsection{Subsection~\ref{ssc:diverse_mortality}: Mortality data and
calibration details}
\label{app:diversified_mortality_calibration}

For stochastic mortality, we work with the same HMD source but use the raw
deaths and exposures required for model calibration.
{\red{Specifically, we extract annual Australian male deaths $D_{a,y}$ and
central exposures-to-risk $E^c_{a,y}$ for ages $a=55,\ldots,95$ and calendar
years $y=1980,\ldots,2021$}} from the HMD tables
\texttt{Deaths\_1x1.txt} and \texttt{Exposures\_1x1.txt}.%
\footnote{Available at
\url{https://www.mortality.org/File/GetDocument/hmd.v6/AUS/STATS/Deaths_1x1.txt}
and
\url{https://www.mortality.org/File/GetDocument/hmd.v6/AUS/STATS/Exposures_1x1.txt}.}

{\red{Following \cite{villegas2018stmomo}, the central exposures are converted
to initial exposures using
$E^0_{a,y}\approx E^c_{a,y}+0.5D_{a,y}$, with empirical one--year death
probabilities given by $q_{a,y}=D_{a,y}/E^0_{a,y}$. Both LC and CBD are then
estimated under the Binomial--logit specification:
$D_{a,y}\sim
\mathrm{Binomial}\bigl(E^0_{a,y},q_{a,y}\bigr)$, with $\mathrm{logit}\,q_{a,y}=\eta_{a,y}$.}}
{\red{Consequently, the fitted and simulated model outputs are one--year death
probabilities directly; no conversion from central death rates to death
probabilities is applied.}}

These historical mortality series are used to calibrate
{\red{the LC and CBD models}} from the generalized age--period--cohort (GAPC)
family, following Subsection~\ref{ssc:mortality_models}.
Calibration and forecasting are implemented in \textsf{R} using the
\textsf{StMoMo} package \cite{villegas2018stmomo}, which casts these
specifications in the GAPC framework. Model--specific identifiability
constraints are those standard in \textsf{StMoMo}; for brevity we do not
reproduce them here, and instead refer the reader to
\cite{villegas2018stmomo} for details.

{\red{For each model, the projected period factors are simulated using the
random-walk-with-drift specifications in
\eqref{eq:lc-rw} and \eqref{eq:cbd-rw}. We generate $256{,}000$ stochastic
mortality sequences using random seed $3$. Each sequence is extracted
diagonally from the projected age--calendar-year surface, beginning with
$q_{65,2022}$ and continuing with $q_{66,2023},q_{67,2024},\ldots$.
The simulation output contains $31$ probabilities corresponding to ages
$65$--$95$; the first $M=30$ probabilities, through $q_{94,2051}$, enter the
mortality-credit and death-time calculations over the 30-year horizon.}}

\section{Subsection~\ref{ssc:mortality_model_robustness}: Mortality-model robustness}
\label{app:mortality_model_robustness}
For each scalarization parameter $\gamma$, one set of learned NN
controls is trained under LC mortality and another under CBD mortality. Both
sets are then evaluated using the same out-of-sample asset, CPI, and CBD
mortality paths, with the LC-trained controls applied without retraining. This
common evaluation specification isolates sensitivity to the mortality model
used during training. For the {\dpurple{representative-contract}} MBG
comparison, we use $\gamma=1.5$, $L_0=1000$, $\alpha_g=5\%$,
$\lambda=0.05$, and $\beta_0=0.05$.

\begin{figure}[!htb]
\centering

\begin{minipage}[t]{0.45\textwidth}
\vspace{0pt}
\centering

\includegraphics[width=0.8\linewidth]
{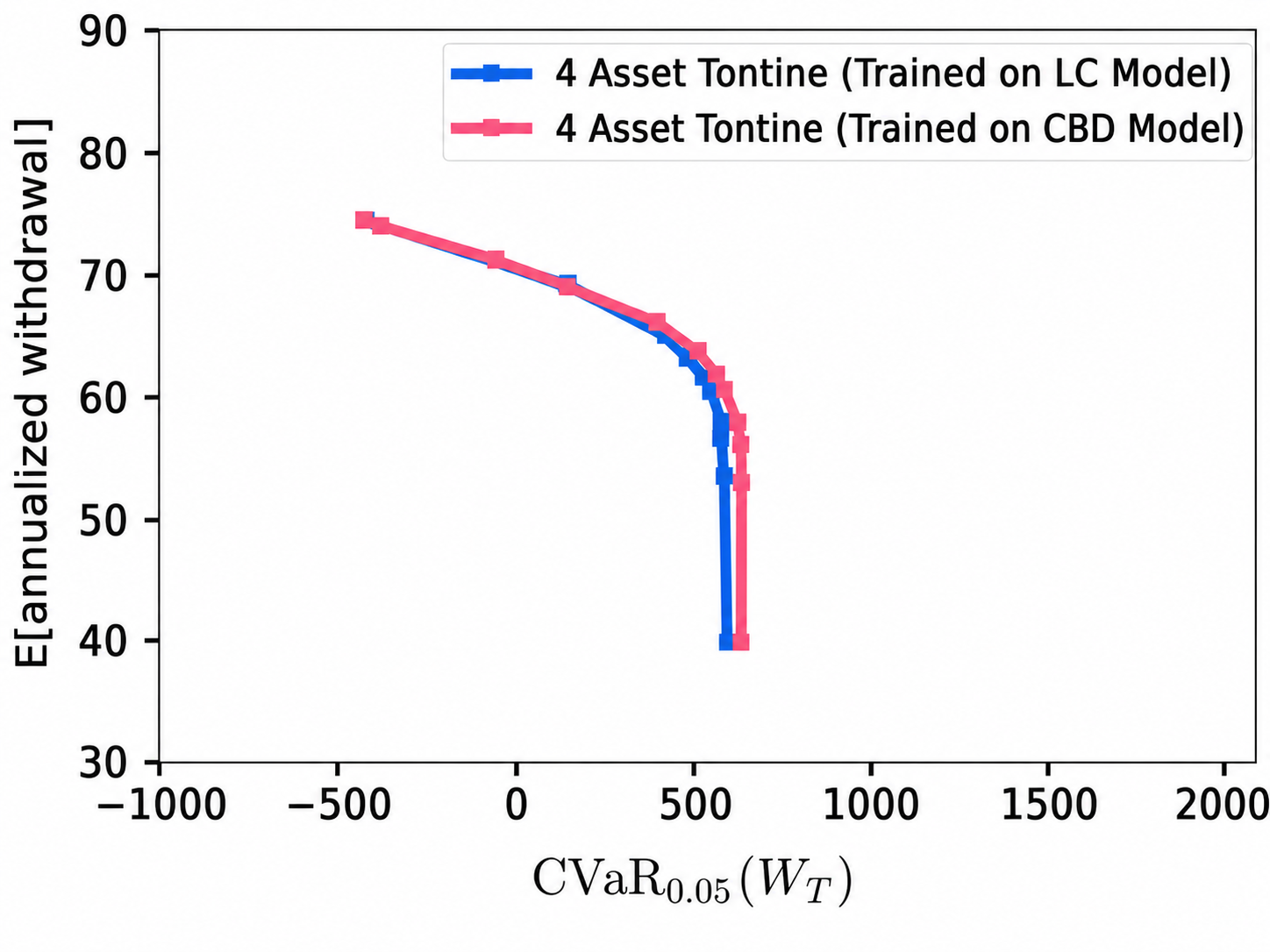}

\captionof{figure}{Mortality-model robustness of the four--asset
tontine EW--CVaR frontier under out-of-sample CBD evaluation.}
\label{fig:mortality_model_robustness}

\end{minipage}
\hfill
\begin{minipage}[t]{0.5\textwidth}
\vspace{0pt}
\centering

\begingroup
\setlength{\abovecaptionskip}{0pt}
\setlength{\belowcaptionskip}{4pt}

\captionof{table}{{\dpurple{Representative-contract MBG pricing}} under
out-of-sample CBD evaluation at $\gamma=1.5$.}
\label{tab:mortality_model_robustness}

\setlength{\tabcolsep}{3.5pt}
\begin{tabular}{|l|c|c|c|c|}
\hline
\shortstack{Training\\model}
& $\widehat{\mathbb{E}}[Z_g]$
& $\widehat{\mathrm{CVaR}}^{+}_{0.05}(Z_g)$
& $\widehat f_g$
& $(1-\widehat f_g)\beta_0$ \\
\hline
LC  & $75.49$ & $760.79$ & $0.114$ & $4.43\%$ \\
CBD & $74.71$ & $759.98$ & $0.113$ & $4.44\%$ \\
\hline
\end{tabular}

\endgroup
\end{minipage}

\end{figure}

Figure~\ref{fig:mortality_model_robustness} shows that the two
frontiers are nearly coincident. Table~\ref{tab:mortality_model_robustness}
similarly shows differences of less than one thousand real AUD in both the
expected MBG payout and its upper-tail CVaR. Because the same evaluation paths
are used, the differences can also be assessed on a paired basis. The paired
LC-minus-CBD differences are $0.786\;(0.012)$ for
$\widehat{\mathbb{E}}[Z_g]$, $0.810\;(0.050)$ for
$\widehat{\mathrm{CVaR}}^{+}_{0.05}(Z_g)$,
$8.26\times10^{-4}\;(1.31\times10^{-5})$ for $\widehat f_g$, and
$-0.413\;(0.007)$ basis points for the implied starting payment
rate, where Monte Carlo standard errors are shown in parentheses. Thus,
although the small differences are estimated precisely, they are economically
negligible. {\dpurple{These results support robustness of both the
retirement-income frontiers and the representative-contract MBG valuation to
the choice between LC and CBD as the training mortality model under the common
CBD evaluation.}}

\section{Subsection~\ref{ssc:effect_mortality}: Learned control heatmaps}
\label{ssc:heatmaps}
\begin{figure}[!htb]
  \centering
  \begin{minipage}{0.48\textwidth}
    \centering
    \includegraphics[width=0.9\textwidth]{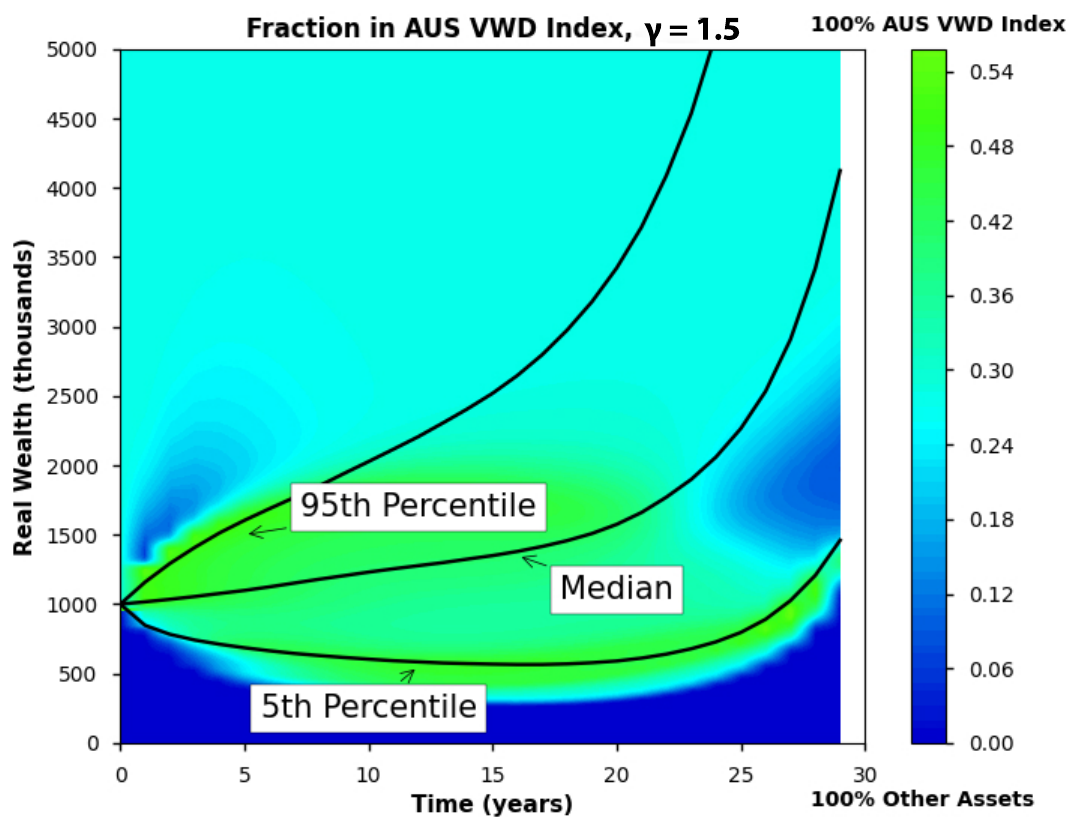}
    \caption*{(a) Australian equity index (domestic)}
  \end{minipage}
  \hfill
  \begin{minipage}{0.48\textwidth}
    \centering
    \includegraphics[width=0.9\textwidth]{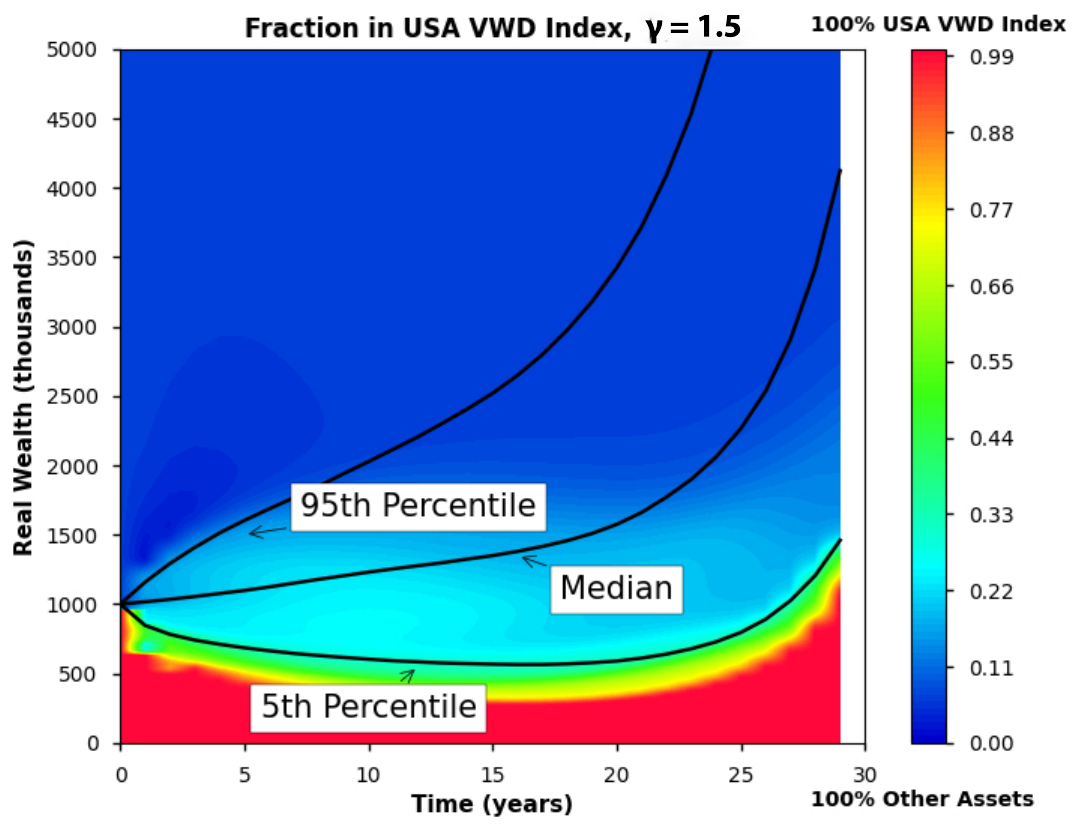}
    \caption*{(b) U.S.\ equity index}
  \end{minipage}

  \medskip

  \begin{minipage}{0.48\textwidth}
    \centering
    \includegraphics[width=0.9\textwidth]{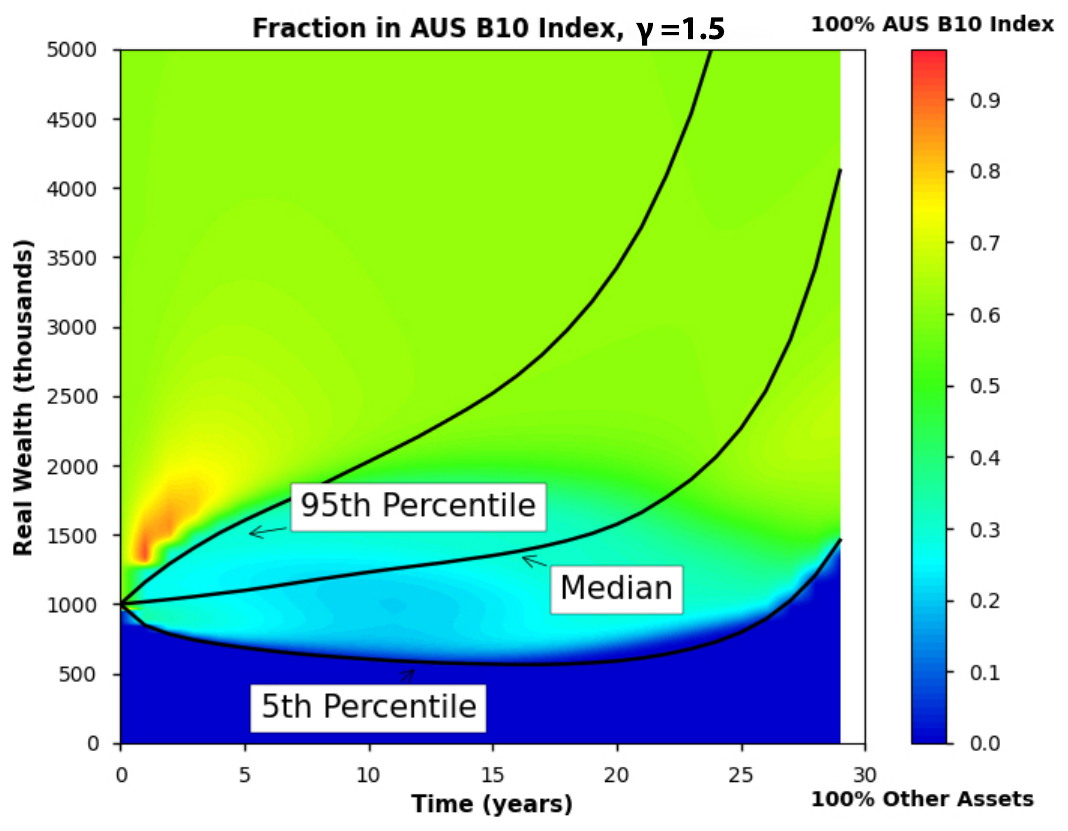}
    \caption*{(c) Australian 10--year government bond index (domestic)}
  \end{minipage}
  \hfill
  \begin{minipage}{0.48\textwidth}
    \centering
    \includegraphics[width=0.9\textwidth]{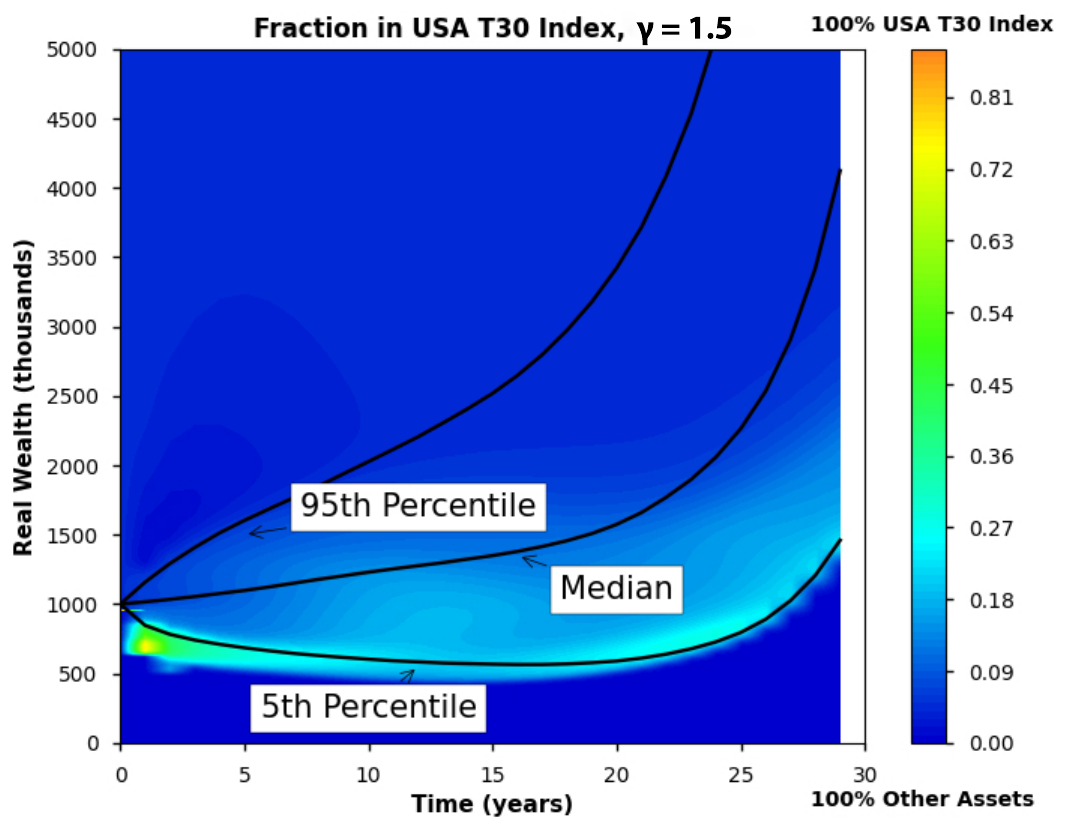}
    \caption*{(d) U.S.\ 30--day T--bill index}
  \end{minipage}

  \caption{Learned rebalancing controls: the four--asset tontine with stochastic
  mortality (LC) for $\gamma = 1.5$. Colours show the fraction of wealth invested in each asset as a function of time
  and real wealth.}
  \label{fig:heatmaps_four_asset}
\vspace*{-0.25cm}
\end{figure}

To interpret the retirement-outcome gains from international diversification, Figure~\ref{fig:heatmaps_four_asset}
plots the learned rebalancing controls for the four--asset tontine with stochastic
mortality (LC model) and a representative scalarization parameter
$\gamma = 1.5$. Each panel shows, as a function of time and real wealth, the
fraction of wealth invested in one of the four indices described in
Subsection~\ref{ssc:data}, together with the 5th, 50th, and 95th percentiles of the
wealth distribution under the learned policy. Control heatmaps under deterministic table mortality and the CBD mortality model are qualitatively very similar and are therefore omitted for brevity.

We make several qualitative observations from Figure~\ref{fig:heatmaps_four_asset}:
\begin{itemize}[noitemsep,topsep=2pt,leftmargin=*]

\item
Along the median and upper wealth trajectories, the strategy exhibits a clear
tilt towards domestic assets: a substantial fraction of wealth is invested in
Australian securities, with the Australian 10--year government bond
(Figure~\ref{fig:heatmaps_four_asset}~(c)) acting as the main stabilising asset. Specifically, for moderate
to high wealth levels, the allocation to the domestic bond is large (typically
well above 50\%), while the U.S.\ 30--day T--bill (Figure~\ref{fig:heatmaps_four_asset}~(d)) is used only
marginally.
This aligns with Table~\ref{tab:return_stats_summary}: the domestic bond index has
substantially lower volatility and markedly milder 5\% monthly tail losses
(VaR/CVaR) than the U.S.\ T--bill once returns are expressed in real AUD, supporting
its role as the primary defensive allocation in the learned policy.

\medskip
Along the 95th--percentile path the equity allocations
(Figures~\ref{fig:heatmaps_four_asset}~(a) and~(b)) are gradually reduced, so that the portfolio becomes
dominated by the Australian bond index, effectively locking in favourable
outcomes; any remaining equity risk is taken primarily via the domestic
market.

\item
The U.S.\ equity allocation (Figure~\ref{fig:heatmaps_four_asset}~(b)) is highly
state dependent. When wealth falls into the lower region of the state space,
particularly along and below the 5th--percentile path, the
learned control assigns a high weight to U.S.\ equity, with weights
close to $100\%$ in that asset and very low exposure to domestic bonds. This
behaviour is concentrated in the extreme low-wealth tail and can be interpreted
as a state-dependent high-growth allocation in a low-probability tail
region, rather than a typical allocation across the state space. When wealth
is low, the learned control allocates more to growth assets
in order to improve the expected withdrawal profile, with foreign equity
providing the main source of additional growth exposure.

\smallskip
\noindent
Tables~\ref{tab:return_stats_summary}--\ref{tab:return_stats_corr} provide a
simple empirical rationale: U.S.\ equity combines the highest historical average
real return in the sample with only moderate correlation to Australian equity,
so it offers both growth upside and diversification potential when deployed as
a state-dependent source of growth exposure in low-wealth regions.
This is consistent with related optimal--decumulation evidence that moderate
caps on the equity share have little effect on the efficient frontier, with the
main control differences concentrated in extreme low-wealth tail states
\cite{forsythnumerical}.

\item
Comparing Figures~\ref{fig:heatmaps_four_asset}~(a) and~(b) shows that Australian and U.S.\ equities play
different roles. At moderate wealth, the strategy holds a sizeable but
balanced allocation to Australian equities, while U.S.\ equity exposure is
more concentrated in low--wealth regions of the state space. This suggests that the learned control uses domestic equity as the
core growth asset, while foreign equity provides a more concentrated source of
growth exposure in low--wealth regions of the state space.
\end{itemize}

Taken together, the four--asset heatmaps show that international
diversification changes how equity exposure is allocated across assets and
wealth states, rather than producing a uniformly higher total equity allocation.
Australian equity remains the core growth asset at moderate wealth, while
U.S.\ equity provides a more concentrated source of growth exposure in
low--wealth regions.

\begin{figure}[hbt!]
\centering
\begin{minipage}[t]{0.5\textwidth}
\vspace{0pt}
\centering
  \includegraphics[width=0.85\textwidth]{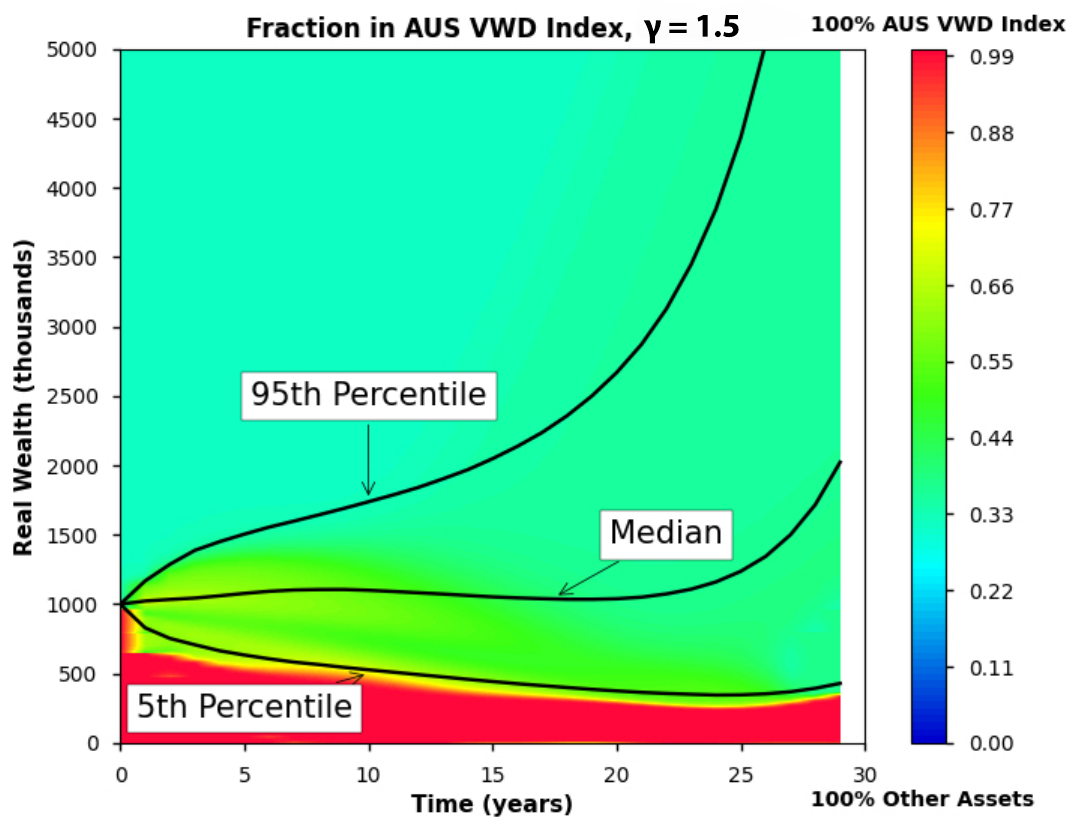}
\end{minipage}
\hfill
\begin{minipage}[t]{0.48\textwidth}
\vspace{0pt}
\raggedright
\begingroup
\setlength{\abovecaptionskip}{0pt}
\captionof{figure}{Learned fraction invested in domestic equity in the
  two--asset domestic-only tontine under stochastic mortality (LC model), for
  the representative frontier point $\gamma=1.5$. Colours show the allocation
  as a function of time and real wealth.}
  \label{fig:heatmap_two_asset}
\endgroup
\end{minipage}
\end{figure}

\vspace*{-0.25cm}
\paragraph{Two-- versus four--asset equity allocation.}
\label{app:two_asset_control_comparison}
To complement the four--asset controls in
Figure~\ref{fig:heatmaps_four_asset}, Figure~\ref{fig:heatmap_two_asset} reports the
learned domestic-equity allocation for the corresponding two--asset
domestic-only tontine under stochastic LC mortality.

Figure~\ref{fig:heatmap_two_asset} shows that, in the domestic-only
case, the equity allocation is primarily wealth-dependent rather than strongly
time-dependent. Along the median wealth trajectory, the learned control remains
in a moderate equity range for most of the horizon, approximately
$25\%$--$45\%$. Near and below the 5th--percentile trajectory, however, the
allocation approaches $100\%$, since domestic equity is the only available
growth asset.

{\apurple{However, as shown in
Figure~\ref{fig:heatmaps_four_asset}, once international diversification is
allowed, the growth role performed solely by domestic equity in the two--asset
setting is divided between Australian and U.S.\ equities. Australian equity
remains part of the core growth allocation at moderate wealth, whereas U.S.\
equity becomes more prominent in low--wealth regions. This control comparison
indicates that the four--asset frontier improvement reflects the
state-dependent allocation of equity exposure across domestic and foreign
markets, rather than a uniformly higher total equity exposure.}}

\section{Subsection~\ref{ssc:mbg_pricing_results}: supplementary results}
\label{app:mbg_pricing_supplement}
\subsection{Representative-contract MBG load sensitivity}
\label{sssc:mbg_sens}
\label{app:mbg_pricing_supplement_sensitivity}
We study the sensitivity of the {\dpurple{representative-contract MBG load}} to
(i) the prudential--buffer parameters $(\lambda,\alpha_g)$ and (ii) the
retiree's risk--reward preference in the EW--CVaR optimization, indexed by the
scalarization parameter~$\gamma$. Table~\ref{tab:mbg_sensitivity} presents the
results for the four--asset (internationally diversified) tontine with
stochastic mortality (LC model) across
$\gamma\in\{1.5,0.5,0.2\}$, with $L_0=1000$ fixed. For each $\gamma$, we
report the Monte Carlo estimates $\widehat{\mathbb{E}}[Z_g]$ and
$\widehat{\mathrm{CVaR}}^{+}_{\alpha_g}(Z_g)$ together with the implied load
estimates $\widehat{f}_g$ computed from~\eqref{eq:widehat_g} for
$\lambda\in\{0,0.05,0.5\}$. Monte Carlo standard errors are reported in
Table~\ref{tab:mbg_mc_se_sensitivity}. For fixed $(\gamma,\alpha_g)$,
$\widehat{f}_g$ depends linearly on $\lambda$, with slope
$\widehat{\mathrm{CVaR}}^{+}_{\alpha_g}(Z_g)/L_0$.

\begin{table}[!htbp]
\vspace*{-0.5cm}
\centering
\caption{Representative-contract MBG load sensitivity: the
four--asset tontine with stochastic mortality (LC).}
\label{tab:mbg_sensitivity}

\emph{(a)} $\alpha_g=5\%$\\
\begin{tabular}{|c|c|c|c|c|c|}
\hline
$\gamma$
& $\widehat{\mathbb{E}}[Z_g]$
& $\widehat{\mathrm{CVaR}}^{+}_{0.05}(Z_g)$
& $\widehat{f}_g$ ($\lambda=0$)
& $\widehat{f}_g$ ($\lambda=0.05$)
& $\widehat{f}_g$ ($\lambda=0.5$) \\
\hline
$1.5$
& $70.69$
& $758.28$
& $0.071$
& $0.109$
& $0.450$ \\

$0.5$
& $58.13$
& $729.37$
& $0.058$
& $0.095$
& $0.423$ \\

$0.2$
& $47.51$
& $664.39$
& $0.048$
& $0.081$
& $0.380$ \\
\hline
\end{tabular}

\vspace{2mm}

\emph{(b)} $\alpha_g=1\%$\\
\begin{tabular}{|c|c|c|c|c|c|}
\hline
$\gamma$
& $\widehat{\mathbb{E}}[Z_g]$
& $\widehat{\mathrm{CVaR}}^{+}_{0.01}(Z_g)$
& $\widehat{f}_g$ ($\lambda=0$)
& $\widehat{f}_g$ ($\lambda=0.05$)
& $\widehat{f}_g$ ($\lambda=0.5$) \\
\hline
$1.5$
& $70.69$
& $917.56$
& $0.071$
& $0.117$
& $0.529$ \\

$0.5$
& $58.13$
& $915.19$
& $0.058$
& $0.104$
& $0.516$ \\

$0.2$
& $47.51$
& $871.95$
& $0.048$
& $0.091$
& $0.483$ \\
\hline
\end{tabular}
\vspace*{-0.25cm}
\end{table}

Table~\ref{tab:mbg_sensitivity} shows that the dominant drivers of the
equivalent load are the prudential--buffer parameters
$(\lambda,\alpha_g)$. In particular, $\lambda=0$ corresponds to actuarially
fair (expected--cost) pricing and yields a modest load
$\widehat{f}_g\approx5\%$--$7\%$ across the reported $\gamma$ values (about
$24$--$35$~bps off a notional $\beta_0=5\%$ starting rate). At
$\alpha_g=5\%$, the illustrative value $\lambda=0.05$ gives loads of
approximately $0.081$--$0.109$, while the stress value $\lambda=0.5$ gives
loads of approximately $0.380$--$0.450$. Reducing the upper-tail probability
from $\alpha_g=5\%$ to $\alpha_g=1\%$ further increases the loads to
approximately $0.091$--$0.117$ for $\lambda=0.05$ and $0.483$--$0.529$ for
$\lambda=0.5$.

Varying $\gamma$ affects the load in a secondary, policy--induced
way: changing $\gamma$ changes the learned policy along the efficient frontier
and thereby shifts the distribution of $Z_g$ through the timing and magnitude
of withdrawals, which determine the cumulative nominal withdrawals made before
death. Over the ranges considered here, this effect is modest compared with
the direct impact of $(\lambda,\alpha_g)$ in \eqref{eq:widehat_g}.

The sensitivity patterns in
Table~\ref{tab:mbg_sensitivity} are not specific to the four--asset LC
setting. At $\gamma=1.5$, the two-- and four--asset portfolios under both table
and LC mortality produce similar upper-tail CVaR estimates and, by the linear
dependence of $\widehat f_g$ on $\lambda$ noted above, similar sensitivity to
$\lambda$. Moreover, reducing $\alpha_g$ from $5\%$ to $1\%$ increases
$\widehat f_g$ by approximately $0.008$--$0.009$ at $\lambda=0.05$ and by
approximately $0.08$--$0.09$ at $\lambda=0.5$ in every setting. Detailed
cross-setting results are reported in
Subsection~\ref{app:mbg_pricing_supplement_cross_setting}.

Finally, because $(\lambda,\alpha_g)$ represent prudential choices
that are generally not observable from public product disclosures, these
sensitivity results should be viewed as pricing stress tests rather than as
estimates of any particular provider's loading rule.

\subsection{Tables~\ref{tab:mbg_basecase_compare}-\ref{tab:mbg_frequency_severity} and Table~\ref{tab:mbg_sensitivity}}
\label{app:mbg_pricing_supplement_error}
All MBG pricing experiments use $K=256{,}000$ Monte Carlo
observations. Tables~\ref{tab:mbg_mc_se_basecase} and
\ref{tab:mbg_mc_se_sensitivity} report standard errors (SE) for the estimates in
Tables~\ref{tab:mbg_basecase_compare}--\ref{tab:mbg_frequency_severity} and
Table~\ref{tab:mbg_sensitivity}, respectively. For the illustrative post--load
starting rate in Table~\ref{tab:mbg_basecase_compare}, its standard error in
percentage points is $5\,\mathrm{SE}(\widehat f_g)$ when $\beta_0=0.05$.

\begin{table}[!htb]
\vspace*{-0.5cm}
\centering
\setlength{\tabcolsep}{4pt}
\caption{Monte Carlo standard errors (SE)
corresponding to estimates in Tables~\ref{tab:mbg_basecase_compare}-\ref{tab:mbg_frequency_severity}.}
\label{tab:mbg_mc_se_basecase}
\begin{tabular}{lccccc}
\hline
\textbf{Setting}
& $\mathrm{SE}(\widehat{\mathbb E}[Z_g])$
& $\mathrm{SE}(\widehat{\mathrm{CVaR}}^{+}_{0.05})$
& $\mathrm{SE}(\widehat f_g)$
& $\mathrm{SE}(\widehat{\mathbb P}(Z_g>0))$
& $\mathrm{SE}(\widehat{\mathbb E}[Z_g\mid Z_g>0])$ \\
\hline
2 assets, table mortality
& $0.36$ & $2.01$ & $4.4\times10^{-4}$
& $0.08$ & $1.3$ \\
2 assets, LC mortality
& $0.37$ & $1.96$ & $4.5\times10^{-4}$
& $0.07$ & $1.3$ \\
4 assets, table mortality
& $0.36$ & $1.96$ & $4.4\times10^{-4}$
& $0.08$ & $1.2$ \\
4 assets, LC mortality
& $0.38$ & $1.86$ & $4.5\times10^{-4}$
& $0.08$ & $1.3$ \\
\hline
\end{tabular}
\end{table}

\vspace*{-0.5cm}
\begin{table}[!htb]
\centering
\caption{Monte Carlo standard errors (SE) corresponding to estimates in
Table~\ref{tab:mbg_sensitivity}.}
\label{tab:mbg_mc_se_sensitivity}
\emph{(a)} $\alpha_g=5\%$\\
\begin{tabular}{cccccc}
\hline
$\gamma$
& $\mathrm{SE}(\widehat{\mathbb E}[Z_g])$
& $\mathrm{SE}(\widehat{\mathrm{CVaR}}^{+}_{0.05})$
& $\mathrm{SE}(\widehat f_g)$ ($\lambda=0$)
& $\mathrm{SE}(\widehat f_g)$ ($\lambda=0.05$)
& $\mathrm{SE}(\widehat f_g)$ ($\lambda=0.5$) \\
\hline
$1.5$ & $0.38$ & $1.86$
& $3.8\times10^{-4}$ & $4.5\times10^{-4}$
& $1.2\times10^{-3}$ \\
$0.5$ & $0.35$ & $2.18$
& $3.5\times10^{-4}$ & $4.4\times10^{-4}$
& $1.4\times10^{-3}$ \\
$0.2$ & $0.31$ & $2.40$
& $3.1\times10^{-4}$ & $4.2\times10^{-4}$
& $1.5\times10^{-3}$ \\
\hline
\end{tabular}

\vspace{2mm}

\emph{(b)} $\alpha_g=1\%$\\
\begin{tabular}{cccccc}
\hline
$\gamma$\vspace*{0.25cm}
& $\mathrm{SE}(\widehat{\mathbb E}[Z_g])$
& $\mathrm{SE}(\widehat{\mathrm{CVaR}}^{+}_{0.01})$
& $\mathrm{SE}(\widehat f_g)$ ($\lambda=0$)
& $\mathrm{SE}(\widehat f_g)$ ($\lambda=0.05$)
& $\mathrm{SE}(\widehat f_g)$ ($\lambda=0.5$) \\
\hline
$1.5$ & $0.38$ & $0.99$
& $3.8\times10^{-4}$ & $4.0\times10^{-4}$
& $7.3\times10^{-4}$ \\
$0.5$ & $0.35$ & $1.11$
& $3.5\times10^{-4}$ & $3.8\times10^{-4}$
& $7.7\times10^{-4}$ \\
$0.2$ & $0.31$ & $1.32$
& $3.1\times10^{-4}$ & $3.5\times10^{-4}$
& $8.5\times10^{-4}$ \\
\hline
\end{tabular}
\vspace*{-0.5cm}
\end{table}

\subsection{Sensitivity across asset and mortality specifications}
\label{app:mbg_pricing_supplement_cross_setting}
This subsection reports the {\dpurple{representative-contract}} cross-setting
sensitivity results summarized in Subsection~\ref{sssc:mbg_sens}. Holding
$\gamma=1.5$ fixed, Table~\ref{tab:mbg_cross_setting_sensitivity} compares the
upper-tail CVaR estimates and equivalent MBG loads for $\alpha_g=5\%$ and
$\alpha_g=1\%$ across the two-- and four--asset portfolios under both table
and LC mortality, using the illustrative loading coefficient $\lambda=0.05$
and the stress value $\lambda=0.5$. Standard errors are shown in parentheses.

\begin{table}[!t]
\centering
\caption{{\dpurple{Sensitivity of representative-contract MBG pricing to the
upper-tail probability across asset and mortality specifications;
$\gamma=1.5$. Standard errors are shown in parentheses.}}}
\label{tab:mbg_cross_setting_sensitivity}

\emph{(a)} $\alpha_g=5\%$\\
\begin{tabular}{lccc}
\hline
\textbf{Setting}
& $\widehat{\mathrm{CVaR}}^{+}_{0.05}(Z_g)$
& $\widehat f_g$ ($\lambda=0.05$)
& $\widehat f_g$ ($\lambda=0.5$) \\
\hline
2 assets, table mortality
& $736.62\;(2.01)$
& $0.102\;(4.4\times10^{-4})$
& $0.434\;(1.3\times10^{-3})$ \\
2 assets, LC mortality
& $755.48\;(1.96)$
& $0.104\;(4.5\times10^{-4})$
& $0.444\;(1.3\times10^{-3})$ \\
4 assets, table mortality
& $736.82\;(1.96)$
& $0.106\;(4.4\times10^{-4})$
& $0.438\;(1.3\times10^{-3})$ \\
4 assets, LC mortality
& $758.28\;(1.86)$
& $0.109\;(4.5\times10^{-4})$
& $0.450\;(1.2\times10^{-3})$ \\
\hline
\end{tabular}

\vspace{2mm}

\emph{(b)} $\alpha_g=1\%$\\
\begin{tabular}{lccc}
\hline
\textbf{Setting}
& $\widehat{\mathrm{CVaR}}^{+}_{0.01}(Z_g)$
& $\widehat f_g$ ($\lambda=0.05$)
& $\widehat f_g$ ($\lambda=0.5$) \\
\hline
2 assets, table mortality
& $910.94\;(1.07)$
& $0.111\;(3.8\times10^{-4})$
& $0.521\;(7.6\times10^{-4})$ \\
2 assets, LC mortality
& $916.91\;(0.96)$
& $0.113\;(3.9\times10^{-4})$
& $0.525\;(7.1\times10^{-4})$ \\
4 assets, table mortality
& $911.07\;(1.12)$
& $0.115\;(3.9\times10^{-4})$
& $0.525\;(7.8\times10^{-4})$ \\
4 assets, LC mortality
& $917.56\;(0.99)$
& $0.117\;(4.0\times10^{-4})$
& $0.529\;(7.3\times10^{-4})$ \\
\hline
\end{tabular}
\end{table}

Reducing $\alpha_g$ from $5\%$ to $1\%$ increases the equivalent
load by approximately $0.008$--$0.009$ when $\lambda=0.05$ and by
approximately $0.080$--$0.087$ when $\lambda=0.5$ across the four settings.
The similar upper-tail CVaR estimates also imply similar sensitivity to
$\lambda$. Thus, international diversification changes the level of the MBG
load only modestly and does not materially alter its sensitivity to either
prudential-buffer parameter.

\section{{\dpurple{Monte Carlo uncertainty for contract-pooling results}}}
\label{app:mbg_pooling_uncertainty}
{\dpurple{The estimated quantities in Table~\ref{tab:mbg_pooling_approx} are obtained
from the pooling approximation and are functions of the representative-contract
estimators $\widehat{\mathbb E}[Z_g]$ and
$\widehat{\mathrm{CVaR}}_{0.05}^{+}(Z_g)$. Using the $K=256{,}000$ pathwise
payouts for the four--asset LC base case, we estimate the variances and
covariance of these two estimators from the same simulated sample. Since the
quantities in \eqref{eq:pooled_mbg_estimates} are linear functions of these
estimators, the variances of the pooled quantities are computed directly from
the estimated variances and covariance, thereby accounting for the dependence
between the two estimators. Table~\ref{tab:mbg_pooling_mc_se} reports the
resulting standard errors for $\lambda=0.05$ and $L_0=1000$. }}
\begin{table}[!t]
\centering
\caption{{\dpurple{Monte Carlo standard errors corresponding to the estimated
approximate quantities in Table~\ref{tab:mbg_pooling_approx}. The CVaR standard
errors are in thousands of real AUD.}}}
\label{tab:mbg_pooling_mc_se}
\begin{tabular}{ccc}
\hline
{\dpurple{$J$}}
&
{\dpurple{\shortstack{
$\mathrm{SE}\!\left(
\widehat{\mathrm{CVaR}}_{0.05}^{+,\mathrm{a}}
(\overline Z_{g,J})\right)$}}}
&
{\dpurple{$\mathrm{SE}(\widehat f_{g,J}^{\mathrm{a}})$}}
\\
\hline
{\dpurple{$1$}}
& {\dpurple{$1.857$}}
& {\dpurple{$4.51\times10^{-4}$}}
\\
{\dpurple{$10$}}
& {\dpurple{$0.803$}}
& {\dpurple{$4.11\times10^{-4}$}}
\\
{\dpurple{$50$}}
& {\dpurple{$0.551$}}
& {\dpurple{$4.02\times10^{-4}$}}
\\
{\dpurple{$100$}}
& {\dpurple{$0.496$}}
& {\dpurple{$3.99\times10^{-4}$}}
\\
{\dpurple{$500$}}
& {\dpurple{$0.426$}}
& {\dpurple{$3.96\times10^{-4}$}}
\\
{\dpurple{$1000$}}
& {\dpurple{$0.410$}}
& {\dpurple{$3.96\times10^{-4}$}}
\\
{\dpurple{$10{,}000$}}
& {\dpurple{$0.386$}}
& {\dpurple{$3.94\times10^{-4}$}}
\\
{\dpurple{$20{,}000$}}
& {\dpurple{$0.383$}}
& {\dpurple{$3.94\times10^{-4}$}}
\\
{\dpurple{$J\to\infty$}}
& {\dpurple{$0.375$}}
& {\dpurple{$3.94\times10^{-4}$}}
\\
\hline
\end{tabular}
\end{table}

These Monte Carlo
standard errors quantify the sampling uncertainty inherited from the
representative-contract mean and CVaR estimates. They do not capture
approximation error associated with the $J^{-1/2}$ pooling rule or uncertainty
in the learned policy and calibrated model inputs.

\end{document}